\documentclass[11pt,a4paper]{article}
\usepackage{graphicx}
\usepackage{url}
\usepackage{wrapfig}
\usepackage{multicol}
\usepackage{multirow}
\usepackage{authblk}
\usepackage{lscape}
\usepackage{lipsum}
\usepackage{esint}

\usepackage{array}

\usepackage{type1cm}
\usepackage{eso-pic}
\usepackage{color}
\usepackage{setspace}

\makeatletter
\newcommand{\thickhline}{%
    \noalign {\ifnum 0=`}\fi \hrule height 1pt
    \futurelet \reserved@a \@xhline
}
\newcolumntype{"}{@{\hskip\tabcolsep\vrule width 1pt\hskip\tabcolsep}}
\makeatother

\usepackage{array}
\newcolumntype{C}[1]{>{\centering\arraybackslash}p{#1}}

\title{\bf A computational framework for bioimaging simulation}
\author{\small Masaki Watabe$^{1}$, Satya N. V. Arjunan$^{1}$, Seiya Fukushima$^{2,4}$, Kazunari Iwamoto$^{1}$, Jun Kozuka$^{2}$, Satomi Matsuoka$^{2}$, Yuki Shindo$^{1,4}$, Masahiro Ueda$^{2,4}$ and Koichi Takahashi\,$^{1,3,\ast}$}
\affil{\small
$^{1}$ Laboratory for biochemical simulation, RIKEN Quantitative Biology Center, Suita-shi Osaka-fu, Japan\\
$^{2}$ Laboratory for cell signaling dynamics, RIKEN Quantitative Biology Center, Suita-shi Osaka-fu, Japan\\
$^{3}$ Institute for Advanced Bioscience Keio University, Tsuruoka-shi Yamagata-ken, Japan\\
$^{4}$ Graduate School of Frontier Bioscience, Osaka University, Suita-shi Osaka-fu, Japan}
\date{\small (Accepted Date) : May 15, 2015}

\begin{document}
\maketitle

\noindent {\small Using bioimaging technology, biologists have attempted to identify and document analytical interpretations that underlie biological phenomena in biological cells. Theoretical biology aims at distilling those interpretations into knowledge in the mathematical form of biochemical reaction networks and understanding how higher level functions emerge from the combined action of biomolecules. However, there still remain formidable challenges in bridging the gap between bioimaging and mathematical modeling. Generally, measurements using fluorescence microscopy systems are influenced by systematic effects that arise from stochastic nature of biological cells, the imaging apparatus, and optical physics. Such systematic effects are always present in all bioimaging systems and hinder quantitative comparison between the cell model and bioimages. Computational tools for such a comparison are still unavailable. Thus, in this work, we present a computational framework for handling the parameters of the cell models and the optical physics governing bioimaging systems. Simulation using this framework can generate digital images of cell simulation results after accounting for the systematic effects. We then demonstrate that such a framework enables comparison at the level of photon-counting units. }
{\small {\it Doi} : 10.1371/journal.pone.0130089}\\
{\small {\it E-mail} : masaki@riken.jp}\\

\newpage

\section*{Introduction}
\paragraph{}
All scientific measurements are subject to some uncertainties. Experimental accuracy and precision must be always estimated to establish the validity of our results \cite{bevington2003, taylor1997}. It is also true for measurements using bioimaging techniques such as fluorescence microscopy. The measurements are generally influenced by systematic effects that arise from the stochastic nature of biological cells, the imaging apparatus, and optical physics. Such systematic effects are always present in all bioimaging systems and hinder the validation of the mathematical models of biological cells. For example, the local precision of reconstructed images obtained by precise localization microscopy, such as stochastic optical reconstruction microscopy (STORM), and photoactivated localization microscopy (PALM) is particularly limited by the systematic effects that are governed by camera specifications and its operating conditions \cite{huang2013, cse0, cse1}. The limitation constrains the validation of the mathematical models of the biological dynamics. 


\paragraph{}
Theory of model validation is often applied to obtain valid mathematical models in physics and engineering fields \cite{vv2007, vv2006, vv2004}. It can be also applied to biological science, because it offers a formal representation of the progressive build-up of trust in the mathematical model of interest. In a standard exercise of model validation, one performs an experiment and in parallel, runs a simulation of the model. Then, using metrics controlled by the parameters embedded in the model and the experimental configuration, the output of the model simulation is iteratively compared and analyzed with the actual experimental output. There are three important parts in the iterative process. (1) The experimental outputs are generally influenced by the systematic effects that arise from various sources in the bioimaging process. The outputs of the model simulation are usually not presented in the most efficient way for comparison with the experimental outputs. Simulations of the experimental techniques and their operating conditions are essential for proper comparison and analysis. (2) The predictive capability of the model is to go beyond the well-known parameter domain and into a new parameter domain of unknown conditions and outcome. Calibration and validation are one of the important processes of parameter adjustment in each domain. Calibration is defined as the process of improving the agreement of a set of simulated outputs with a set of actual outputs obtained under well-controlled experimental systems. Validation is defined as the process of quantifying our confidence in the predictive capability for a given application. (3) Analyses of parameter sensitivity and limitation are also important to reduce the size of the parameter domain.

\paragraph{}
In this article, we focus on the first (comparison) issue/part. In order to properly compare spatial models of biological cell with actual cell images, we propose a computational framework for managing parameter dependences by defining a uniform interface and common organizational principles governing the systematic effects. Such a framework allows us to efficiently handle the parameters defined in a spatial cell model and the physical principles governing the bioimaging techniques and their operating conditions. Using this framework, we program bioimaging simulation modules to generate digital images of the cell simulation results after accounting for the systematic effects. The intensity of the simulated images corresponds to the number of photons detected in a light-sensitive device. Thus, the framework streamlines the comparison at the level of photon-counting units. In particular, we implement the simulation modules for relatively simple microscopy systems: total internal reflection fluorescence microscopy (TIRFM) and laser-scanning confocal microscopy (LSCM). We then evaluate the performance of the simulation modules by comparing a simulated image with an actual image for simple particle models of fluorescent molecules. Therefore, these images are comparable at the level of photon-counting units. Each simulated image is visually similar to the corresponding real one. In addition, using the LSCM simulation module, we compared a more complex cell model with real cell images obtained by the actual LSCM system. We construct the following spatial cell models for the comparison: (i) the ERK nuclear translocation model for the epidermal growth factor (EGF) signaling pathway, and (ii) the self-organizing wave model of phosphatase and a tensinin homolog (PTEN) for the chemotactic pathway of {\it Dictyostelium discoideum}. Using a test version of the TIRFM simulation module, we compared the oscillation model of the Min proteins of {\it Escherichia coli} with actual cell images \cite{arjunan2010}.

\section*{Method}
\subsection*{Computational framework}

\begin{wrapfigure}{R}{9cm}
\vspace{-30pt}
  \centering
      \includegraphics[width=9cm]{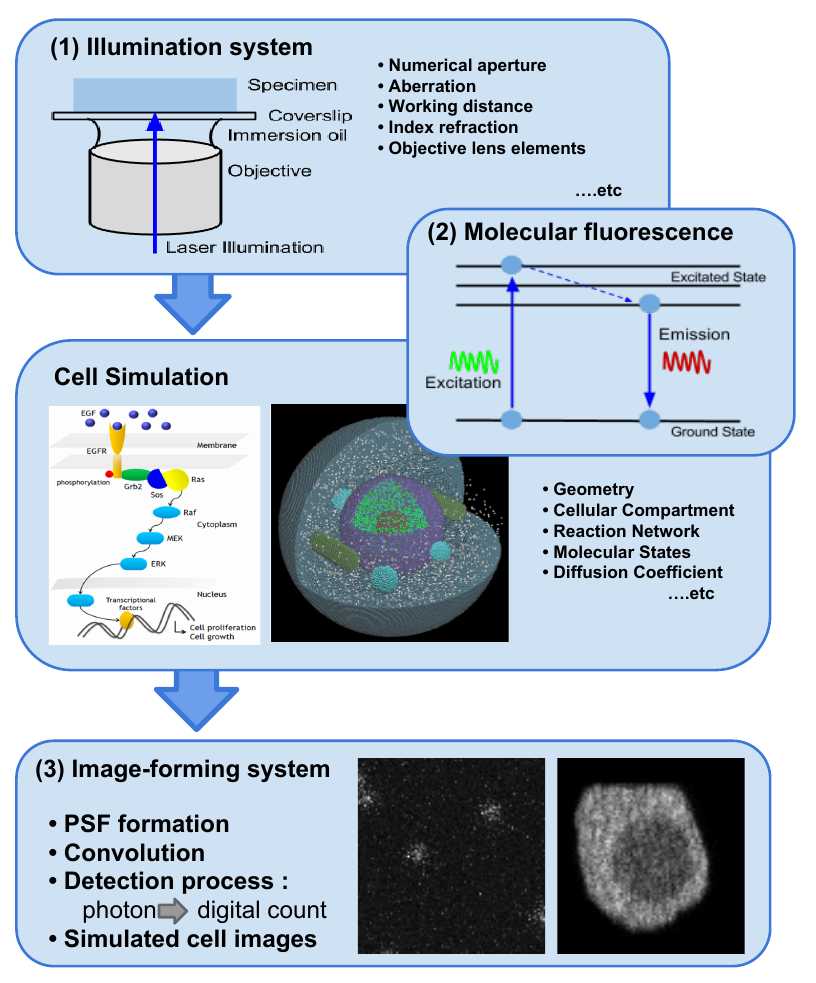}
  \caption{Schematic overview of the computational framework. Direction of photon propagation is presented by thick blue arrows.}
  \label{fig;framework}
\vspace{-10pt}
\end{wrapfigure}

\paragraph{}
To render the simulated output of a spatial cell model well suited for comparison at the level of photon-counting units, we propose a computational framework for simulating the passage of photons through fluorescent molecules and the optical system. Simulations using this framework can generate simulated digital images after accounting for the systematic effects that are governed by the parameters embedded in spatial cell model and optics system. An overview of the computational framework is schematically shown in Figure \ref{fig;framework}. The simulation of the optical system is composed of three components: (1) an illumination system, (2) molecular fluorescence, and (3) an image-forming system. The illumination system transfers photon flux from a light source to the spatial cell model to create a prescribed photon distribution and maximize the photon flux delivered to the cell model. Fluorophores defined in the cell model absorb photons from the distribution and are quantum mechanically excited to higher energy states. Molecular fluorescence is the result of physical and chemical processes in which the fluorophores emit photons from the excited states \cite{valeur2012, lakowicz2006}. Finally, the image-forming system relays a nearly exact image of the cell model to a light-sensitive detector.


\subsubsection*{Simulation of cell model}
\paragraph{}
In particular, the bioimaging simulation system requires the space-time trajectory of each simulated molecule of interest to generate realistic digital images. However, many cell simulation systems have been designed to model and simulate both deterministic and stochastic biochemical processes, assuming that simulated molecules are dimensionless and homogeneously distributed in a compartment \cite{takahashi2005}. Here, we use spatial simulation methods that can provide accurate space-time trajectories of molecules \cite{vanzon2005, takahashi2010, arjunan2010, smoldyn, mcell, vcell}. For a given cell system, simulations using these methods include a statistical model of biological fluctuation that arises from stochastic changes in the cellular compartment geometry, number of molecules, type of molecule, molecular state, and translational and rotational diffusion. 


\subsubsection*{Simulation of optical system}
\paragraph{}
Simulations of an optical system particularly require the computation of the photon counting, propagation, and distribution. The optics simulations are based on geometric optics (or wave optics) and the Monte Carlo method. Each optics simulation includes a statistical model of the systematic effects that are influenced by the parameters defined in optical devices such as the light source, objective lens, special filter, and detector. The classical theory of geometric optics is applied to simulate the photon propagation and distribution through the illumination and image-forming systems, including optical aberrations. Geometric optics approximates the photon propagation as a ray (paraxial approximation), and provides the procedures to compute the numerical or analytical forms of the photon distributions for a given photon wavelength. It is an excellent approximation when the photon wavelength is very small compared with the size of the structure with which the photon interacts. However, it introduces normalization constant as an input parameter, and is formalized without counting the number of photons propagating through the optical system. The Monte Carlo method is applied to the simulation of the stochastic process of counting photons for a given probability density function. The details for each optics simulation are described below.

\begin{enumerate}
\item [(1)]  Illumination system \cite{mansuripur2009, pawley2008}: The bioimaging system requires intense, near-monochromatic, illumination by a widely spreading light source, such as lasers. Incident photons from such a light source can illuminate a specimen. The surviving photons after passing the excitation filters interact with the fluorophores in the cell model, and excite the fluorophores to electrically excited states. The optics simulations of the focusing of the incident photons through the objective lens include a statistical model of the systematic effects due to the numerical aperture (NA), magnification, working distance, degree of aberration, correction refracting surface radius, thickness, refractive index, and details of each lens element.

\item [(2)] Molecular fluorescence : The incident photons propagating through the illumination system are absorbed by the fluorophores in the cell model. Fluorescence is the result of physical and chemical processes in which the fluorophores emit photons from electronically excited states \cite{valeur2012, lakowicz2006}. The Monte Carlo simulation of the overall fluorescence process includes a statistical model of the systematic effects that are influenced by the absorption and emission spectra, quantum yield, lifetime, quenching, photobleaching and blinking, anisotropy, energy transfer, solvent effect, diffusion, complex formation, and a host of environmental variables.

\item [(3)] Image-forming system \cite{mansuripur2009, pawley2008}: In an optical system that employs incoherent illumination of the cell model, the image-forming process can be considered as a linear system \cite{gaskill1987}. Impulse response of the image-forming system to a point-like fluorophore is described by the point spread function (PSF) of the wavelength and position. When all fluorophores in the cell model are imaged simultaneously, the distribution of emitted photons of longer wavelengths that passed through the use of the objective lens and special filters, is computed as the sum of the PSFs of all fluorophores. The optics simulations of PSF formation and convolution include a statistical model of the systematic effects that are ruled by the parameters embedded in the objective lens, the special filters, and each details of lens elements. \\
The emitted photons are finally detected by light-sensitive devices, and digitized as an image at detection time. The properties of the final image depend on the detector specifications and conditions during the readout process that converts an incident photon signal into a digital signal. The Monte Carlo simulation for the detection process includes a statistical model of the systematic effects that arise from signal and background shot noises, and detector specifications and conditions, such as pixel size, quantum efficiency (QE), readout noise, dark current, excess noise factor, gain, offset, exposure time, and binning.

\end{enumerate}

\subsection*{Implementation}

\begin{wrapfigure}{R}{9cm}
\vspace{-30pt}
  \centering
      \includegraphics[width=9.0cm]{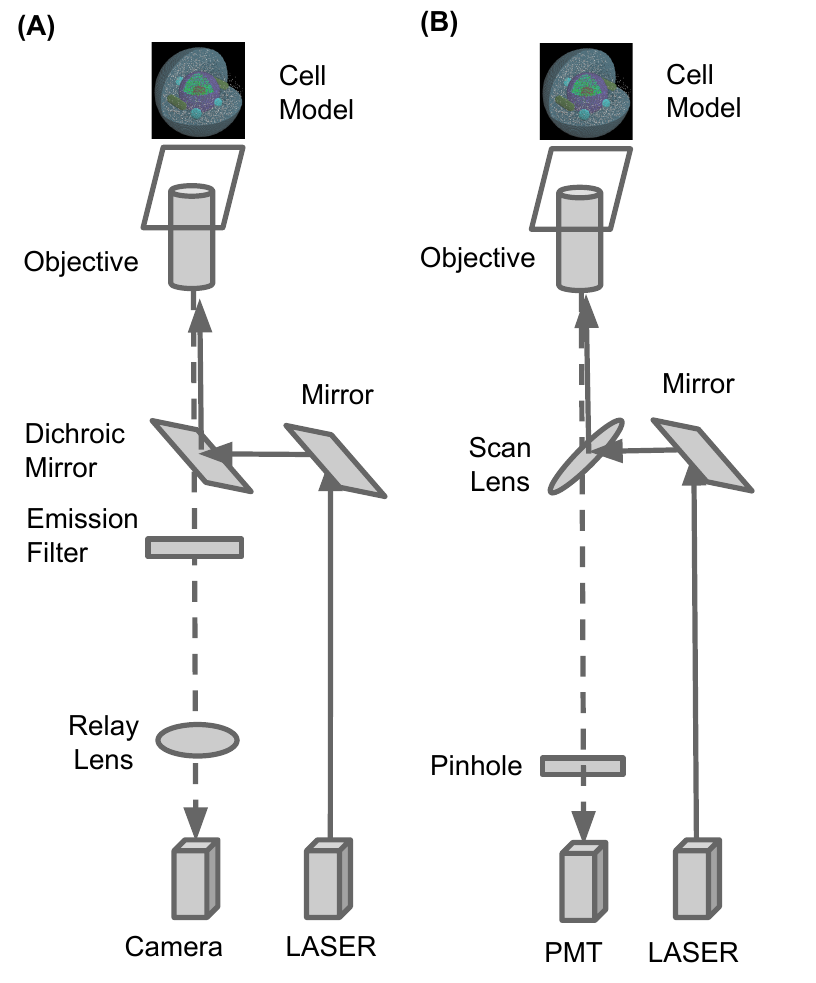}
  \caption{Optical configuration for the TIRFM (A) and LSCM (B) simulation modules. Grey arrows represent direction of photon propagation.}
  \label{fig;config}
\vspace{-10pt}
\end{wrapfigure}

\paragraph{}
We provide a standard computational framework to simulate various different types of bioimaing systems. In particular, we implemented the simulation modules for relatively simple microscopy systems: TIRFM and LSCM. Optical configurations are shown in Figure \ref{fig;config}. The modules are designed to generate digital images of the cell simulation results after accounting for the systematic effects that are governed by the parameters defined in the TIRFM and LSCM systems. A cell simulation method with Spatiocyte is used to construct the spatial cell models \cite{arjunan2010}. For a given cell system, Spatiocyte can provide a statistical model of biological fluctuation that arises from stochastic changes in the cellular compartment geometry, number of molecules, type of molecule, molecular state, and translational diffusion. The method can be used to model complex reaction-diffusion mediated cellular processes occurring on the surface and in the volume compartments of the cell at a single-molecule resolution. To represent cell compartments and rapidly resolve molecular collisions, the method discretizes space into a hexagonal closed-packed lattice. Each molecule randomly walks from voxel to voxel. Molecular collisions occur between walks. Immobile lipid molecules represent surface compartments, such as cellular and nuclear membranes. Implementation details are described in ref. \cite{arjunan2009}. 


\paragraph{}
The three dimensional point spreading function (3D-PSF) model plays a key role in the bioimaing simulations \cite{kirshner2013}. Each point-like source of a fluorophore gives rise to a 3D-PSF pattern in the image-forming systems. The normalization constant of the PSF is usually considered as an user input parameter. However, the bioimaging simulations  requires the counting of the number of photons emitted from a single fluorophore, and spatial PSF integration to be unity within infinite volume region ($\int^{\infty}_{0} PSF\ d^3r = 1$). The PSF decays in an oscillatory manner at tails along the radial and axial axes. Such damping characteristics hinders the estimation of an exact or approximate form of the PSF normalization constant. A wrong estimation can easily lead to the miscounting of the number of photons, and provide a wrong intensity of the final images. Such problematic normalization has not been well discussed in the literature. In addition, optical aberrations can lead to a non-uniform distribution of the 3D-PSF. The aberrations are deviations in an image that occur when photons from one point of an object does not converge into a single point after propagating through an optical system. They can be caused by artifacts that arise from the interaction of photons with glass lenses. Using first order paraxial approximation, makers of optical instruments typically correct the optical systems to compensate for the optical aberrations. 


\paragraph{}
Assuming the first order paraxial approximation, and the spatial PSF integration to be unity within a limited volume region ($\int^{\Lambda}_{0} PSF\ d^3r = 1$), we implement the TIRFM and LSCM simulation modules. Step-by-step instructions are provided below. More details are discussed in the supporting information. Simulation studies to estimate the errors that arise from the PSF normalization and the optical aberrations are required for the future implementation. 

\begin{enumerate}
\item[A1.] The TIRFM simulation module enables selective visualization of the basal surface regions of the cell model. Incident beam photons of the excitation wavelength ($\lambda$) can uniformly illuminate the specimen. Evanescent field is generated along z-axes as perpendicular to the total internal reflection surface, and capable of exciting the fluorophores near the surface. The incident photon flux density at the level of photon-counting unit is defined by 
\begin{eqnarray}
\left| {\bf A}_{I} \right|^2 \cong \frac{\phi}{E_{\lambda}}\ {\rm \left[ \frac{\# photons}{sec \cdot cm^2} \right]}
\end{eqnarray}
where $\phi$ and $E_{\lambda} = \frac{hc}{\lambda}$ are the incident beam flux density (${\rm W/cm^2}$) and single photon energy, respectively. $h$ and $c$ are Planck constant and a speed of light. ${\bf A}_I$ is the amplitude of the incident photon flux density.
\item[A2.] Because of the desperate timescales of the quantum transitions, we simply assume that the fluorescence molecules subsequently emit single photon of longer wavelength while they absorb one million photons of excitation wavelength, and the cross-section of photon-molecule interaction is roughly $10^{-14}\ {\rm cm^2}$ \cite{olympus_url}. No other physical processes is simulated. The expected number of photons emitted from a single fluorophore is defined by
\begin{eqnarray}
n_{emit} & \cong & \frac{\sigma\ \delta T}{4 \pi}\ \left| {\bf A}_{T} \right|^2 \times 10^{-6}\ {\rm \left[ \# photons \right]}
\label{eqn:conv}
\end{eqnarray}
where $\left| {\bf A}_{T} \right|^2$, $\sigma$, and $\delta T$ are the transmitted beam flux density, the cross-section, and detection time. The detector is located in a specific direction. We expect to observe the number of photons devided by an unit surface area of a sphere ($4\pi$). The amplitude of the transmitted beam flux density depends on the index refraction, and the incident beam angle, amplitude, and polarization.
\item[A3.] When all the fluorophores in the cell model are imaged simultaneously, the distribution of the emitted photon of longer wavelengths that passed through the use of objective lens and special filters is computed as the sum of the PSFs of all the fluorophores. In particular, we use the Born-Wolf PSF model \cite{kirshner2013}. For an optimal wavelength ($\lambda'$) of a fluorophore, we estimated that $55\%$ of the emitted photons that passed through the Dichroic mirror and emission filter survive ($n_{emit} \rightarrow n'_{emit}$). The expected image plane at the focal point ($z=z_0$) is given by the convolution of the PSF and written in the form of
\begin{eqnarray}
Exp.\ Image (\vec{r}, z) = \sum^{N}_{k = 0} n'_{emit}\ PSF_{\lambda'}(\vec{r} - \frac{\vec{r}_k}{M}, z - z_k)
\end{eqnarray}
where $N$ and $M$ are the total number of fluorophores, and optical magnification, respectively. $(\vec{r}_k, z_k)$ is the position of the $k$-th fluorophore. $(\vec{r},z)$ is the position in an image plane. The PSF is normalized within a $\pm 1.0\ {\rm \mu m}$ range of radial and axial axes. In addition, polarization of the evanescent field is non-isotropic, which means that dipoles of different orientations are excited with different probabilities per unit time. In order to accurately simulate image-formation process, the polarized form of the PSF is required for the future implementations.
\item[A4.] The emitted photons are finally detected by CMOS or EMCCD cameras, and digitized as an image at a detection time. The readout process can convert expected incident photon signals to digital signals relies on camera specifications and camera operating conditions to carry out the properties for final images. The observed image of the cell model can be obtained using the Monte Carlo method in the presence of systematic sources, including statistical fluctuations in photon counting (photon shot noise), and camera specification and camera operating conditions. Finally, photoelectron signals can be linearly converted to digital signals. Unit conversions are given by
\[
Exp.\ Image\ {\rm \left[\# photons \right]} \longrightarrow Obs.\ Image\ {\rm \left[\# photoelectrons \right]} \longrightarrow Digital\ Image\ {\rm \left[ A/D\ counts \right]} 
\]
\end{enumerate}

\begin{enumerate}
\item[B1.] The LSCM simulation module can visualize focal regions of the cell model. In general, laser beam propagation of excitation wavelength can be approximated by assuming that the laser beam has an ideal Gaussian beam profile . The incident beam flux of excitation wavelength ($\lambda$) and continuously illuminates specimen, and is focused into a confocal volume at a given scan time and beam position. Incident photon flux is defined by
\begin{eqnarray}
P' \cong \frac{\Phi}{E_{\lambda}}\ {\rm \left[ \frac{\# photons}{sec} \right]}
\end{eqnarray}
where $\Phi$ and $E_{\lambda} = \frac{hc}{\lambda}$ are the incident beam flux (W) and single photon energy. $h$ and $c$ are Planck constant and speed of light, respectively. 
\item[B2.] We also assume that the linear conversion of photon emission is by $10^{-6}$, and the cross-section of photon-molecule interaction is roughly $10^{-14}\ {\rm cm^2}$ \cite{olympus_url}. No other physical processes are simulated. For a given position and time, the expected number of photons emitted from a single fluorophore is defined by
\begin{eqnarray}
n_{emit}(\vec{r},z) & \cong & \frac{\sigma\ \delta T}{4 \pi}\ I(\vec{r}, z) \times 10^{-6}\ {\rm \left[ \# photons \right]}
\label{eqn:conv}
\end{eqnarray}
where $I(r,z)$, $\sigma$, and $\delta T$ are the transmitted beam flux density, cross-section, and scan time per pixel, respectively. The detector is located in a specific direction. We expect to observe the number of photons divided by an unit surface area of a sphere ($4\pi$). The transmitted beam flux density depends on the incident photon flux, and the beam waist radius at the focal plane where the wavefront is assumed to be flat. 
\item[B3.] When all the fluorophores in the cell model are imaged simultaneously, the distribution of the emitted photon of longer wavelengths that passed through the use of objective lens and pinhole is computed as the sum of the PSFs of all the fluorophore. In particular, we use the Born-Wolf PSF model  \cite{kirshner2013}. As an incident beam is scanned across the cell model in horizontal and vertical axes, a digital image is generated at a time. For a given scan time and beam central position, the expected image plane at the focal point ($z = z_0$) is given by the integration of the image plane obtained from the PSF convolution. It is written in the form of
\begin{eqnarray}
Exp.\ Image(\vec{r}, z) & = & \iint \delta(\vec{r}_b-\vec{r}, z_b-z) \left[ \iint_{|\vec{r'} - \vec{r}_b| < R} I'(\vec{r'}-\vec{r}_b, z'-z_b)\ dx'dy' \right] dx_bdy_b \\
& & {\rm where}\ \ I'(\vec{r''}, z'') \cong \sum^{N}_{k = 0} n_{emit}(\vec{r''}, z'')\ PSF_{\lambda'}(\vec{r''} - \frac{\vec{r}_k}{M}, z'' - z_k) \nonumber
\end{eqnarray}
where $N$, $R$ and $M$ are the total number of fluorophores, pinhole radius, and optical magnification, respectively. $(\vec{r}_k,z_k)$ is the position of the $k$-th fluorophore. $(\vec{r}_b,z_b)$ is the position of beam center. $(\vec{r},z)$ is position in the image plane. The PSF is normalized within a $\pm 1.0\ {\rm \mu m}$ range of radial and axial axes. 
\item[B4.] The emitted photons are finally detected by a photomultipliers tube (PMT), and digitized as an image at a given scan time. The observed image of the cell model can be obtained using the Monte Carlo method in the presence of systematic sources, including statistical fluctuations in photon counting (photon shot noise), and PMT specifications and PMT operating conditions. Finally, photoelectron signals can be linearly converted to digital signals. Unit conversions are given by
\[
Exp.\ Image\ {\rm \left[\# photons \right]} \longrightarrow Obs.\ Image\ {\rm \left[\# photoelectrons \right]} \longrightarrow Digital\ Image\ {\rm \left[ A/D\ counts \right]} 
\]
\end{enumerate}


\begin{figure}[!h]
      \centering
      \includegraphics[width=14cm]{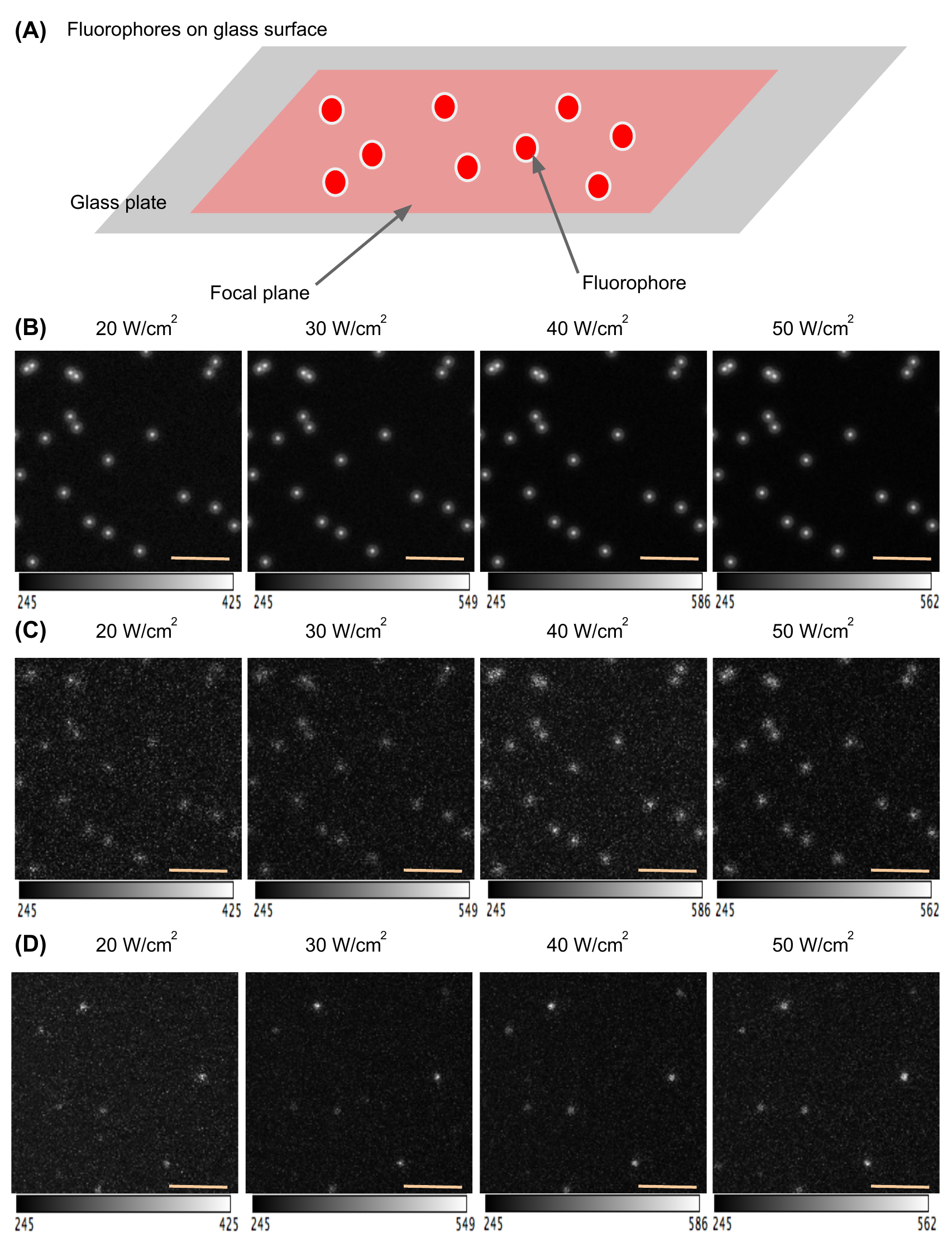}

      \caption{{\bf Using HaloTag-TMR molecules distributed on a glass surface to evaluate the performance of TIRFM simulation module.} (A) $100$ stationary HaloTag-TMR molecules are distributed on a glass surface. (B) Expected images of the simple particle model at various beam flux densities ($20, 30, 40$ and $50\ {\rm W/cm^2}$). The expected image is obtained by averaging $100$ images over {\rm 3\ {\rm sec}} exposure period. (C) Simulated digital images of the simple particle model are shown at various beam flux densities ($20, 30, 40$ and $50\ {\rm W/cm^2}$). Size of each images is $152 \times 156$ pixel. Orange scalebar represents $3.15\ {\rm \mu m}$.  (D) Real captured images obtained from {\it in vitro} experiment are shown at various beam flux densities ($20, 30, 40$ and $50\ {\rm W/cm^2}$). The maximum value of the grayscale is adjusted to improve visualization of each image.}
      \label{fig;tmr_tirfm}
\end{figure}

\begin{figure}[!h]
      \centering
      \includegraphics[width=14cm]{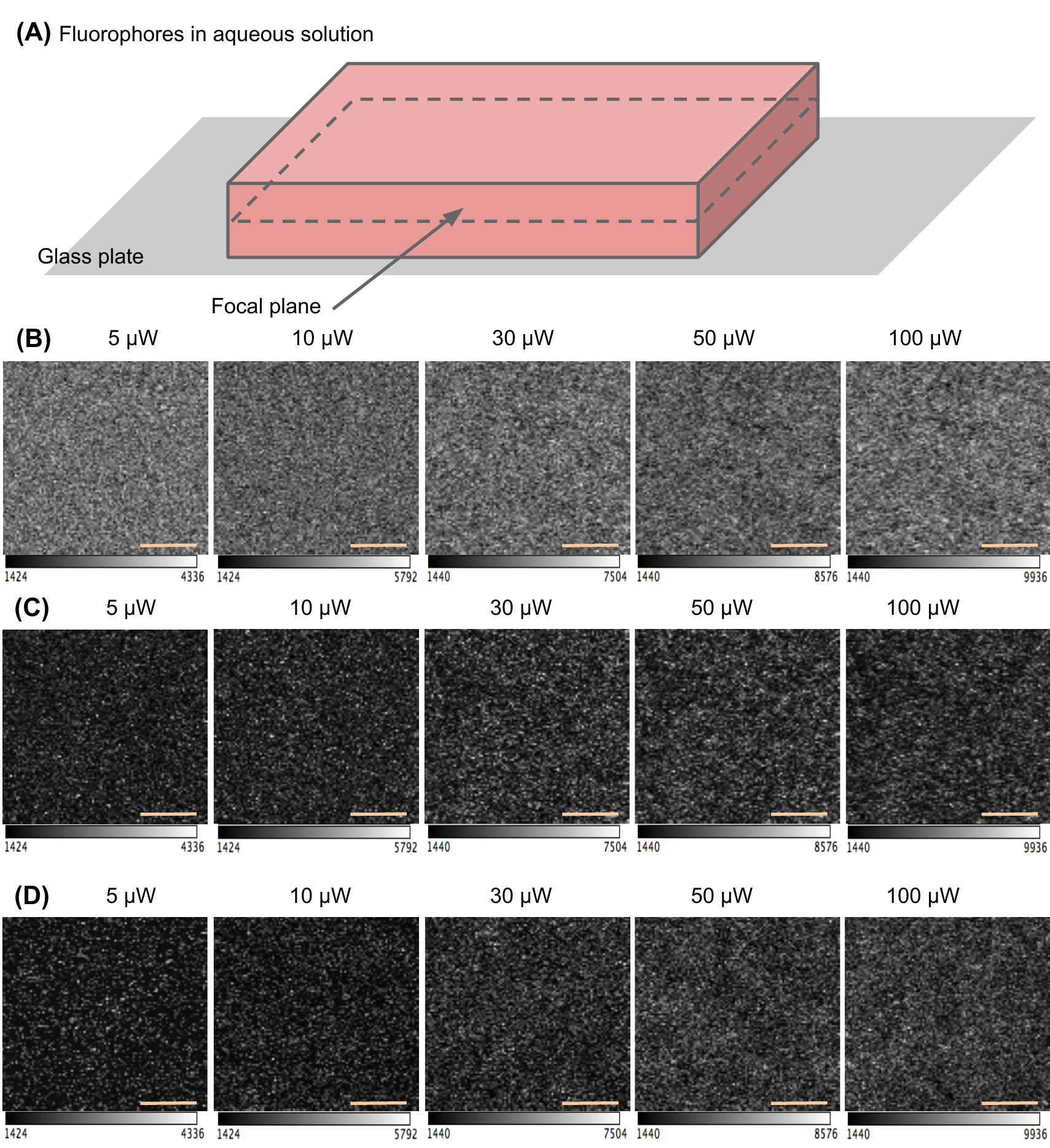}

      \caption{{\bf Using HaloTag-TMR molecules to evaluate the performance of LSCM simulation modules.} (A) $19,656$ HaloTag-TMR molecules are distributed in a $30 \times 30 \times 6\ {\rm \mu m^3}$ box of aqueous solution ($= 5\ {\rm nM}$), and rapidly diffuse at $100\ {\rm \mu m^2/sec}$. (B) Expected images of the simple particle model at various beam flux ($5, 10, 30, 50,$ and $100\ {\rm \mu W}$). Each expected image is generated by averaging $26$ images over {\rm 30\ {\rm sec}} exposure period. (C) Simulated digital images of the simple particle model are shown for various beam flux ($5, 10, 30, 50,$ and $100\ {\rm \mu W}$). Size of each images is $100 \times 100$ pixel. Orange scalebar represents $5.18\ {\rm \mu m}$.  (D) Real captured images obtained from {\it in vitro} experiment are shown for various beam flux ($5, 10, 30, 50,$ and $100\ {\rm \mu W}$). Size of each images is $100 \times 100$ pixel. The maximum value of the grayscale is adjusted to improve visualization of each image.}
  \label{fig;tmr_lscm}
\end{figure}

\section*{Results}
\subsection*{Comparison of {\it in vitro} images}
\paragraph{}
We evaluated the performance of our simulation modules by comparing the simulated images with the actual photographed ones for simple particle models of fluorescent molecules. We simulated imaging of the focal region of those simple models for the optical system with the detector specifications and detector  operating conditions, as shown in Tables \ref{tab;tirfm_tmr} and \ref{tab;lscm_tmr}. Evaluation details are described in the supporting information. The results are shown in Figures \ref{fig;tmr_tirfm} and \ref{fig;tmr_lscm}. The intensity of the simulated images corresponds to the number of photons detected in the digital cameras or the PMT. Each simulated image is visually similar to the corresponding real ones. Thus, the simulated images were compared with images obtained using actual microscopy systems at the level of photon-counting units. However, differences still remain in the resulting images owing to calibration. Calibration is the process of improving the agreement of the code calculation with a chosen set of benchmarks through the adjustment of the parameters implemented in the simulation modules \cite{vv2007, vv2006, vv2004}. Such a calibration process is required in all experiments to improve the agreement of the simulated outputs with the {\it in vitro} data sets. Even though the results of a simple calibration were used, the first version of our simulation modules was capable of generating images that closely reproduce images obtained with an actual microscopy system. A more elaborate set of calibration is required in the future. More details are described below.

\begin{enumerate}
\item[(1)] To test the performance of the TIRFM simulation module, we constructed a simple particle model of $100$ stationary HaloTag-with-tetramethylrhodamine (HaloTag-TMR) molecules distributed on a glass surface, as shown in Figure \ref{fig;tmr_tirfm}A. We simulated imaging of the basal region of the simple model for the TIRFM specifications and TIRFM operating conditions shown in Table \ref{tab;tirfm_tmr}. Figure \ref{fig;tmr_tirfm}B shows the expected optical distribution used for the simulation, which was generated by averaging $100$ images over a {\rm 3\ {\rm sec}} exposure period. Figure \ref{fig;tmr_tirfm}C and D show the simulated images and the real captured ones at various beam flux densities. The intensity of the simulated images corresponded to the number of photons detected in the EMCCD camera. Increasing the beam flux density results in a relatively brighter image. Each simulated image is visually similar to the corresponding real one. Thus, the simulated images were compared with the images obtained using the actual TIRFM systems at the level of photon-counting units. In addition, intensity histograms of each images are shown in Figure \ref{fig;comparison_hist_tirfm_expected}, \ref{fig;comparison_hist_tirfm} and \ref{fig;comparison_linearity_tirfm}. 

\item[(2)] To evaluate the performance of the LSCM simulation module, we constructed a simple particle model of $19,656$ HaloTag-TMR molecules diffused in an aqueous solution as shown in Figure \ref{fig;tmr_lscm}A. We simulated imaging of the middle region of the simple model for the LSCM specifications and LSCM operating conditions as shown in Tables \ref{tab;lscm_tmr}. Figure \ref{fig;tmr_tirfm}B shows the expected optical distribution used for the simulation, which was obtained by averaging $26$ images over a {\rm 30\ {\rm sec}} exposure period. Figure \ref{fig;tmr_lscm}C and D show the simulated images and the real captured ones at various beam fluxes. The intensity of the simulated images corresponds to the number of photon pulses detected in the PMT. Increasing the beam flux results in relatively brighter image. Each simulated image is visually similar to the corresponding real ones. Thus, the simulated images were compared with the images obtained using the actual LSCM systems at the level of photon-counting units. In addition, intensity histograms of each images are shown in Figure  \ref{fig;comparison_hist_lscm_expected}, \ref{fig;comparison_hist_lscm} and \ref{fig;comparison_linearity_lscm}. 

\end{enumerate}

\subsection*{Comparison of {\it in vivo} images}
\begin{figure}[!h]
      \centering
      \includegraphics[width=14cm]{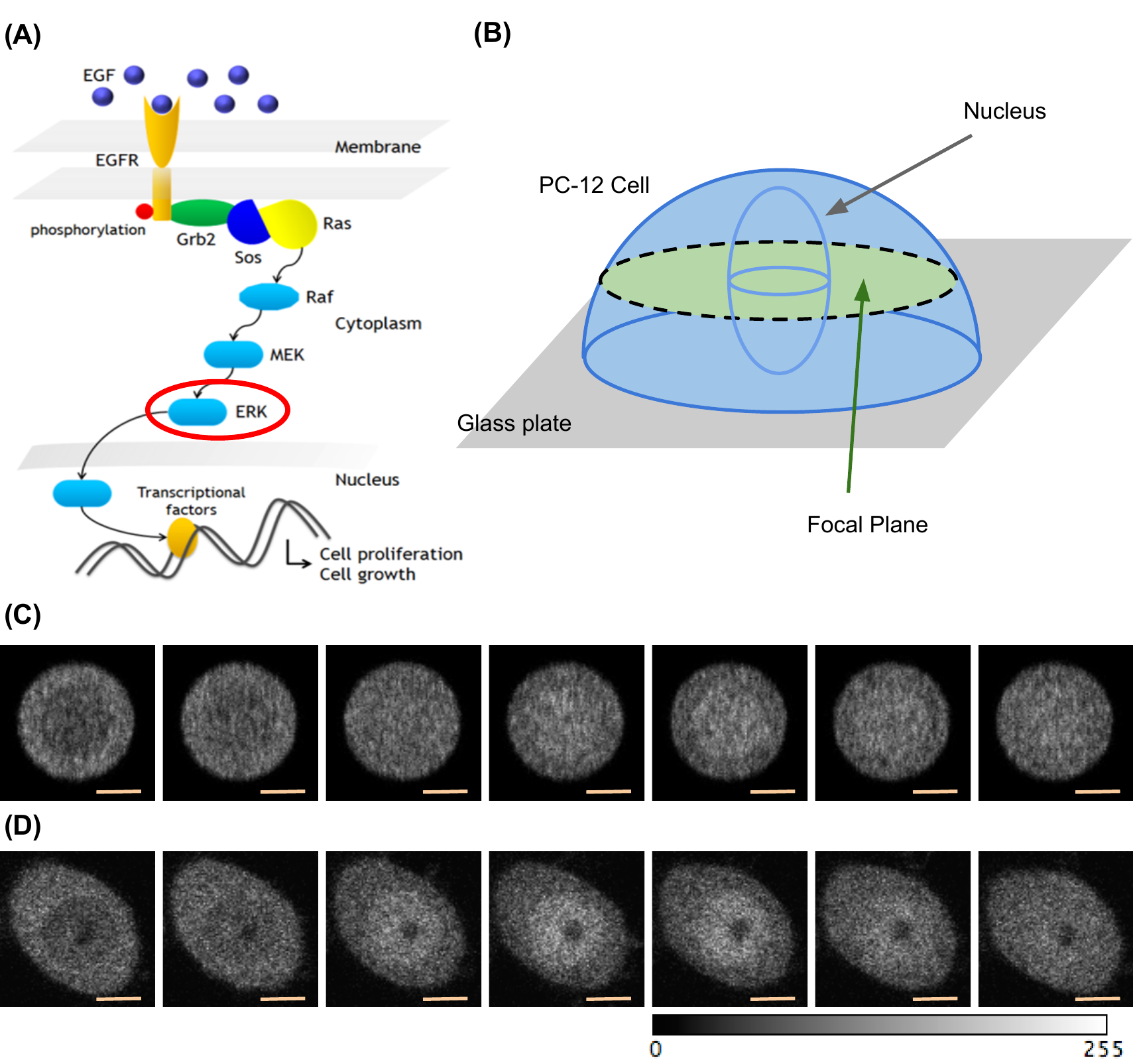}
      \caption{{\bf ERK nuclear translocation model of EGF signaling pathway.} (A) Reaction network. (B) Geometry of PC-12 cell model. A hemispherical cell measuring $20\ {\rm \mu m}$ in diameter and $7\ {\rm \mu m}$ in height is assumed. (C) Time-lapse images of the ERK nuclear translocation model observed using the LSCM simulation module. Size of each images is $90 \times 90$ pixel. Orange scalebar represents $4.66\ {\rm \mu m}$.  (D) TIme-lapse images obtained from the experiment. The grayscale of each images is adjusted in the range of $0$ to $225$.}
      \label{fig;egf}
\end{figure}

\begin{figure}[!h]
      \centering
      \includegraphics[width=14cm]{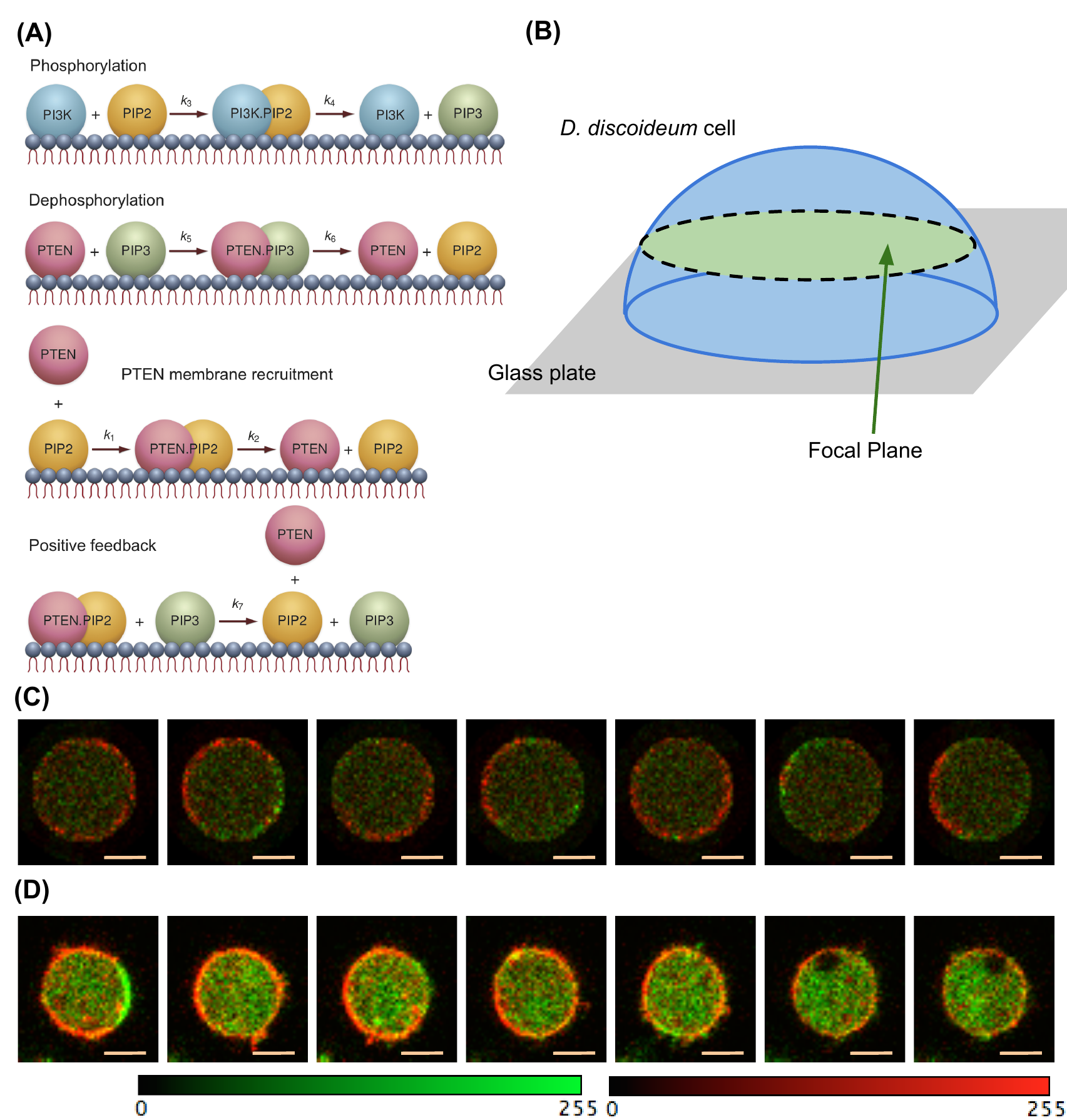}
      \caption{{\bf Self-organizing wave model of PTEN for the chemotactic pathway of {\it D. discoideum}.} (A) Reaction network. (B) Geometry of {\it D. discoiduem} cell model. A hemispherical cell measuring $25\ {\rm \mu m}$ in diameter and $5\ {\rm \mu m}$ in height is assumed. (C) Simulated image of the self-organizing wave model observed using the LSCM simulation module. Size of each images is $52 \times 51$ pixel. Orange scalebar represents $5.39\ {\rm \mu m}$.  (D) Real captured images obtained from the experiment. Red and green indicate PTEN-TMR and PH-EGFP, respectively. The colorscale of each images is adjusted in the range of $0$ to $225$.}
      \label{fig;pten}
\end{figure}

\paragraph{}
Using the LSCM simulation module, we compared a more complex cell model with real cell images obtained using the actual LSCM system. We constructed the following spatial cell models: (i) the ERK nuclear translocation model for the EGF signaling pathway, and (ii) the self-organizing wave model of PTEN for the chemotactic pathway of {\it D. discoideum}. We developed these cell models, which are not available in the literature. We assumed that the parameters of each cell model and the LSCM system are well evaluated with {\it in vitro} data sets. We then simulated imaging of the focal region of those cell models for the LSCM specifications and LSCM operating conditions, as shown in Table \ref{tab;lscm_egf} and \ref{tab;lscm_dicty}. Details of the {\it in vivo} comparison are described in the supporting information. The results are shown in Figures \ref{fig;egf} and \ref{fig;pten}. The intensity of the simulated images corresponds to the number of photon pulses detected in the PMT. Thus, the simulated cell images were compared with the images obtained by the actual microscopy systems at the level of photon-counting units. Significant new insight on the cell models will be published in the future.

\begin{enumerate}
\item[(i)] We constructed the cell model of ERK nuclear translocation for the EGF signaling pathway. We assumed the PC12 cell model that represents the ERK molecules tagged with the enhanced green fluorescent protein (ERK-mEGFP). Figure \ref{fig;egf}A and B show the main reaction network and the geometry of the model, respectively. The cell was placed on the glass surface, and was nearly hemispherical. The size of the hemispherical cell was estimated by experimentalists. A cell measuring $20\ {\rm \mu m}$ in diameter and $7\ {\rm \mu m}$ in height was assumed. The model consisted of $75$ chemical species, $143$ reactions, and $85$ kinetic parameters. A maximum of $100,000$ ERK molecules were distributed in the cell cytoplasm and rapidly diffuse at $1.00\ {\rm \mu m^2/sec}$. The input of the EGF ligand could drive the transport of $30 \%$ of the ERK molecules into the nucleus and back to the initial condition in $10$ min. We simulated imaging of the middle regions of the cell model for the LSCM specifications and LSCM operating conditions, as shown in Tables \ref{tab;lscm_egf}. Figure \ref{fig;egf}C and D show the simulated cell images and the cell images obtained using the actual LSCM system. The intensity of the simulated images corresponds to the number of photon pulses detected in the PMT. Therefore, the simulated images were compared with images obtained using the actual LSCM system at the level of photon-counting units. Each simulated image was visually similar to the corresponding real one, but differences still remain in the resulting images owning to calibrations. A more elaborate set of calibration is required in the future. 

\item[(ii)] We also constructed a self-organizing wave model of PTEN for the chemotactic pathway of {\it D. discoideum} to validate the performance of two-color imaging for the LSCM simulation module. We assumed a {\it D. discoideum} cell model that expresses the fluorescently labeled pleckstrin homology domain of Akt/PKB (PH) and PTEN, where PH and PTEN are indicators for phosphorylates  phosphatidylinositol 3,4,5-trisphosphate (PIP3) metabolism. PH can bind to PIP3 at the membrane, whereas PTEN catalyzes the degradation of PIP3 and has a binding motif for phosphatidylinositol 4,5-biphosphate (PIP2). PH was tagged with EGFP (PH-EGFP), whereas PTEN was tagged with HaloTag with TMR (PTEN-TMR). A maximum of $10,000$ molecules of PTEN-TMR and PH-EGFP were homogeneously distributed in the cell cytoplasm. On the membrane, PI3K catalyzed PIP2 phosphorylation to PIP3, whereas PTEN dephosphorylated PIP3 into PIP2. Cytosolic PTEN was recruited to the membrane regions containing PIP2. Nonetheless, PIP3 could dislodge PTEN from PIP2 into the cytosol when they came in contact with each other. This last reaction acted as a positive feedback for PIP3 accumulation. Figure \ref{fig;pten}A and B show the main reaction network and the geometry of the model, respectively. A cell was placed on the glass surface, and was nearly hemispherical. The size of the hemispherical cell was estimated by experimentalists. The cell measuring $25\ {\rm \mu m}$ in diameter and $5\ {\rm \mu m}$ in height was assumed. The model involved $8$ chemical species, $12$ reactions, and $12$ kinetic parameters. Lattice-based particle simulation of the cell model enabled of the reproduction of the local oscillatory dynamics of PTEN-TMR and PH-EGFP. We simulated imaging of the middle region of the cell model for the LSCM specifications and LSCM operating conditions, as shown in Tables \ref{tab;lscm_dicty}.  Figure \ref{fig;pten}C and D show the simulated cell images and the cell images obtained by the actual LSCM system. The intensity of the simulated images corresponds to the number of photon pulses detected in the PMT. Therefore, the simulated images were compared with the images obtained using the actual LSCM system at the level of photon-counting units. Each simulated image was visually similar to the corresponding real one, but intensity differences still remained in the resulting images. The number of PTEN-TMR and PH-EGFP in the wave model are approximately 4,000 for each, but we expect more ($\sim 30,000$) in the observed images. A more elaborate set of calibration is required in the future. 

\end{enumerate}

\section*{Conclusion and discussion}
\paragraph{}
Measurements using bioimaging techniques are generally influenced by systematic effects that arise from the stochastic nature of biological cells, the photon-molecule interaction, and the optical configuration. Such systematic effects are always present in all bioimaging systems and hinder the comparison between the cell model and the real cell image. Combining optics and cell simulation technologies, we proposed a computational framework for handling the parameters embedded in the cell model and the optical principles governing the bioimaging systems. The simulation using this framework generated digital images from cell simulation results after accounting for the systematic effects. In particular, we demonstrated that the simulated digital images closely reproduce the images obtained using actual TIRFM and LSCM systems. Each pixel intensity corresponded to the number of photon pulses detected in the camera or the PMT. Thus, the framework streamlines the comparison at the level of photon-counting units. However, the image comparison is insufficient to check the validity of the simulation modules. Verification is the process of confirming the simulation modules are correctly implemented with respect to conceptual description and analytical solutions \cite{vv2007, vv2006, vv2004}. During the verification process, the simulation modules must be tested to find and estimate numerical errors in the implementations. The simulation modules are designed to count the number of photons that passed through the optical configurations. A wrong estimation of the numerical errors that arise from the photon-counting principle can provide a wrong intensity of the final images. For example, a wrong PSF normalization can miscount the number of photons, and lead to wrong final images. Furthermore, the simulated images can be also compared with a chosen set of experimental benchmarks defined in calibration and validation parameter domains \cite{vv2007, vv2006, vv2004}. Systematic variance and covariance that arise from various different parameter settings must be estimated to establish the validity of the simulation modules. Analyses to quantify the systematic uncertainties are required for the future implementation.

\paragraph{}
One of the key challenges of transforming biology from a phenomenological science to a predictive one is how to bridge the gap between a cell model and an actual biological cell \cite{szekely2014, cvijovic2014, qu2011, kitano2002_sci, kitano2002_nat}. Over the last two decades, large-scale, accurate, and comprehensive simulations of cell models have greatly improved our understanding of many cellular networks and processes  \cite{sanghvi2013, karr2012, tomita2001}. However, we are still far away from having predictive cell models for actual applications in medicine and biotechnology. In this work, we focused on the "comparison" part of the model validation and demonstrated the single cell-to-cell image comparison at the level of photon-counting units. For future implementation, it is important to fully simulate optical systems and to demonstrate other important parts of the model validation \cite{vv2007, vv2006, vv2004}. Within this framework, the functionality and capability of the cell models will be more easily seen and understood. Future tasks required for the model validation include studying diversity in cell populations and obtaining the nominal and predicted probability distributions of the cell model. The behavior of individual cells depends on the internal variables and the environmental conditions. The nominal and predicted probability distributions of those variables are characterized by their statistical quantities. A likelihood that quantifies the discrepancy between the predicted distribution and the observed one can be evaluated by using a statistical test of significance. If the result of the statistical test satisfies a certain confidence level, then the cell model is either rejected or accepted with respect to real cell images. Consequently, such model fitting will support discoveries in biological science.  

\paragraph{}
Bioimaging simulation using the computational framework presented here is not meant to replace biological experiments. It provides a realistic estimate of the output that would be obtained in specific biological applications. Biologists often use commercial bioimaging systems for their own biological interests. Optical properties of biological molecules and/or phenomena uniquely change, according to the experimenter's skills and experiences in handling biological samples and optical equipments. The commercial systems are designed for general usage, and are not optimized to measure the optical properties of all biological samples. Although some biologists assemble specialized optical imaging systems for a particular application, it is still difficult for them to adjust systems parameters without expected outputs. Such an approach is quite inefficient since it depends on the experimenter's skills and experiences. A more systematic approach is required to reduce or eliminate unintended experimenter's bias. In order to objectively handle biological and physical principles in an organized manner, it is important to develop an object-oriented simulation toolkit of biological imaging. The simulation toolkit is constructed on the basis of a set of numerous biological and physical processes to handle diverse interactions of photons with molecules over a wide energy range. The toolkit provides a complete set of software components for all area of bioimaging simulations: optical configuration, spatial cell models, run, parameter management, visualization and user interface. Such a multi-disciplinary nature of the toolkit allows a user to easily design, customize and extend bioimaging and/or experimental systems well optimized for specific biological applications. For example, the computational framework can also be applied to simulate other bioimaging techniques including fluorescence recovery after photobleaching (FRAP), fluorescence correlation spectroscopy (FCS), Forster resonance energy transfer (FRET) and localization microscopy. All simulation modules can be objectively handled in a uniform software platform.

\paragraph{}
However, there are two problems in constructing such a software platform. (1) Computational speed is not well optimized for the TIRFM and LSCM simulation modules. The speed of generating a simulated image is proportional to the number of fluorophores embbeded in a cell models. Bioimaging simulation of a cell model containing $100,000$ fluorophores, requires about one day to obtain the final image. Optimization is required in the near future. (2) The optical properties of many commercial materials are not publicly available. In particular, information on the objective lens used is important for predicting an exact PSFs in a wide field. A question is how we can overcome such nonscience-related problems (probably, it is a matter of business model). In conventional approaches to biological research, biologists and optical physicists work independently, and do not interact much technologically. In order to properly design and customize the bioimaging and experimental systems well optimized for the specific biological applications, collaborative work with optical physicists and engineers will be required for the future biological research. Clearly, the bioimaging simulation toolkit allows us to better communicate with optical physicists and engineers, and to perform the simulation studies of bioimaging systems and their operating conditions. Optical materials are well designed by optical physicists and engineers, and their performance is generally validated by simulation studies of physical principles and their boundary conditions. Simulation studies are essential for the objective examination of the response of the optical equipments. However, such simulation studies have not been well performed for biological samples. Without the results of simulation studies for biological samples, the collaboration could easily fail. Then, information on the optical materials could not be shared. Using whatever form of PSF as realistically as possible, it is important to estimate experimental accuracy and precision for valuable discussion. We believe that the simulation toolkit can bridge the gap between biology and optics.

\section*{Acknowledgements}
We would like to thank Dr. Yasushi Okada, Dr. Tomonobu M. Watanabe, Dr. Kazunari Kaizu, Dr. Kozo Nishida, Dr. Yukihiro Miyanaga, Dr. Stephen Young, Dr. Yuko Nakane and Yosuke Onoue for their guidance and support throughout this research work, and Dr. Kenneth H. L. Ho for critical reading of the manuscript. We would also like to thank Dr. Yasushi Sako for providing the rat PC12 pheochromocytoma cells that stably express EGFP-tagged ERK2. We also acknowledge the valuable contribution of Hamamatsu Photonics in programming the detection processes. We also wish to extend our gratitude to Dr. Takeharu Nagai and the minority biology research group of new arts and science domain in Japan for guidance and support throughout this research work.

\section*{Author Contributions}
 MW and KT conceived and designed the computational framework. MW wrote the software. MW, SNVA, and KI constructed cell models. SF, SM and YS performed the experiments. MW wrote the paper. JK, MU, and KT provided support and guidance.

\newpage
\appendix

\setcounter{figure}{0} \renewcommand{\thefigure}{S\arabic{figure}}
\setcounter{table}{0} \renewcommand{\thetable}{S\arabic{table}}

\leftline{\bf\Large Supporting Information :}
\vspace{0.3cm}
\hrule
\vspace{0.3cm}

\paragraph{}
This supporting information provides implementation details of the TIRFM and LSCM simulation modules. Implementation is generally not practical and requires much time. For the first implementation, we often applied simple theoretical formulas to simulate the illumination system, molecular fluorescence and PSF formation. We are planning to fully simulate the optical systems for future implementation. The complete source code of these simulation modules was written in Python and released as an open-source framework at \url{https://github.com/ecell/bioimaging}. The package is freely available for Linux and Mac OS X.

\tableofcontents
\addtocontents{toc}{\protect\rule{\textwidth}{.2pt}\par}

\newpage

\section{Implementation}
\paragraph{}
We proposed a standard computational framework to simulate the passage of photons through fluorescent molecules and the optical system, and to generate a digital image that closely represents the image obtained using an actual fluorescence microscopy system. The computational framework included a statistical model of the systematic effects that are influenced by the parameters defined in the cell model and optical system. Using this framework, we specifically implemented the simulation modules for relatively simple bioimaging systems: TIRFM and LSCM techniques. Optical configurations are shown in Figures \ref{fig;config_tirfm} and \ref{fig;config_lscm}. Those modules were designed to generate digital images that closely represent the actual digital images obtained using actual TIRFM and LSCM systems. The optics simulation of the passage of photon through the microscopy systems was based on geometric optics and the Monte Carlo method. The optics simulation is composed of three components; (1) illumination system, (2) molecular fluorescence and (3) the image-forming system. The illumination system transfers the photon flux from a light source to a cell model, to generate a prescribed photon distribution and maximize the flux delivered to the cell model. Fluorophores defined in the cell model absorb photons from the distribution, and are quantum-mechanically excited to higher energy states. Molecular fluorescence is the result of physical and chemical processes in which the fluorophores emit photons in the excited state. Finally, the image-forming system relays a nearly exact image of the cell model to the light-sensitive detector. 

\paragraph{}
Implementations of the simulation modules are generally not practical and require much time. Assuming the first-order paraxial approximation and the spatial PSF integration to be unity within a limited volume region ($\int^{\Lambda}_{0} PSF\ d^3r = 1$), we implement the TIRFM and LSCM simulation modules. Theoretical formulas are often applied in the first implementation. Simulation studies to estimate the errors that arise from the PSF normalization and optical aberrations are required for future implementation. 

\newpage

\subsection{TIRFM simulation module}
\begin{figure}[!h]
  \centering
      \includegraphics[width=9cm]{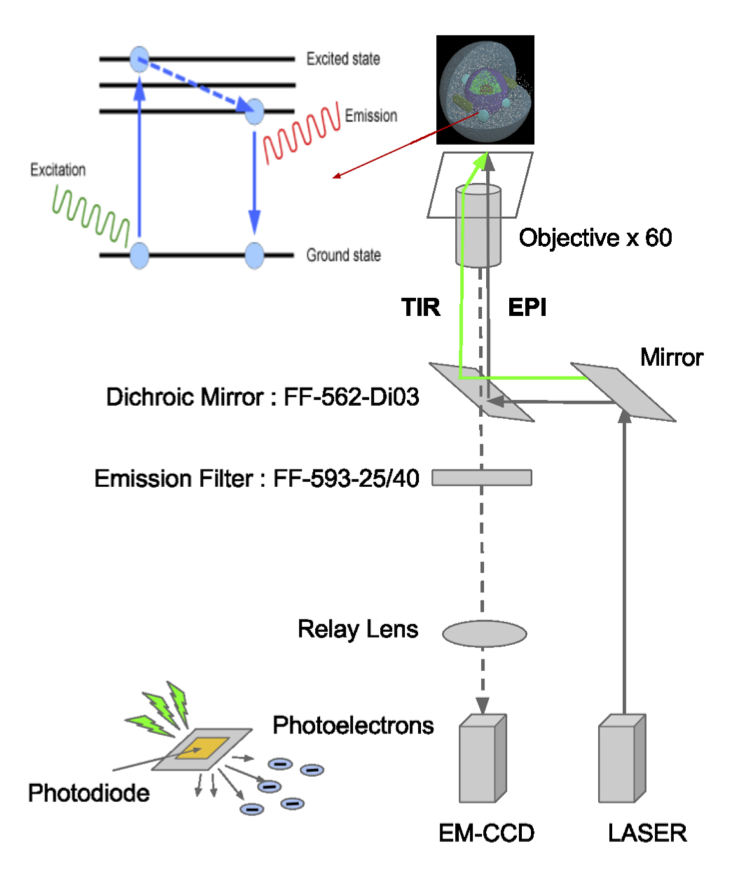}
  \caption[Optical configurations]{Optical configurations of the TIRFM simulation module}
  \label{fig;config_tirfm}
\end{figure}

\paragraph{}
The TIRFM simulation module enables a selective visualization of basal surface regions of a cell model. Its optical configuration is shown in Figure \ref{fig;config_tirfm} \cite{wazawa2005, miyanaga2009}. Implementation assumption are summarized in Table \ref{tab:char_tirfm}. 

\begin{table}[!h]
    \centering
    \begin{tabular}{|c||c|}
    \hline
    \multirow{2}{*}{Principle} & Photon-counting \\
    & 1st-order paraxial approximation (Linear term) \\ \hline
    \multirow{2}{*}{Illumination} & Epi-fluorescent or Evanescent fields \\
    & Continuous / Uniform / Linearly-polarized \\ \hline
    \multirow{2}{*}{Fluorescence} & Linear convertion ($\times 10^{-6}$)\\
    & Cross-section ($\sigma \cong 10^{-14}\ {\rm cm}^2$) \\ \hline
    \multirow{2}{*}{Image-forming} & 3-D Airy PSF (Unpolarized analytical form) \\
    & CMOS or EMCCD cameras \\ \hline
    \end{tabular}
    \caption{Implementation assumptions for the TIRFM simulation module. The detection process for the cameras is performed with Monte Carlo simulation, where CMOS and EMCCD stand for complementary metal-oxide semiconductor and electron multiplication charge coupled device, respectively. }
    \label{tab:char_tirfm}
\end{table}

\subsubsection{Illumination system}
\paragraph{}
An incident beam of excitation wavelength ($\lambda$) that passed through the objective lens is assumed to uniformly illuminate specimen. The survived photons through the use of excitation filters interact with the fluorophores in the cell model, and excite the fluorophores to the electrically excited state. The optics simulations for the focusing of the incident photons through the objective lens include a statistical model of systematic parameter rules by specifications including numerical aperture (NA), magnification, working distance, degree of aberration, correction of refracting surface radius, thickness, refractive index and details of each lens element. Details of the illumination optics are described in refs. \cite{mansuripur2009, pawley2008}. 

\begin{figure}[!h]
  \centering
	\includegraphics[width=14.0cm]{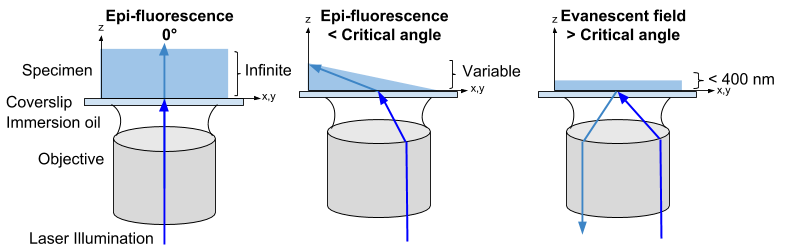}
  \caption{Epifluorescence (Left/Middle) and evanescent field (Right)}
  \label{fig:tirf_beam}
\end{figure}

\paragraph{}
The incidence angle of the beam is particularly important for the TIRFM system. Figure \ref{fig:tirf_beam} illustrates schematic views of the evanescent field and epifluorescence fields with respect to different incidence beam angles. If the incidence angles are less than the critical angles given by $\sin\theta_c = n_2/n_1$, then most of the incidence beam propagates through the interface into the lower index material with a refraction angle given by Snell's Law. However, if the incidence angle is $\theta > \theta_c$, then the incidence beam undergoes total internal refraction (TIR). The evanescent field is generated along the z-axis as perpendicular to the TIR surface, and is capable of exciting the fluorescent molecules near the surface. The intensity of the evanescent field at any position exponentially decays with $z$, and is written in the form of 
\begin{eqnarray}
I(z) & = & \left| {\bf E}_T \right|^2 = \left| {\bf A}_T \right|^2 \exp{\left( -\frac{z}{d} \right)} \\
d & = &\frac{\lambda}{4 \pi \sqrt{n_1^2 \sin^2\theta - n_2^2}}
\end{eqnarray}
where ${\bf E}_T$ and ${\bf A}_T$ are the transmitted electric field and amplitude of the incident beam as a function of incident beam angle, respectively. $d$ and $\lambda$ are the penetration depth of the evanescent field and the wavelength of the incident beam in vacuum, respectively. 

\paragraph{}
The polarization of the evanescent field depends on the incident beam polarization, which can be either p-pol (polarized in the plane of the incidence formed by the incident and reflected rays, denoted here as the x-z plane) or s-pol (polarized normal to the plane of incidence; here, in the y-direction). In both polarizations, the evanescent field fronts travel parallel to the surface in the x-direction. The p-pol evanescent field is a mixture of the transverse (z) and longitudinal (x) components; this distinguishes the p-pol evanescent field from freely propagating subcritical refracted light, which has no component longitudinal to the direction of travel. The longitudinal x component of the p-pol evanescent field range diminishes back toward the critical angle. 
\begin{eqnarray}
A_{Tx} & = & \frac{2 \cos\theta \sqrt{\sin^2\theta - n^2}}{\sqrt{n^4 \cos^2\theta + \sin^2\theta - n^2}}\ A_{Ip}\ e^{-i \left( \delta_p + \pi/2 \right)}\\
A_{Tz} & = & \frac{2 \cos\theta \sin\theta}{\sqrt{n^4 \cos^2\theta + \sin^2\theta - n^2}}\ A_{Ip}\ e^{-i \delta_p}
\end{eqnarray}
For the s-pol evanescent field, the evanescent electric field vector direction remains purely normal to the plane of incidence (y-direction).
\begin{eqnarray}
A_{Ty} & = & \frac{2 \cos\theta}{\sqrt{1 - n^2}}\ A_{Is}\ e^{-i \delta_s}
\end{eqnarray}
where AI is the field amplitude of the polarized incident beam. The phases relative to the incident beam are written as follows.
\begin{eqnarray}
\delta_p & = &\tan^{-1} \left[ \frac{\sqrt{\sin^2\theta - n^2}}{n^2 \cos\theta} \right] \\
\delta_s & = & \tan^{-1} \left[ \frac{\sqrt{\sin^2\theta - n^2}}{\cos\theta} \right] 
\end{eqnarray}
The incident electric field amplitude in the substrate is normalized to unity for each polarization. More details are described in refs. \cite{mortensen2010, miyanaga2009, axelrod2008, wazawa2005, axelrod2003}.

%

\subsubsection{Molecular fluorescence}
\paragraph{}
Incident photons of specific wavelengths are absorbed by the fluorophores in the cell model. Fluo-rescence is the result of physical and chemical processes in which the fluorophores emit photons in the electronically excited state. The excitation of the fluorophores by an incident beam photon occurs in femtoseconds. Vibrational relaxation of excited-state electrons to the lowest energy state is much slower and can be measured in picoseconds. The fluorescence process (that is, emission of a longer-wavelength photon and the return of the molecule to the ground state) occurs in a relatively long time of nanoseconds. However, the process of phosphorescence from the triplet state and back to the ground state occurs in a much longer time of microseconds. A Jablonski diagram of the fluorescence process is shown in Figure S3. The Monte Carlo simulation of the overall fluorecence process includes a statistical model of systematic parameters ruled by the observable changes in absorption and emission spectra, quantum yield, lifetime, quenching, photobleaching and blinking, anisotropy, energy transfer, solvent effects, diffusion, complex formation, and a host of environmental variables. Details are described in refs.\cite{valeur2012, lakowicz2006}.

\begin{figure}[!h]
  \centering
	\includegraphics[width=10cm]{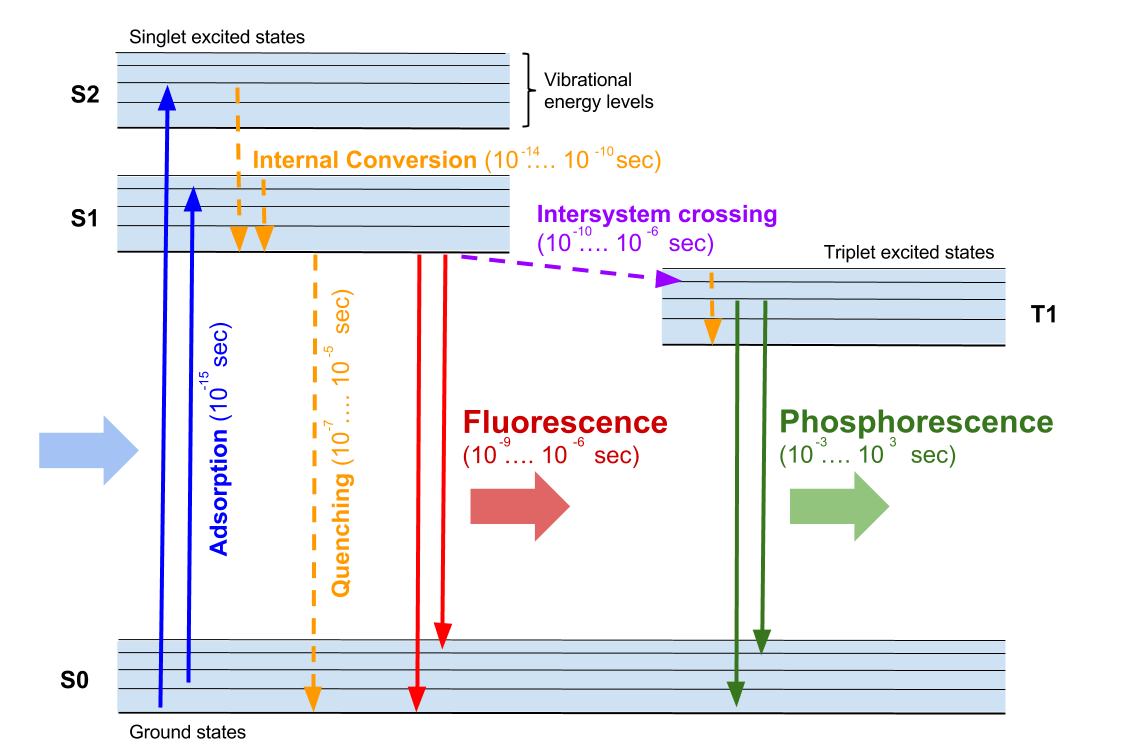}
  \caption{Jablonski diagram of molecular fluorescence and phosphorescence}
  \label{fig:diagram}
\end{figure}

\subsubsection{Image-forming system}
\begin{figure}[!h]
  \centering
	\includegraphics[width=12cm]{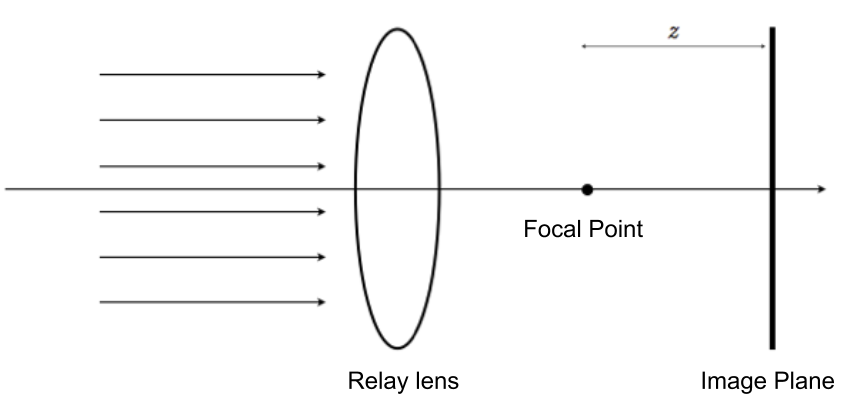}
  \caption{Boundary condition for Born-Wolf PSF model}
  \label{fig:psf_model}
\end{figure}

\paragraph{}
In an optical system that employs incoherent illumination of the cell model, the image-forming process can be considered as a linear system. The impulse response of an image-forming system to a pointlike fluorophore is described by the point spread function (PSF) of wavelength and position. The optics simulations of PSF formation and convolution include a statistical model of systematic parameters ruled by the observational changes in specifications of the objective lens and special filters. When all the fluorophores in the cell model are imaged simultaneously, the distribution of emitted photons of longer wavelengths passing through the objective lens and special filters used is computed as the sum of the PSFs of all the fluorophores. In the TIRFM system, the incident light that excites the fluorescent molecules is an evanescent field generated under the total internal reflection. The polarization of this light is non-isotropic, which means that dipoles of different orientations are excited with different probabilities per unit time. Therefore, the PSF of a fluorescent molecule should be written in the polarized form of the weighted average over orientations. Here, we use a simple analytical form of the unpolarized PSF models. The number of outgoing photons can spread depending on the PSF. The analytical forms of the PSF models are well studied in ref. \cite{kirshner2013} and is generally written as
\begin{eqnarray}
{\rm PSF}_{\lambda'} \left(r, z \right) = \left| C \int^{1}_{0} J_0(\alpha \rho r) \exp{\left( -j \psi \right)}\rho d\rho \right|^2
  \label{eqn:psf_model}
\end{eqnarray}
where $\alpha = \frac{2 \pi}{\lambda'} \frac{\rm N.A.}{n}$. $\lambda'$and $n$ are fluorophore wavelength and the refractive index of the immersion layer. The phase factor, $\psi = \psi(r, z, \rho)$ enables generating the second Airy peak along the z-axis. In particular, we use Born-Wolf PSF model given by
\begin{eqnarray}
{\rm PSF}_{\lambda'} \left(r, z \right) = \left| C \int^{1}_{0} J_0(\alpha \rho r) \exp{\left( - j \beta \rho^2 z \right)}\rho d\rho \right|^2
  \label{eqn:bw_psf_model}
\end{eqnarray}
where $\alpha = \frac{2 \pi}{\lambda'} \frac{\rm N.A.}{n}$, $\beta = \frac{1}{2} \frac{2 \pi}{\lambda} \left( \frac{\rm N.A.}{n} \right)^2$ and $C$ is the normalisation factor. In this PSF model, when the particle is in focus, the scalar-based diffraction can occur in the microscopy system, but the imaging plane is not required to be in focus. The model is shift invariant in all directions. A schematic view of the model condition is shown in Figure \ref{fig:psf_model}.

\paragraph{Detection :}
The Monte Carlo simulation of a camera system includes a statistical model of a systematic source and generates digital images that closely represents the actual image obtained using an actual camera. We particularly simulate the detection process for CMOS and EMCCD cameras. Details of the camera simulations are described as follows.

\newpage

\begin{enumerate}
\item[(1)] Uncertainty sources: Uncertainty sources of the camera systems are ruled by camera specifications and conditions shown in Table \ref{tab:camera_spec} \cite{cse0, cse1}. First, shot noise arises from statistical fluctuations in the number of photons incident to the camera. This noise source is a fundamental property of the quantum nature of light and is always present in imaging systems. The incident photons interact with the photodiode placed on a pixel plate. Photoelectric effects can convert the incident photon signals to photoelectrons. The probability for such a conversion is the so-called quantum efficiency (QE). As both photons and electrons are quantized, the detection process is characterized by binomial distributions. Finally, readout noise is generated whereas the photoelectron signals can be linearly digitized as an image in terms of the count 16-bit analog-to-digital converter (ADC). For CMOS and EMCCD cameras, the linear relationships of photoelectrons outputs with ADC outputs are shown in Figure \ref{fig:adc_linear}.

\paragraph{}
n addition, the EMCCD camera has excess noise that increases the standard deviation of the output signal by $\sqrt{2}$ \cite{mortensen2010, hirsch2013, emccd2009, robbins2003}], whereas the CMOS and CCD cameras have no excess noise ($1.0$). The EMCCD camera uses the multiplication process, and each stage has a small gain to multiply the number of photoelectrons. Such process is stochastic and characterized by multistage binomial distributions, which increased noise.

\begin{table}[!h]
    \centering
	\begin{tabular}{|l||p{5.0cm}|p{5.0cm}|}
	\hline
	Camera type & EMCCD & CMOS\\ \hline
	Image size & $100 \times 100$ & $100 \times 100$ \\ \hline
	QE & $92\ \%$ & $70\ \%$ \\ \hline
	EM Gain & $\times\ 1$, $\times\ 100$, $\times\ 300$ & N/A \\ \hline
	Exposure time & $30\ {\rm msec}$ & $30\ {\rm msec}$ \\ \hline
	Readout noise & $100\ {\rm electrons}$ & $1.3\ {\rm electrons}$ \\ \hline
	Excess noise & $\sqrt{2}$ & $1$ \\ \hline
	A/D Converter & $16\ {\rm bit}$ & $16\ {\rm bit}$ \\ \hline
	Gain & $5.82\ {\rm electrons/count}$ & $0.458\ {\rm electrons/count}$ \\ \hline
	Offset  & $2000\ {\rm counts}$ & $100\ {\rm counts}$ \\ \hline
	Full well & $370,000\ {\rm electrons}$ & $30,000\ {\rm electrons}$ \\ \hline
	Dynamic range & $71.3\ {\rm dB}$ & $87.2\ {\rm dB}$ \\ \hline
	\end{tabular}
	\caption{Camera specification and condition}
	\label{tab:camera_spec}
\end{table}

\begin{figure}[!h]
  \centering
	\includegraphics[width=7.5cm]{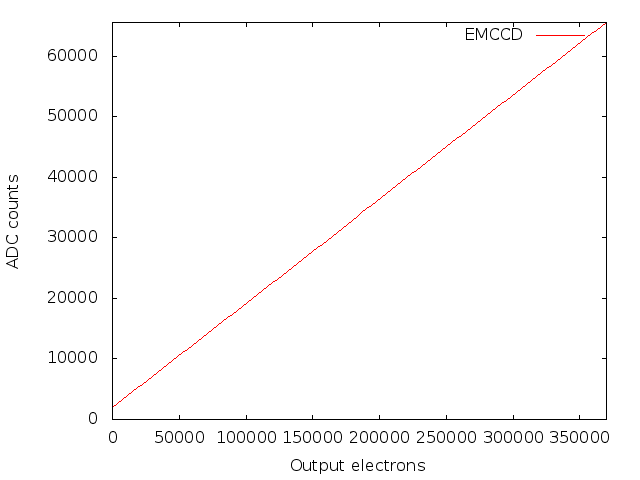}
	\includegraphics[width=7.5cm]{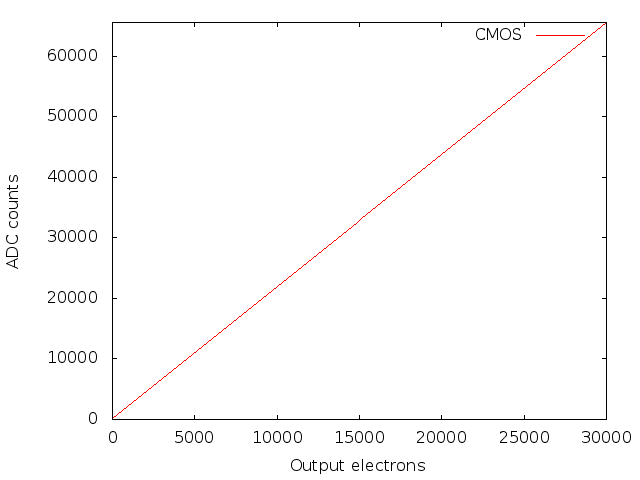}
  \caption{A/D converter linearity for EMCCD (left) and CMOS (right)}
  \label{fig:adc_linear}
\end{figure}

\item[(2)] Probability density function (PDF) per pixel: The camera pixel output is the convolution of the probability distributions of each of the systematic sources. The PDF of CMOS camera pixels is given by the Poisson distribution and written in the form of
\begin{eqnarray}
q{\left( S_i | E_i \right)} & = & \frac{E_i^{S_i} e^{-E_i}}{S_i !}
\end{eqnarray}
where $S_i$ and $E_i$ are the random number of output electrons and expectations in the $i$-th pixel, respectively. The left panel of Figure \ref{fig;pdf_camera} shows the PDF with respect to the number of incident photons. The PDF of EMCCD camera pixels \cite{mortensen2010, hirsch2013} is written in the form of
\begin{eqnarray}
q{\left( S_i | E_i \right)} & = & \exp{\left(-E_i \right)} \delta{\left(S_i\right)} + \sqrt{\frac{\alpha E_i}{S_i}} \exp{\left( -\alpha S_i - E_i \right)}\  I_1{\left( 2\sqrt{\alpha E_i S_i} \right)}
\end{eqnarray}
where $I_1$ is the modified Bessel function of the first kind of order one. $\alpha$ is the inverse of the EM gain. Figure \ref{fig;pdf_emgain} and the right panel of Figure \ref{fig;pdf_camera} show the PDF with respect to the number of incident photons. 

\begin{figure}[!h]
  \centering
	\includegraphics[width=7.5cm]{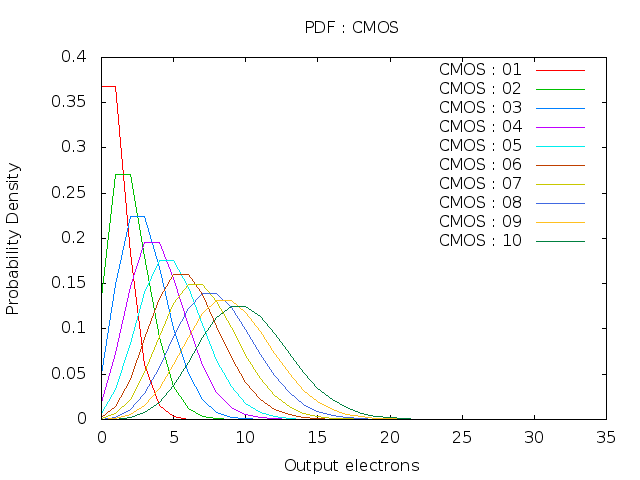}
	\includegraphics[width=7.5cm]{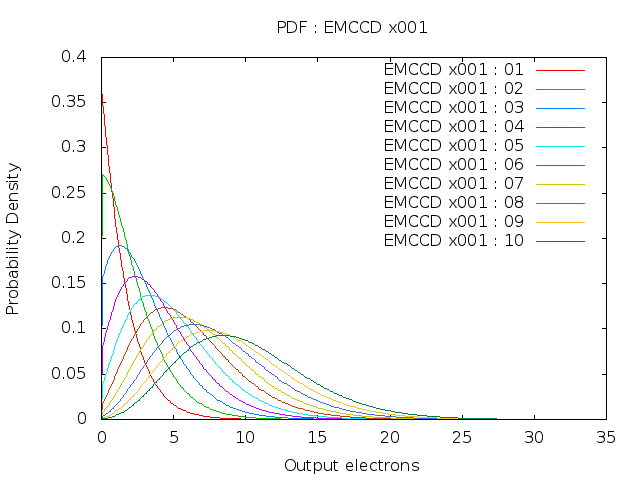}
  \caption{Probability density function for CMOS (left) and EMCCD  (right)}
  \label{fig;pdf_camera}
%
  \centering
	\includegraphics[width=7.5cm]{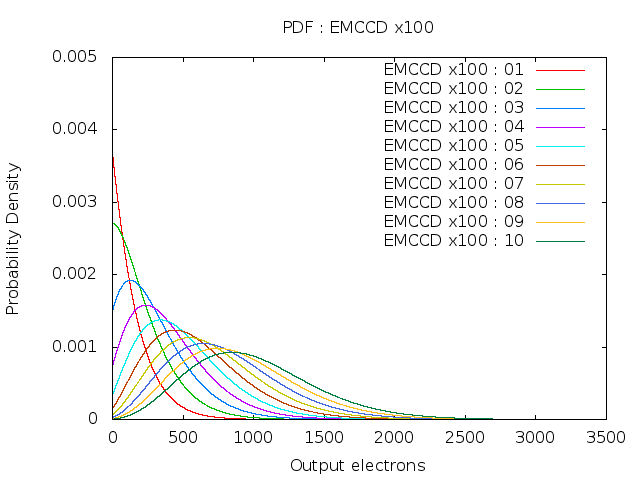}
	\includegraphics[width=7.5cm]{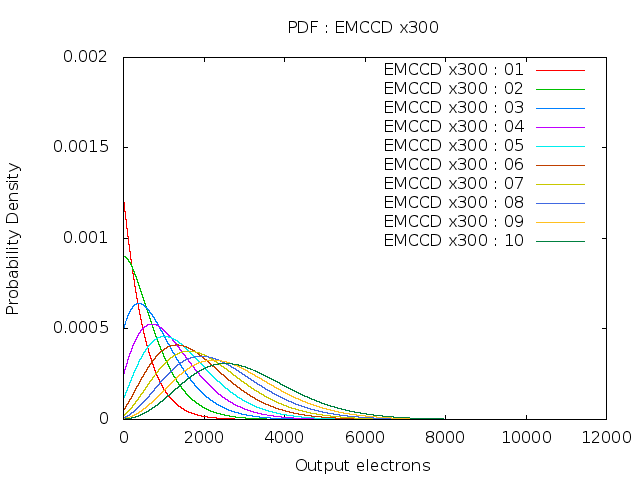}
  \caption{Probability density function for EMCCD : EM gain $\times 100$ (left) and $\times 300$ (right)}
  \label{fig;pdf_emgain}
\end{figure}

\item[(3)] Readout noise: Noise triggered by the readout electronics is typically dominated by the noise on the floating diffusion amplifier and the A/D converter. It increases with clocking speed or frame readout speed. This noise is the result of the statistical uncertainty that occurs when the amplifier attempted to reset itself to zero before the next image. The readout noise distribution for the EMCCD camera is usually Gaussian, as shown on the left panel of Figure S8 and Figure S9. However, the readout noise distribution for the CMOS camera is uneven, because of the differences in characteristics of the amplifiers in each pixel. The distribution is shown on the right panel of Figure \ref{fig;pdf_readout} and Figure \ref{fig;cmos_readout}.

\begin{figure}[!h]
  \centering
	\includegraphics[width=7.5cm]{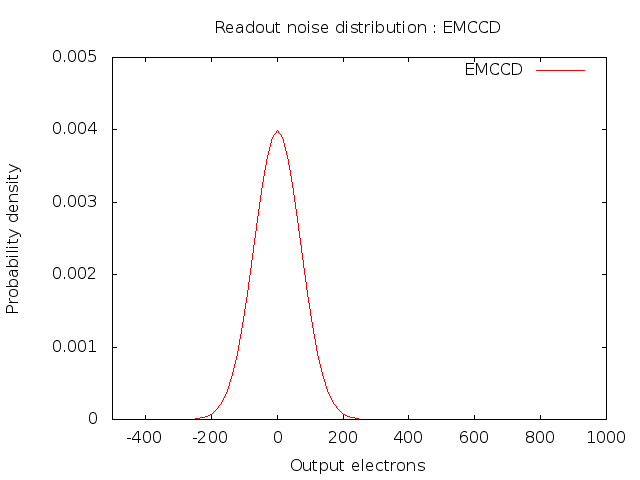}
	\includegraphics[width=7.5cm]{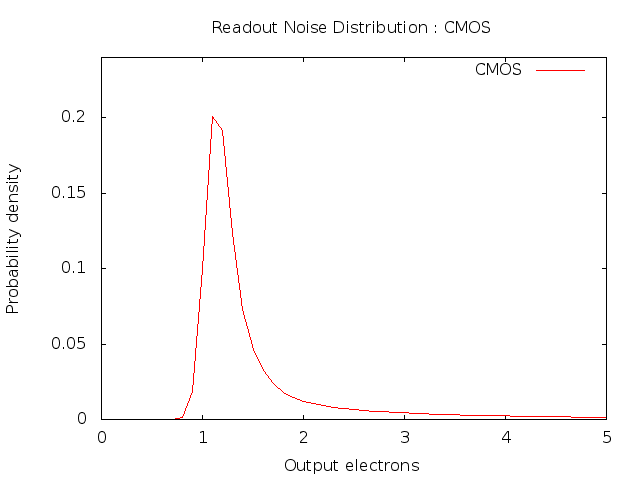}
  \caption{Readout noise distributions for EMCCD (left) and CMOS (right)}
  \label{fig;pdf_readout}

  \centering
	\includegraphics[width=4.0cm]{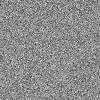}
	\includegraphics[width=6.0cm]{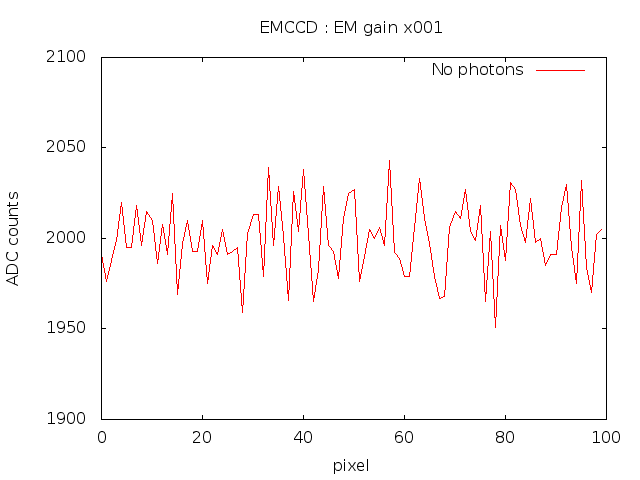}
  \caption{Readout noise for EMCCD. Image (left) and its intensity graph that depict the readout noise of horizontal line at vertical center (right)}
  \label{fig;emccd_readout}

  \centering
	\includegraphics[width=4.0cm]{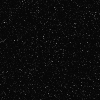}
	\includegraphics[width=6.0cm]{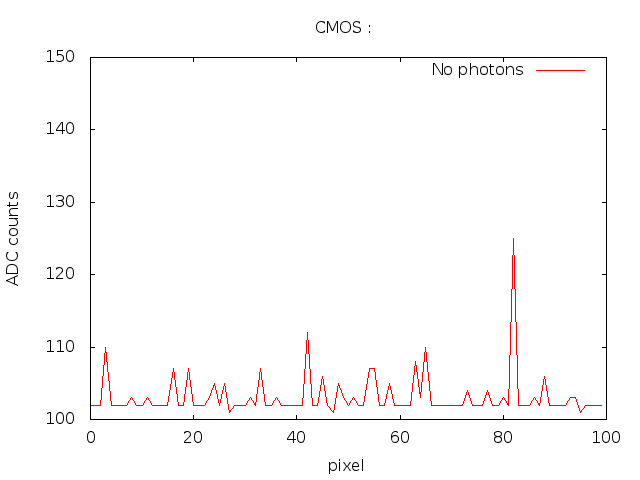}
  \caption{Readout noise for CMOS. Image (left) and its intensity graph that depict the readout noise of horizontal line at vertical center (right)}
  \label{fig;cmos_readout}
\end{figure}

\item[(4)] SNR per pixel: The variance of the camera pixel output is given by the sum of the variances of all uncertainty sources. The SNR, which is the ratio of the output signal to the standard deviation of the signal, is written in the form of
\begin{eqnarray}
SNR & = & \frac{QE \cdot S}{\sqrt{F_n^2 \cdot QE \cdot \left( S + I_b \right) + \left( N_r/M \right)^2}}
\end{eqnarray}
The input photon signal ($S$) and optical background ($I_b$) falling on the photodiodes have average photon flux per pixel. The fluctuations at this rate are governed by Poisson statistics and therefore have a standard deviation that is the square root of the number of photons ($\sqrt{S + I_b}$). The quantum efficiency ($QE$) of the camera is the wavelength dependent probability that photon is converted to a photoelectron. Since the QE is predominated in the SNR equation, high QE is a fundamental attribute for obtaining high SNR. Readout noise ($N_r$) is a statistical expression of the variability within the electrons that convert the charge of the photoelectrons in each pixel to the number of ADC counts.  EM gain ($M$) is a factor of electron multiplication. It occurs in voltage dependent, stepwise manner and the total amour is a combination of the voltage applied and number of steps in EM register. The EM gain also has a statistical distribution and an associated variance accounted for the excess noise factor ($F_n$). The SNR and relative SNR for three cameras specifications are shown in Figure \ref{fig:snr_nobg} and \ref{fig:snr_05bg}. 

\end{enumerate}

\begin{figure}[!h]
  \centering
	\includegraphics[width=7.5cm]{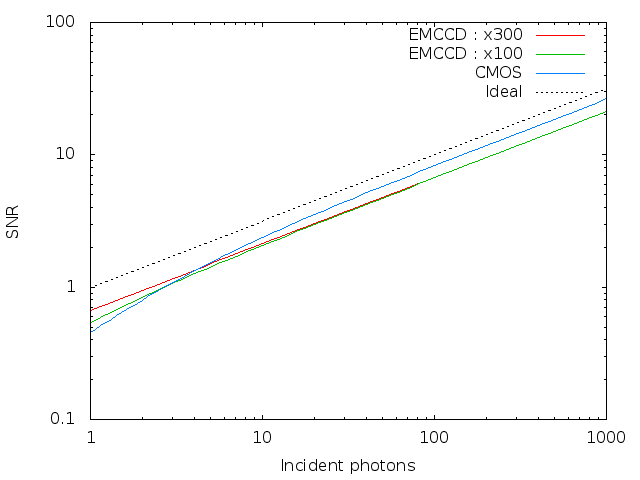}
	\includegraphics[width=7.5cm]{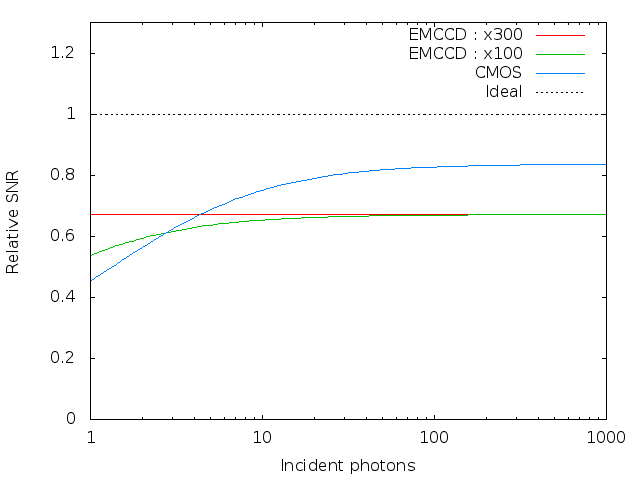}
  \caption{No background photons, SNR (left) and relative SNR (right) for CMOS and EMCCD}
  \label{fig:snr_nobg}
\end{figure}

\begin{figure}[!h]
  \centering
	\includegraphics[width=7.5cm]{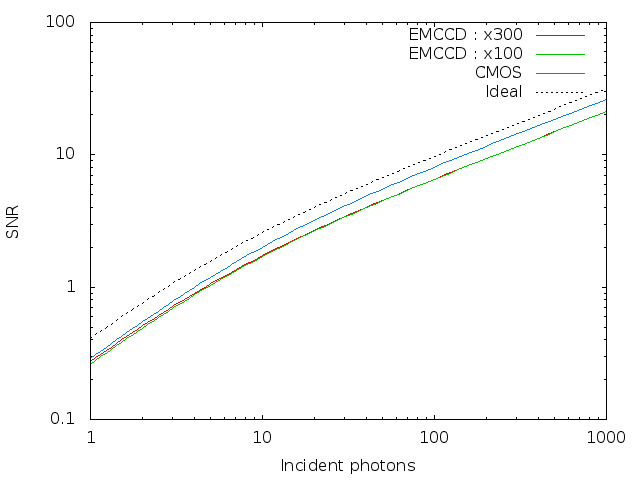}
	\includegraphics[width=7.5cm]{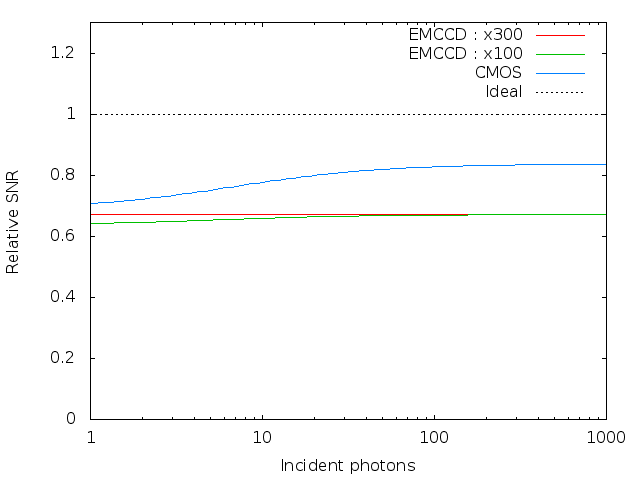}
  \caption{Five background photons, SNR (left) and relative SNR (right) for CMOS and EMCCD}
  \label{fig:snr_05bg}
\end{figure}

\subsubsection{Examples of images}
\begin{figure}[!h]
  \centering
  	{\bf No background photon} \hspace{3.0cm} {\bf 5 background photons}
	\vspace{0.1cm}

  \centering
	\includegraphics[width=2.8cm]{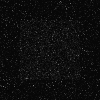}
	\includegraphics[width=3.8cm]{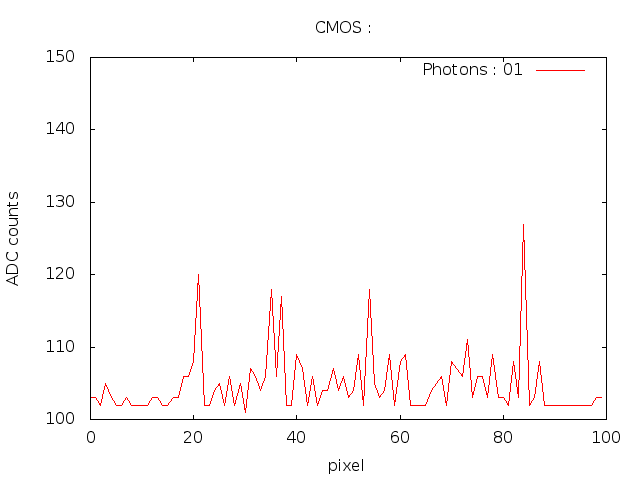}
	\includegraphics[width=2.8cm]{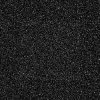}
	\includegraphics[width=3.8cm]{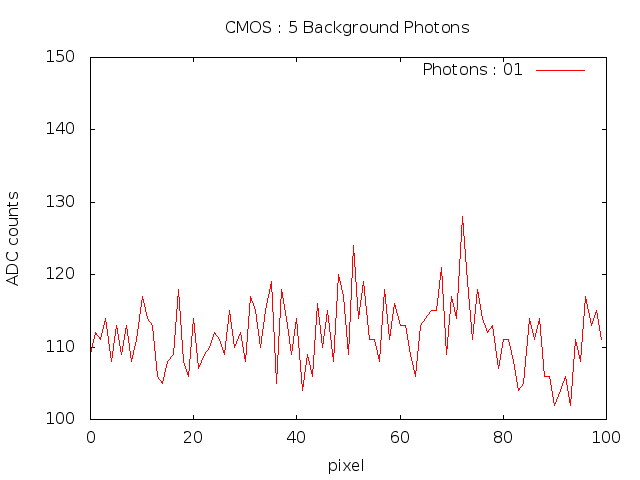}
	
  \centering
	\includegraphics[width=2.8cm]{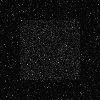}
	\includegraphics[width=3.8cm]{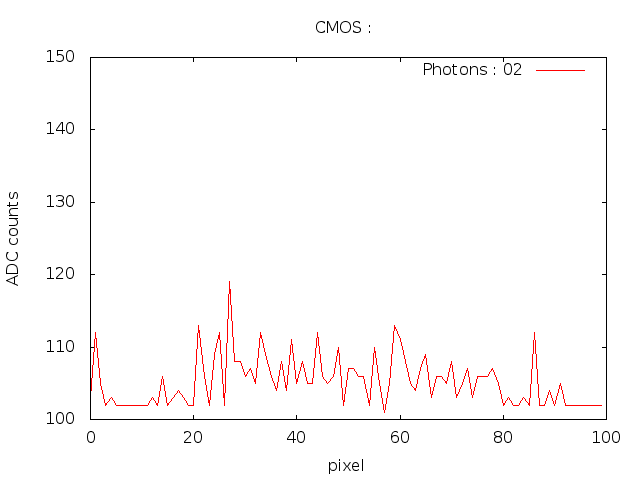}
	\includegraphics[width=2.8cm]{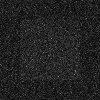}
	\includegraphics[width=3.8cm]{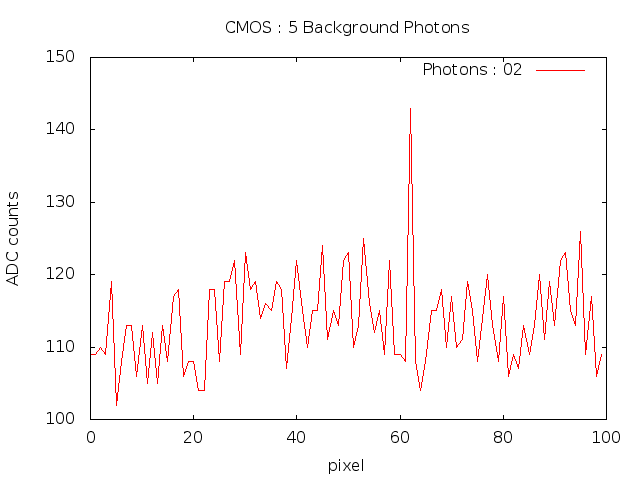}
	
  \centering
	\includegraphics[width=2.8cm]{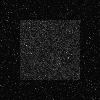}
	\includegraphics[width=3.8cm]{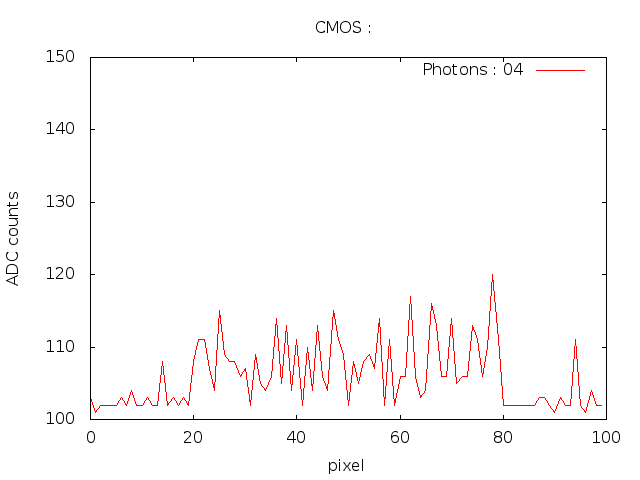}
	\includegraphics[width=2.8cm]{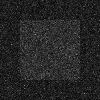}
	\includegraphics[width=3.8cm]{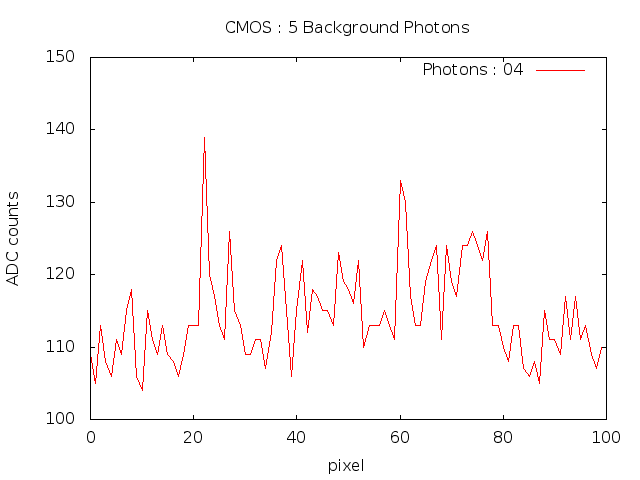}
	
  \centering
	\includegraphics[width=2.8cm]{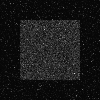}
	\includegraphics[width=3.8cm]{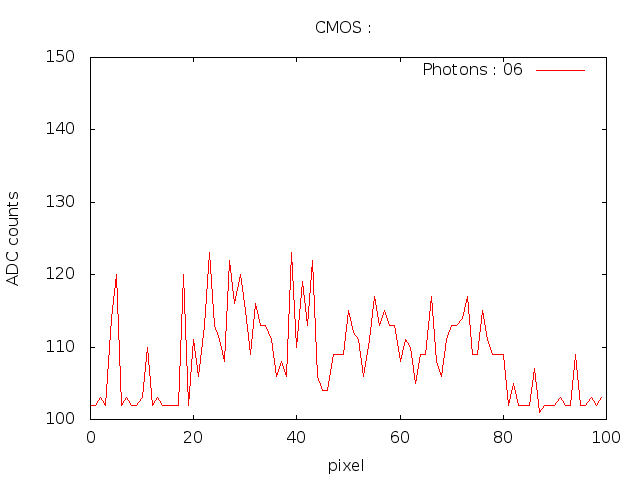}
	\includegraphics[width=2.8cm]{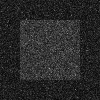}
	\includegraphics[width=3.8cm]{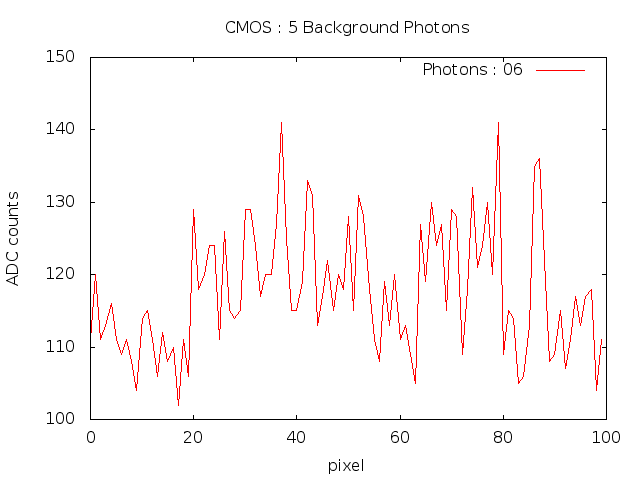}
	
  \centering
	\includegraphics[width=2.8cm]{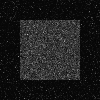}
	\includegraphics[width=3.8cm]{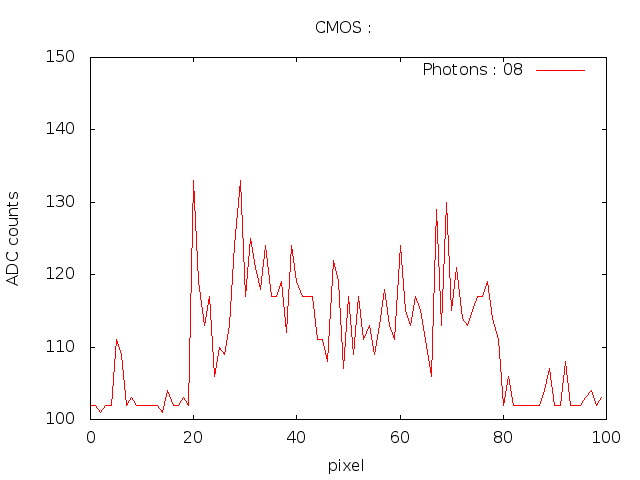}
	\includegraphics[width=2.8cm]{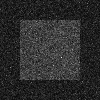}
	\includegraphics[width=3.8cm]{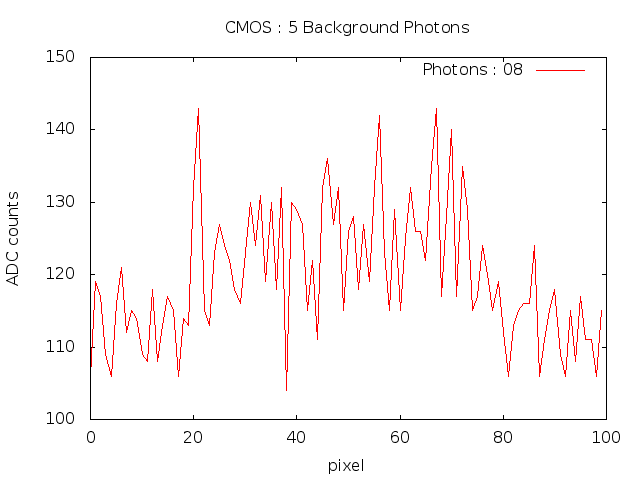}
	
  \centering
	\includegraphics[width=2.8cm]{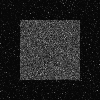}
	\includegraphics[width=3.8cm]{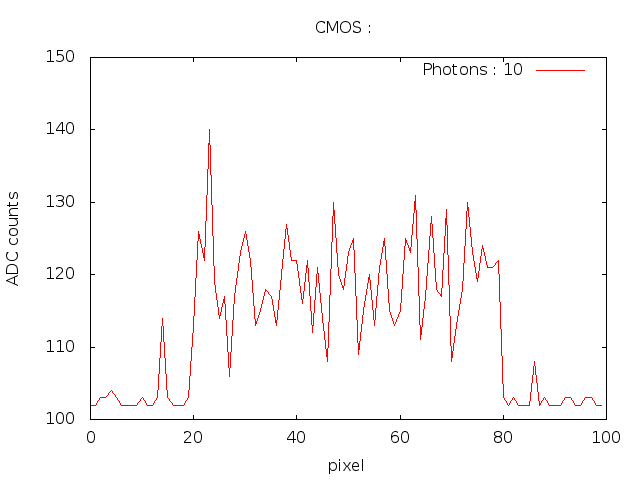}
	\includegraphics[width=2.8cm]{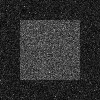}
	\includegraphics[width=3.8cm]{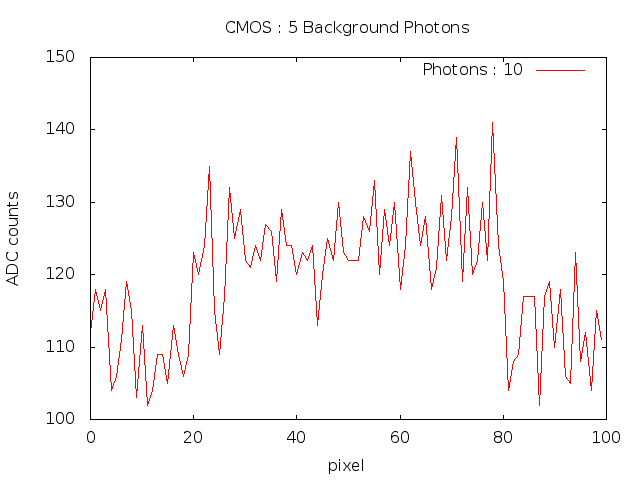}
	
  \caption{CMOS}
  \label{fig:cmos_images}
\end{figure}

\begin{figure}[!h]
  \centering
  	{\bf No background photon} \hspace{3.0cm} {\bf 5 background photons}
	\vspace{0.1cm}

  \centering
	\includegraphics[width=2.8cm]{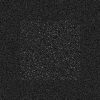}
	\includegraphics[width=3.8cm]{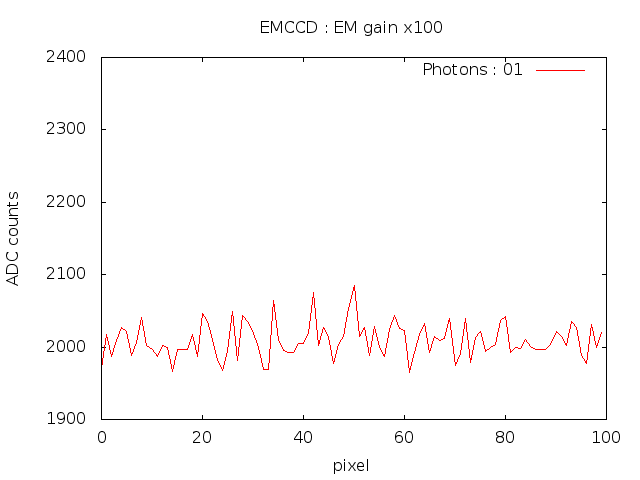}
	\includegraphics[width=2.8cm]{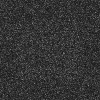}
	\includegraphics[width=3.8cm]{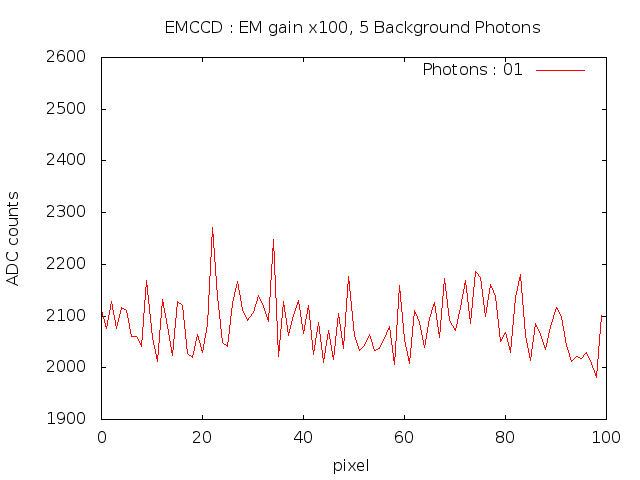}
	
  \centering
	\includegraphics[width=2.8cm]{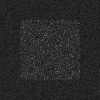}
	\includegraphics[width=3.8cm]{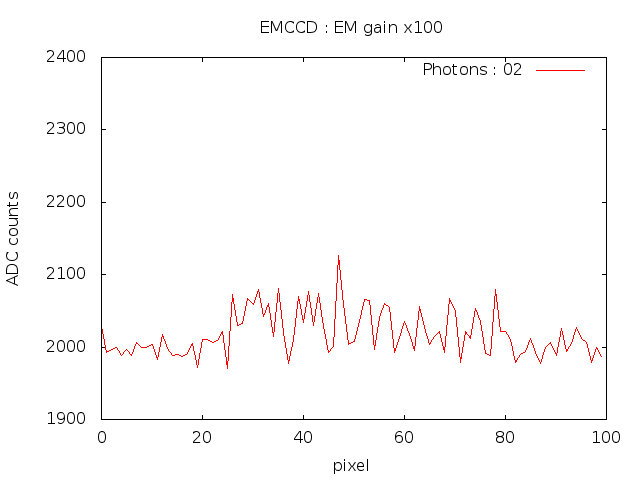}
	\includegraphics[width=2.8cm]{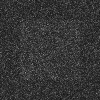}
	\includegraphics[width=3.8cm]{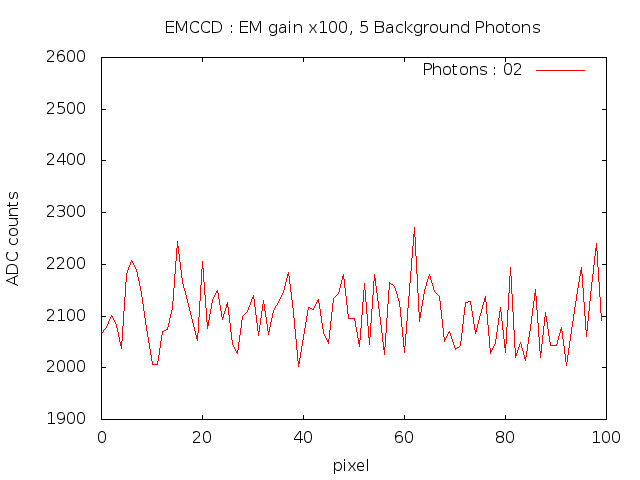}
	
  \centering
	\includegraphics[width=2.8cm]{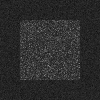}
	\includegraphics[width=3.8cm]{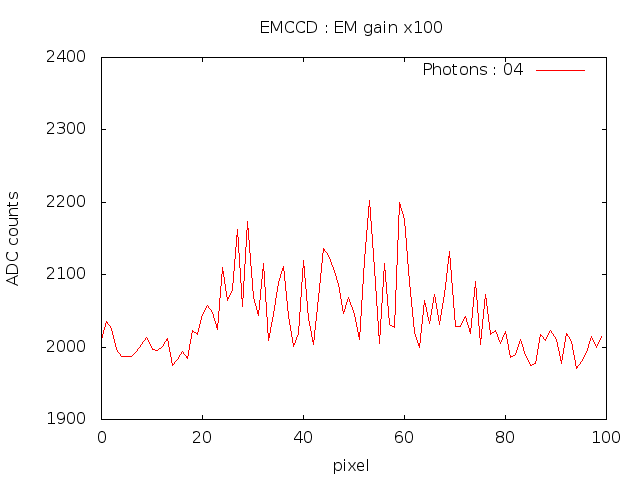}
	\includegraphics[width=2.8cm]{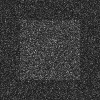}
	\includegraphics[width=3.8cm]{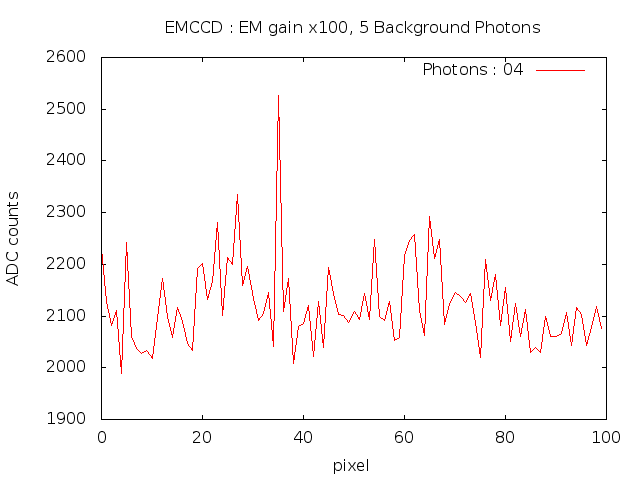}
	
  \centering
	\includegraphics[width=2.8cm]{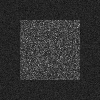}
	\includegraphics[width=3.8cm]{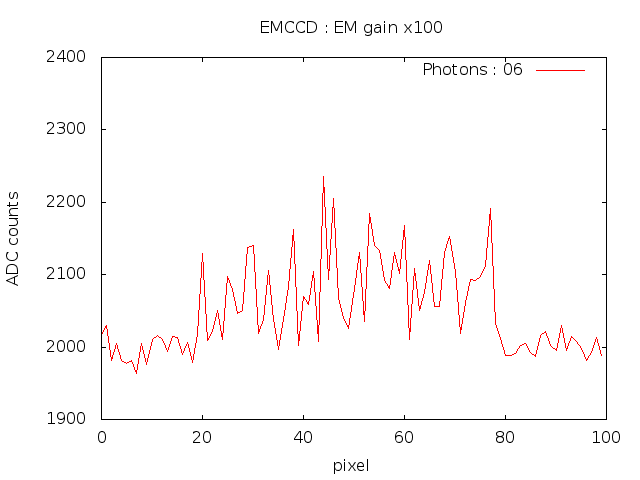}
	\includegraphics[width=2.8cm]{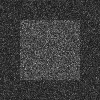}
	\includegraphics[width=3.8cm]{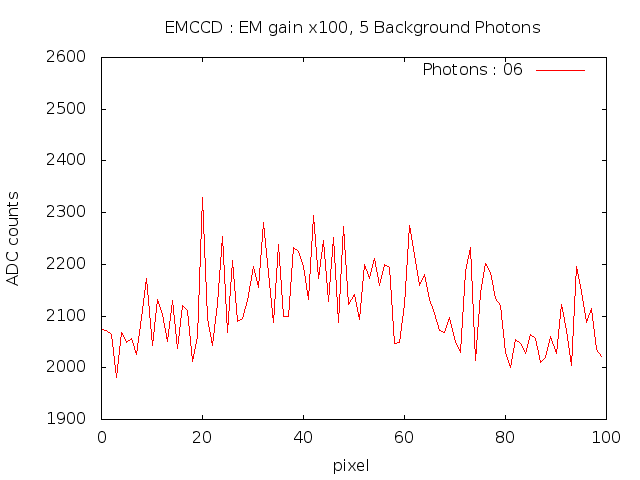}
	
  \centering
	\includegraphics[width=2.8cm]{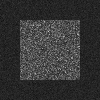}
	\includegraphics[width=3.8cm]{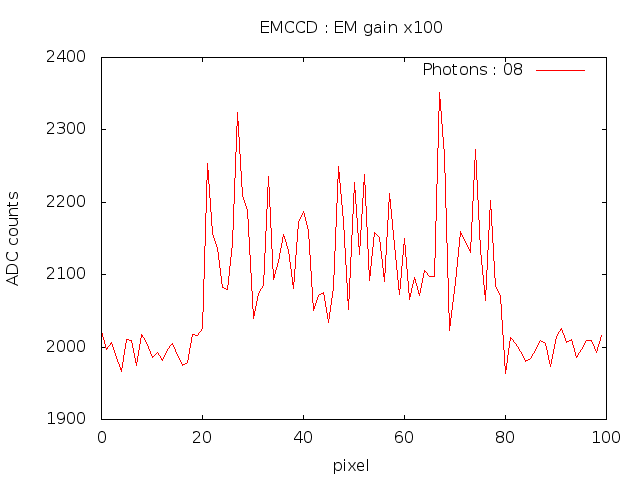}
	\includegraphics[width=2.8cm]{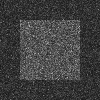}
	\includegraphics[width=3.8cm]{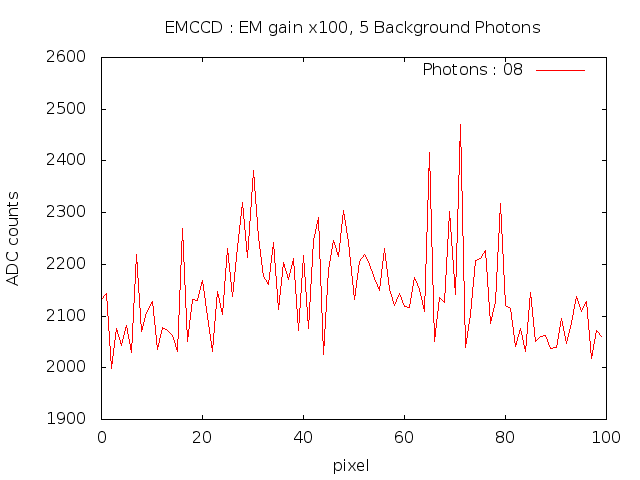}
	
  \centering
	\includegraphics[width=2.8cm]{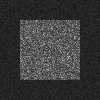}
	\includegraphics[width=3.8cm]{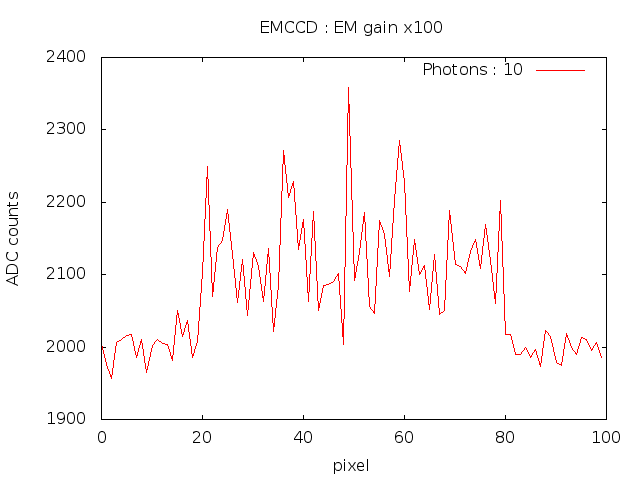}
	\includegraphics[width=2.8cm]{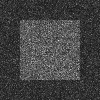}
	\includegraphics[width=3.8cm]{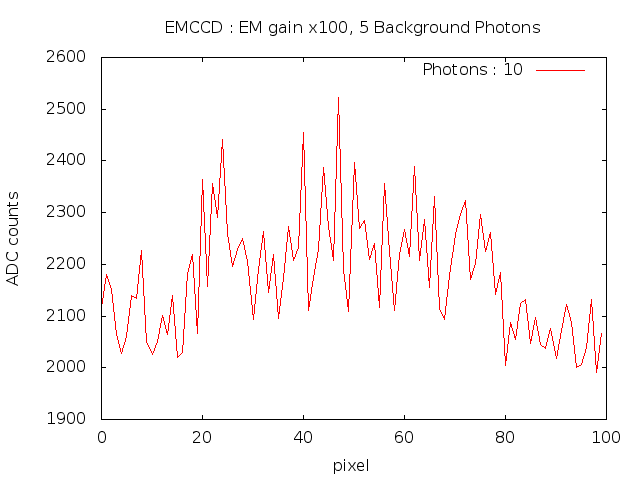}
	
  \caption{EMCCD EM gain $\times 100$}
  \label{fig:emccd100_images}
\end{figure}

\begin{figure}[!h]
  \centering
  	{\bf No background photon} \hspace{3.0cm} {\bf 5 background photons}
	\vspace{0.1cm}

  \centering
	\includegraphics[width=2.8cm]{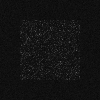}
	\includegraphics[width=3.8cm]{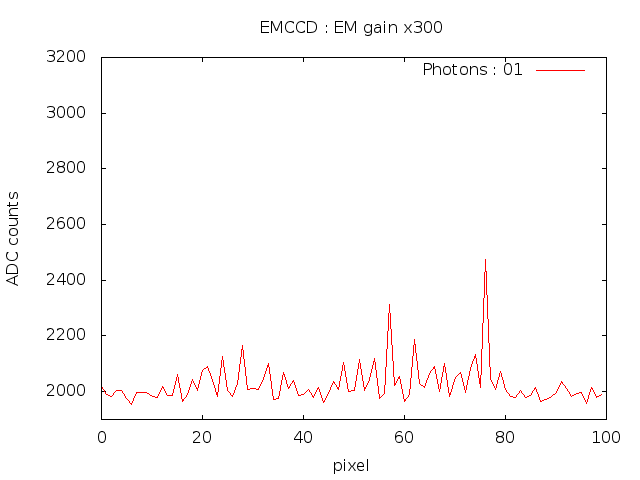}
	\includegraphics[width=2.8cm]{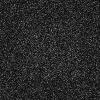}
	\includegraphics[width=3.8cm]{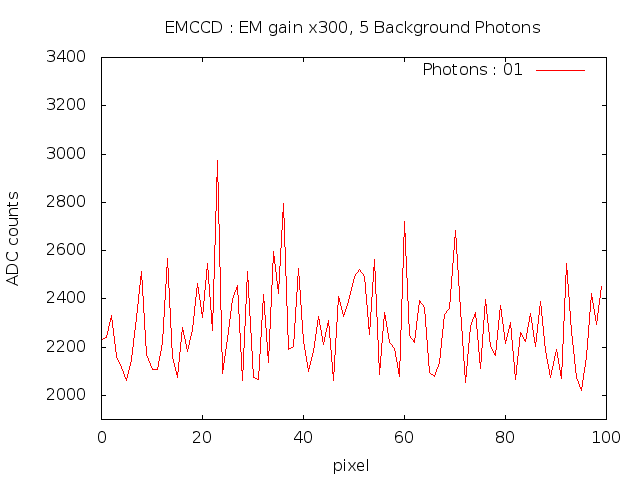}

  \centering
	\includegraphics[width=2.8cm]{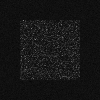}
	\includegraphics[width=3.8cm]{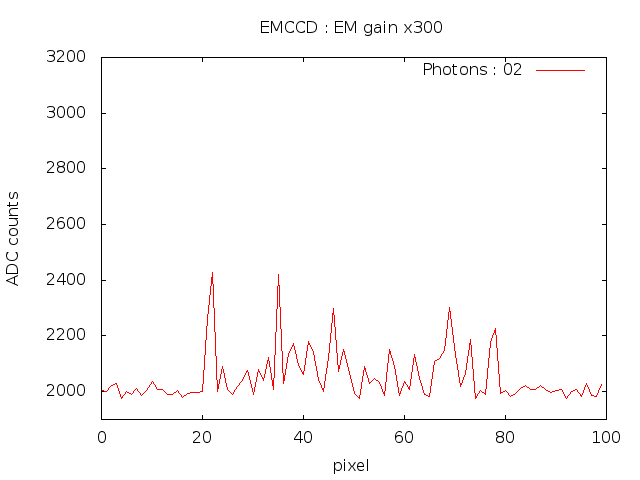}
	\includegraphics[width=2.8cm]{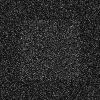}
	\includegraphics[width=3.8cm]{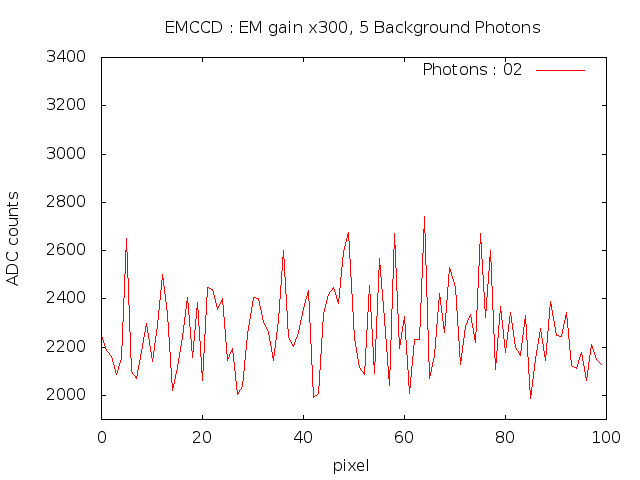}

  \centering
	\includegraphics[width=2.8cm]{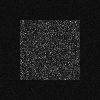}
	\includegraphics[width=3.8cm]{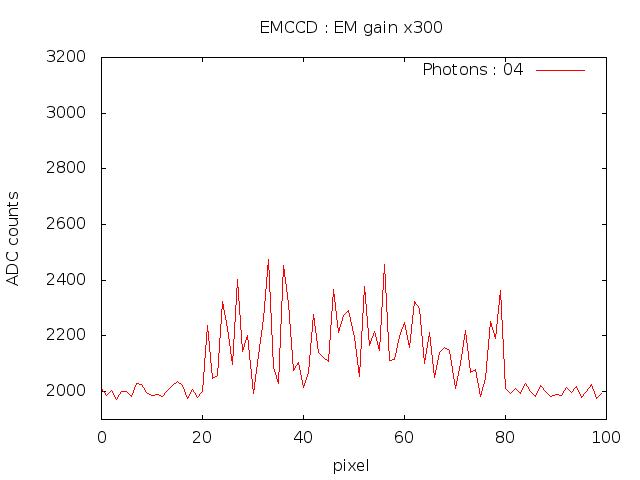}
	\includegraphics[width=2.8cm]{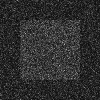}
	\includegraphics[width=3.8cm]{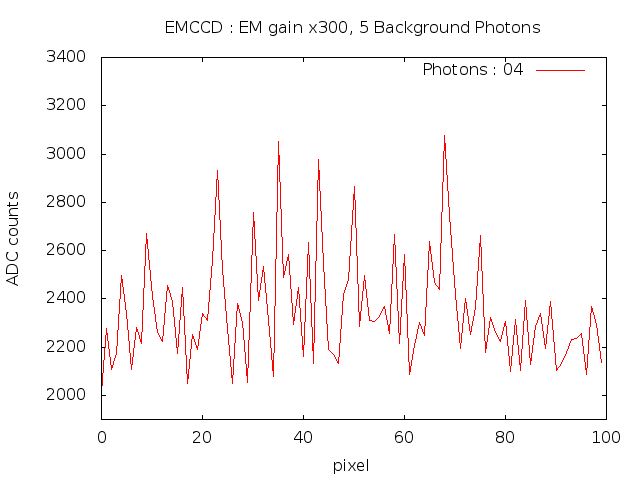}

  \centering
	\includegraphics[width=2.8cm]{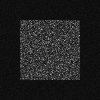}
	\includegraphics[width=3.8cm]{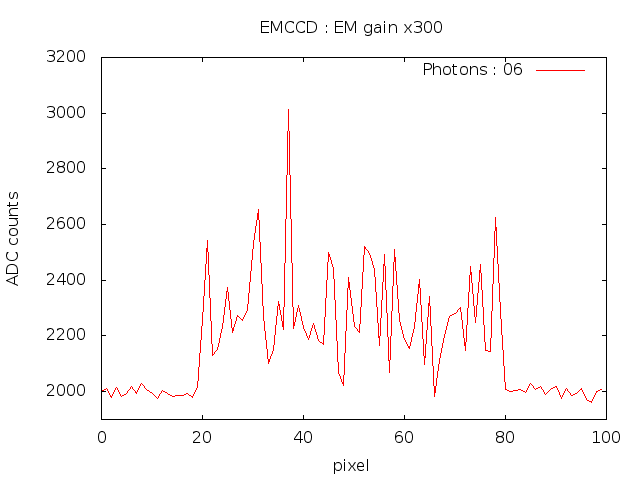}
	\includegraphics[width=2.8cm]{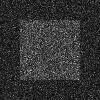}
	\includegraphics[width=3.8cm]{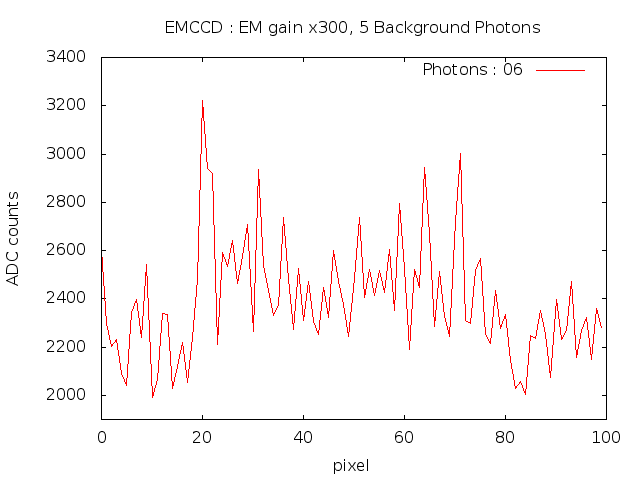}

  \centering
	\includegraphics[width=2.8cm]{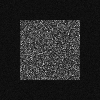}
	\includegraphics[width=3.8cm]{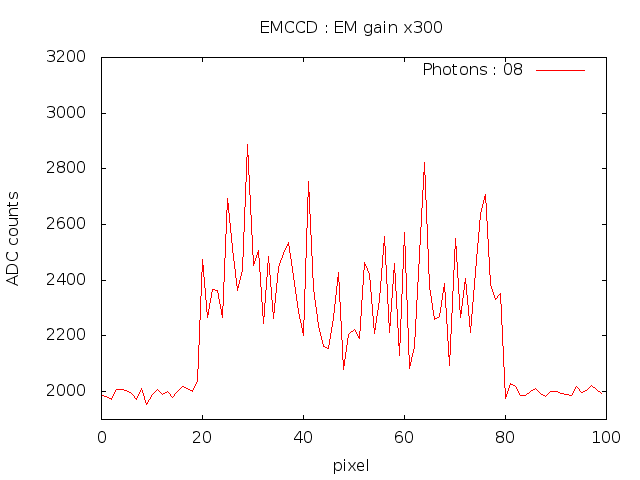}
	\includegraphics[width=2.8cm]{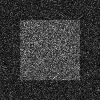}
	\includegraphics[width=3.8cm]{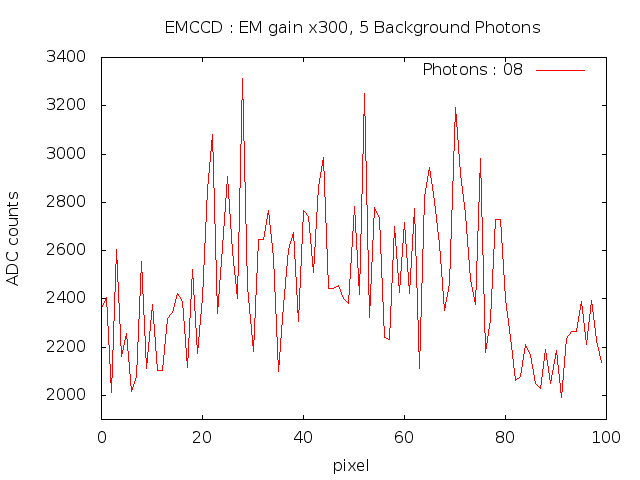}

  \centering
	\includegraphics[width=2.8cm]{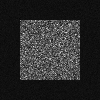}
	\includegraphics[width=3.8cm]{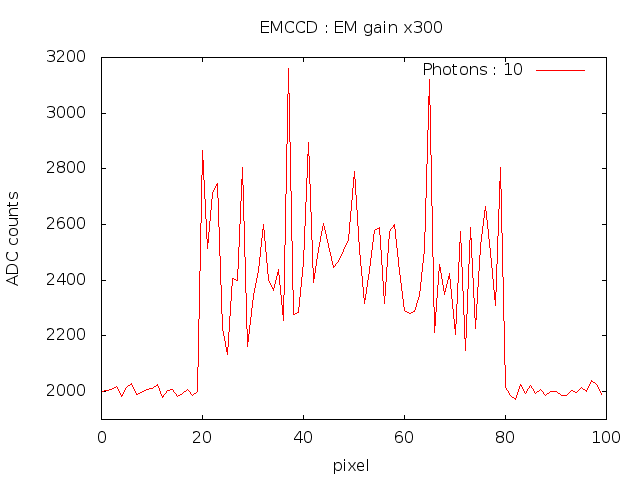}
	\includegraphics[width=2.8cm]{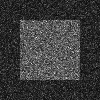}
	\includegraphics[width=3.8cm]{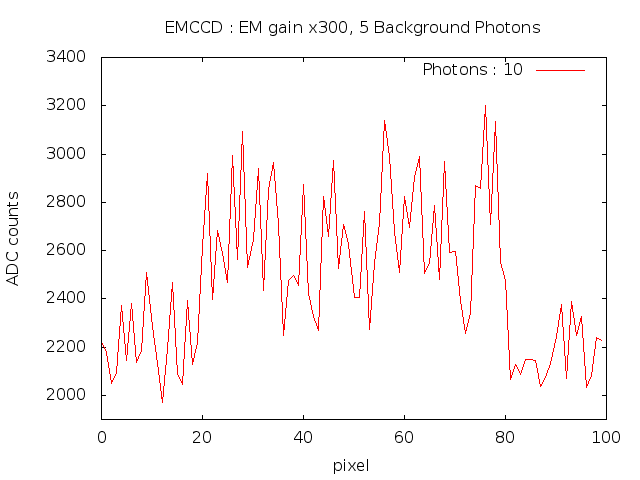}
	
  \caption{EMCCD EM gain $\times 300$}
  \label{fig:emccd300_images}
\end{figure}

\paragraph{}
Figure \ref{fig:cmos_images}, \ref{fig:emccd100_images} and \ref{fig:emccd300_images} show images and their intensity graphs showing the signal intensity and noise of the horizontal line at the vertical center. From the top to bottom rows in each Figure, $1, 2, 4, 6, 8$, and $10$ incident photons per pixel are expected in $80 \times 80$ pixel squares at the image center.

\newpage

\newpage

\newpage

\newpage

\begin{figure}[!h]
  \centering
	\includegraphics[width=7cm]{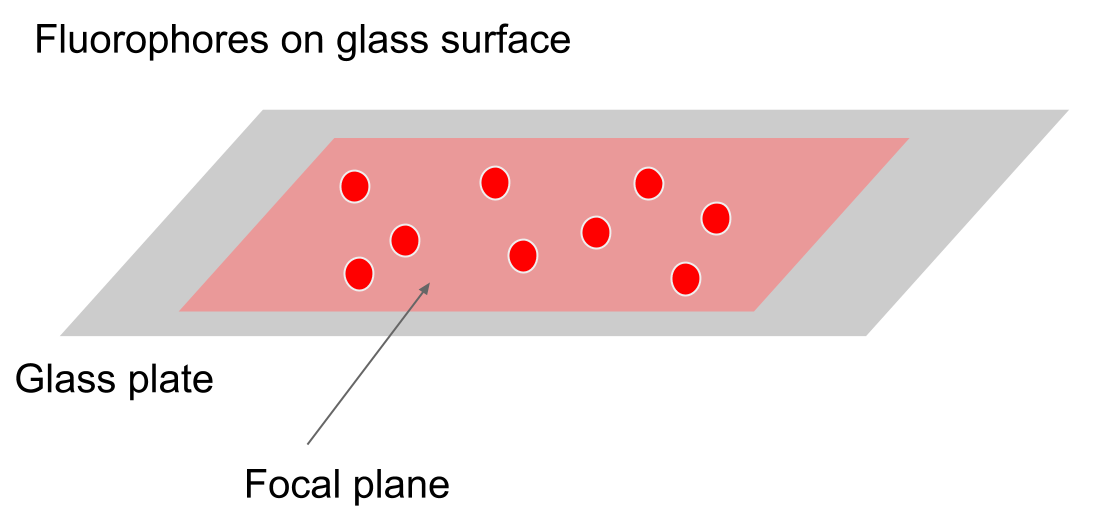}
	\includegraphics[width=3cm]{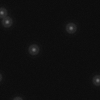}

  \caption{Fluorophores on glass surface (left) and the expected image (right).}
  \label{fig;simple_model_00}
\end{figure}

\begin{figure}[!h]
  \centering
  	{\bf CMOS} \hspace{2.2cm} {\bf EMCCD $\times 300$} \hspace{1.0cm} {\bf EMCCD $\times 500$}
	\vspace{0.1cm}


  \centering
	\includegraphics[width=4.0cm]{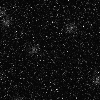}
	\includegraphics[width=4.0cm]{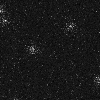}
	\includegraphics[width=4.0cm]{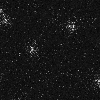}

  \centering
	\includegraphics[width=4.0cm]{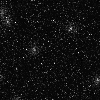}
	\includegraphics[width=4.0cm]{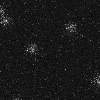}
	\includegraphics[width=4.0cm]{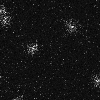}

  \centering
	\includegraphics[width=4.0cm]{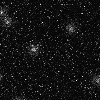}
	\includegraphics[width=4.0cm]{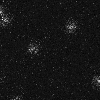}
	\includegraphics[width=4.0cm]{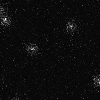}

  \centering
	\includegraphics[width=4.0cm]{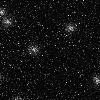}
	\includegraphics[width=4.0cm]{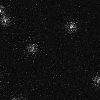}
	\includegraphics[width=4.0cm]{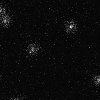}
	
  \caption{Comparison of single molecule images ($100 \times 100$ pixels at image center). Increasing the beam flux density results in relatively smaller noise. Grayscale is the count number of 16-bit ADC. }
  \label{fig:images_tirfm_simple_model_00}
\end{figure}

\paragraph{Simple model 1 :}
We constructed relatively simple particle model of TMR on glass surface as shown in left of Figure \ref{fig;simple_model_00}. We assumed that $100$ TMR molecules are stationary, and randomly distributed on the surface ($30 \times 30\ {\rm \mu m^2}$). Images are simulated for the optical specification and condition of the TIRFM simulation module shown in Table \ref{tab;spec_tirfm_simple_model_00}. Right of Figure \ref{fig;simple_model_00} is the expected image averaged by $160$ images captured with CMOS camera. Results are shown in Figure \ref{fig:images_tirfm_simple_model_00}. Figures from top row to bottom one correspond to the beam inputs of $20, 30, 40$ and $50\ {\rm W/cm^2}$, respectively.

\begin{table}[!h]
\centering
\begin{tabular}{|l||p{5.0cm}|p{5.0cm}|}
\hline
Beam flux density & \multicolumn{2}{c|}{$20,\ 30,\ 40,\ 50\ {\rm W/cm^2}$} \\ \hline
Beam wavelength & \multicolumn{2}{c|}{$488\ {\rm nm}$} \\ \hline
Refraction index & \multicolumn{2}{c|}{$1.33\ ({\rm glass})$, $1.27\ ({\rm water})$} \\ \hline
Critical angle & \multicolumn{2}{c|}{$65.6^{\circ}$} \\ \hline
Fluorophore & \multicolumn{2}{c|}{TRITC\ (${\rm Abs.}\ 548\ {\rm nm} /\ {\rm Em.}\ 608\ {\rm nm}$)} \\ \hline
Objective & \multicolumn{2}{c|}{$\times\ 60\ /\ {\rm N.A.}\ 1.40$} \\ \hline
Dichroic mirror & \multicolumn{2}{c|}{Semrok FF-562-Di03} \\ \hline
Emission filter & \multicolumn{2}{c|}{Semrok FF-593-25/40} \\ \hline
Linear conversion & \multicolumn{2}{c|}{$10^{-6}$} \\ \hline
Tube lens & $\times\ 4.2$ & $\times\ 1.67$ \\ \hline
Optical magnification & $\times\ 250$ & $\times\ 100$ \\ \hline
Camera type & EMCCD & CMOS\\ \hline
Image size & $512 \times 512$ & $600 \times 600$ \\ \hline
Pixel size & $16\ {\rm \mu m}$ & $6.5\ {\rm \mu m}$ \\ \hline
QE & $92\ \%$ & $70\ \%$ \\ \hline
EM Gain & $\times 300$, $\times 500$ & N/A \\ \hline
Exposure time & $30\ {\rm msec}$ & $30\ {\rm msec}$ \\ \hline
Readout noise & $100\ {\rm electrons}$ & $1.3\ {\rm electrons}$ \\ \hline
Full well & $370,000\ {\rm electrons}$ & $30,000\ {\rm electrons}$ \\ \hline
Dynamic range & $71.36\ {\rm dB}$ & $87.2\ {\rm dB}$ \\ \hline
Excess noise & $\sqrt{2}$ & $1$ \\ \hline
A/D Converter & $16$-bit & $16$-bit \\ \hline
Gain & $5.82\ {\rm electrons/count}$ & $0.47\ {\rm electrons/count}$ \\ \hline
Offset  & $2000\ {\rm counts}$ & $100\ {\rm counts}$ \\ \hline
Optical background & \multicolumn{2}{c|}{$0.1\ {\rm photons/pixel}$} \\ \hline
\end{tabular}
\vspace{0.3cm}
\caption{TIRFM specifications and condition to image the simple model 1.}
\label{tab;spec_tirfm_simple_model_00}
\end{table}

\newpage

\newpage

\begin{figure}[!h]
  \centering
	\includegraphics[width=10.5cm]{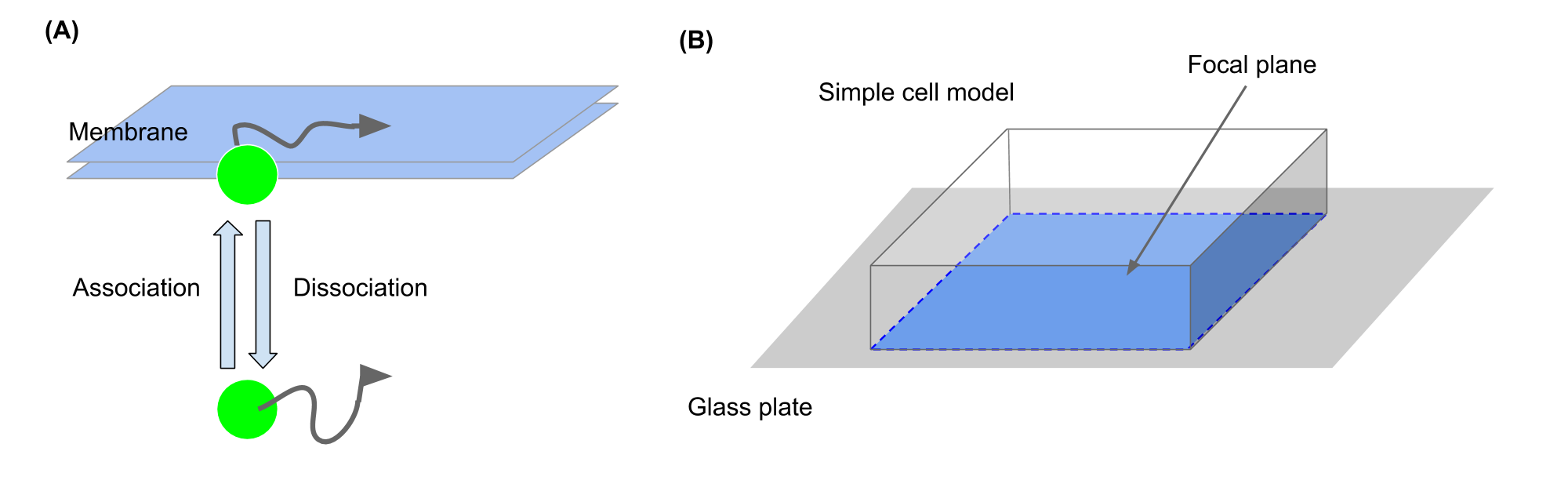}

  \caption{(A) schematics of network and (B) geometry of the simple cell model.}
  \label{fig;simple_model_02}
\end{figure}

\paragraph{Simple model 2 :}
Using the TIRFM simulation module, we simulated imaging the basal region of the simple cell model of TMR-tagged molecules diffusing on membrane and in cytoplasm. The images are simulated for the optical system and detector specification and conditions shown in Table \ref{tab;spec_tirfm_simple_model_02}. Results are shown in Figure \ref{fig;simple_model_02}. We assume the simple cell that express the molecules tagged with TMR fluorescent protein. Figure \ref{fig;simple_model_02}A and \ref{fig;simple_model_02}B show reaction and geometry of the model ($20 \times 20 \times 4\ {\rm \mu m^3}$), respectively. The model consists of $2$ chemical species, $2$ reactions and $4$ kinetic parameters. $100$ TMR-tagged molecules are distributed on the cell membrane and diffuse with $0.1\ {\rm \mu m^2/sec}$. $2,000$ TMR-tagged molecules are distributed in the cell cytoplasm and diffuse with $5.00\ {\rm \mu m^2/sec}$. Association rate from the cytoplasm to the membrane of those molecule is $3.35\ {\rm \mu m/sec}$. Dissociation rate from the membrane to the cytoplasm is $1.00\ {\rm sec^{-1}}$.

\begin{table}[!h]
\centering
\begin{tabular}{|l||p{5.0cm}|p{5.0cm}|}
\hline
Beam flux density & \multicolumn{2}{c|}{$40\ {\rm W/cm^2}$} \\ \hline
Beam wavelength & \multicolumn{2}{c|}{$488\ {\rm nm}$} \\ \hline
Refraction index & \multicolumn{2}{c|}{$1.33\ ({\rm glass})$, $1.27\ ({\rm water})$} \\ \hline
Critical angle & \multicolumn{2}{c|}{$65.6^{\circ}$} \\ \hline
Fluorophore & \multicolumn{2}{c|}{TRITC\ (${\rm Abs.}\ 548\ {\rm nm} /\ {\rm Em.}\ 608\ {\rm nm}$)} \\ \hline
Objective & \multicolumn{2}{c|}{$\times\ 60\ /\ {\rm N.A.}\ 1.40$} \\ \hline
Dichroic mirror & \multicolumn{2}{c|}{Semrok FF-562-Di03} \\ \hline
Emission filter & \multicolumn{2}{c|}{Semrok FF-593-25/40} \\ \hline
Linear conversion & \multicolumn{2}{c|}{$10^{-6}$} \\ \hline
Tube lens & $\times\ 4.2$ & $\times\ 1.67$ \\ \hline
Optical magnification & $\times\ 250$ & $\times\ 100$ \\ \hline
Camera type & EMCCD & CMOS\\ \hline
Image size & $512 \times 512$ & $600 \times 600$ \\ \hline
Pixel size & $16\ {\rm \mu m}$ & $6.5\ {\rm \mu m}$ \\ \hline
QE & $92\ \%$ & $70\ \%$ \\ \hline
EM Gain & $\times 300$ & N/A \\ \hline
Exposure time & $30\ {\rm msec}$ & $30\ {\rm msec}$ \\ \hline
Readout noise & $100\ {\rm electrons}$ & $1.3\ {\rm electrons}$ \\ \hline
Full well & $370,000\ {\rm electrons}$ & $30,000\ {\rm electrons}$ \\ \hline
Dynamic range & $71.36\ {\rm dB}$ & $87.2\ {\rm dB}$ \\ \hline
Excess noise & $\sqrt{2}$ & $1$ \\ \hline
A/D Converter & $16$-bit & $16$-bit \\ \hline
Gain & $5.82\ {\rm electrons/count}$ & $0.47\ {\rm electrons/count}$ \\ \hline
Offset  & $2000\ {\rm counts}$ & $100\ {\rm counts}$ \\ \hline
Optical background & \multicolumn{2}{c|}{$0.1\ {\rm photons/pixel}$} \\ \hline
\end{tabular}
\vspace{0.3cm}
\caption{TIRFM specifications and condition to image the simple model 2.}
\label{tab;spec_tirfm_simple_model_02}
\end{table}

\newpage

\begin{figure}[!h]
  \centering
  	{\bf EMCCD $\times 300$} \hspace{4.0cm} {\bf CMOS}
	\vspace{0.2cm}

  \centering
	\includegraphics[width=3.4cm]{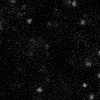}
	\includegraphics[width=3.4cm]{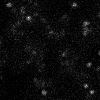}
	\hspace{0.1cm}
	\includegraphics[width=3.4cm]{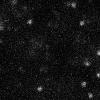}
	\includegraphics[width=3.4cm]{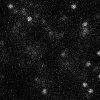}

  \centering
	\includegraphics[width=3.4cm]{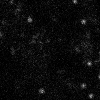}
	\includegraphics[width=3.4cm]{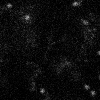}
	\hspace{0.1cm}
	\includegraphics[width=3.4cm]{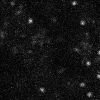}
	\includegraphics[width=3.4cm]{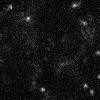}

  \centering
	\includegraphics[width=3.4cm]{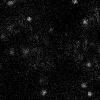}
	\includegraphics[width=3.4cm]{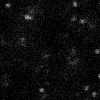}
	\hspace{0.1cm}
	\includegraphics[width=3.4cm]{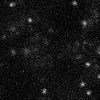}
	\includegraphics[width=3.4cm]{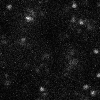}

  \centering
	\includegraphics[width=3.4cm]{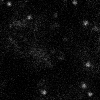}
	\includegraphics[width=3.4cm]{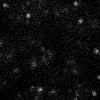}
	\hspace{0.1cm}
	\includegraphics[width=3.4cm]{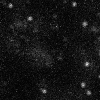}
	\includegraphics[width=3.4cm]{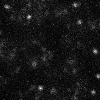}

  \centering
	\includegraphics[width=3.4cm]{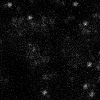}
	\includegraphics[width=3.4cm]{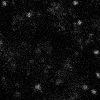}
	\hspace{0.1cm}
	\includegraphics[width=3.4cm]{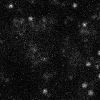}
	\includegraphics[width=3.4cm]{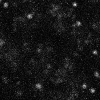}

  \caption{Example images of continuous $10$ frames ($100 \times 100$ pixels).}
  \label{fig;images_tirfm_simple_model_02}
\end{figure}

\newpage

\begin{figure}[!h]
  \centering
	\includegraphics[width=10.5cm]{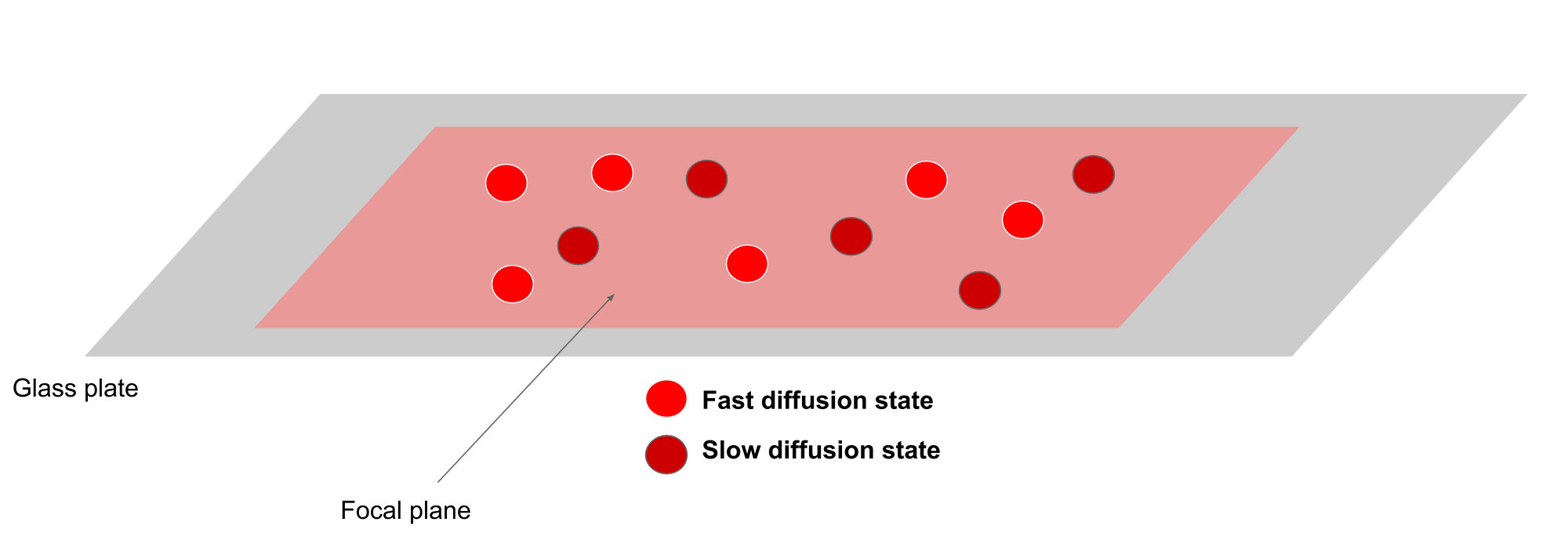}

  \caption{geometry of the two state model.}
  \label{fig;simple_model_03}
\end{figure}

\paragraph{Simple model 3 : }
Using the TIRFM simulation module, we simulated imaging the basal region of the two state model of TMR-tagged molecules diffusing on membrane. The images are simulated for the optical system and detector specification and conditions shown in Table \ref{tab;spec_tirfm_simple_model_03}. Results are shown in Figure \ref{fig;simple_model_03}. We assume the simple cell that express the molecules tagged with TMR fluorescent protein. Figure \ref{fig;simple_model_03} shows geometry of the two state model ($20 \times 20 \times 4\ {\rm \mu m^3}$). The model consists of $1$ chemical species and $2$ kinetic parameters. $200$ TMR-tagged molecules are distributed on the cell membrane and fast diffuse with $0.2\ {\rm \mu m^2/sec}$. $300$ TMR-tagged molecules are distributed on the cell membrane and slow diffuse with $0.02\ {\rm \mu m^2/sec}$.

\begin{table}[!h]
\centering
\begin{tabular}{|l||p{5.0cm}|p{5.0cm}|}
\hline
Beam flux density & \multicolumn{2}{c|}{$10\ {\rm W/cm^2}$} \\ \hline
Beam wavelength & \multicolumn{2}{c|}{$488\ {\rm nm}$} \\ \hline
Refraction index & \multicolumn{2}{c|}{$1.33\ ({\rm glass})$, $1.27\ ({\rm water})$} \\ \hline
Critical angle & \multicolumn{2}{c|}{$65.6^{\circ}$} \\ \hline
Fluorophore & \multicolumn{2}{c|}{TRITC\ (${\rm Abs.}\ 548\ {\rm nm} /\ {\rm Em.}\ 608\ {\rm nm}$)} \\ \hline
Objective & \multicolumn{2}{c|}{$\times\ 60\ /\ {\rm N.A.}\ 1.40$} \\ \hline
Dichroic mirror & \multicolumn{2}{c|}{Semrok FF-562-Di03} \\ \hline
Emission filter & \multicolumn{2}{c|}{Semrok FF-593-25/40} \\ \hline
Linear conversion & \multicolumn{2}{c|}{$10^{-6}$} \\ \hline
Tube lens & $\times\ 4.2$ & $\times\ 1.67$ \\ \hline
Optical magnification & $\times\ 250$ & $\times\ 100$ \\ \hline
Camera type & EMCCD & CMOS\\ \hline
Image size & $512 \times 512$ & $600 \times 600$ \\ \hline
Pixel size & $16\ {\rm \mu m}$ & $6.5\ {\rm \mu m}$ \\ \hline
QE & $92\ \%$ & $70\ \%$ \\ \hline
EM Gain & $\times 300$ & N/A \\ \hline
Exposure time & $30\ {\rm msec}$ & $30\ {\rm msec}$ \\ \hline
Readout noise & $100\ {\rm electrons}$ & $1.3\ {\rm electrons}$ \\ \hline
Full well & $370,000\ {\rm electrons}$ & $30,000\ {\rm electrons}$ \\ \hline
Dynamic range & $71.36\ {\rm dB}$ & $87.2\ {\rm dB}$ \\ \hline
Excess noise & $\sqrt{2}$ & $1$ \\ \hline
A/D Converter & $16$-bit & $16$-bit \\ \hline
Gain & $5.82\ {\rm electrons/count}$ & $0.47\ {\rm electrons/count}$ \\ \hline
Offset  & $2000\ {\rm counts}$ & $100\ {\rm counts}$ \\ \hline
Optical background & \multicolumn{2}{c|}{$0.1\ {\rm photons/pixel}$} \\ \hline
\end{tabular}
\vspace{0.3cm}
\caption{TIRFM specifications and condition to image the two state model.}
\label{tab;spec_tirfm_simple_model_03}
\end{table}

\newpage

\begin{figure}[!h]
  \centering
  	{\bf EMCCD $\times 300$} \hspace{4.0cm} {\bf CMOS}
	\vspace{0.2cm}

  \centering
	\includegraphics[width=3.4cm]{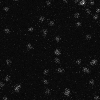}
	\includegraphics[width=3.4cm]{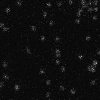}
	\hspace{0.1cm}
	\includegraphics[width=3.4cm]{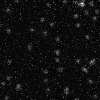}
	\includegraphics[width=3.4cm]{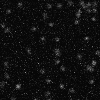}

  \centering
	\includegraphics[width=3.4cm]{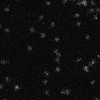}
	\includegraphics[width=3.4cm]{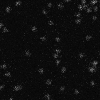}
	\hspace{0.1cm}
	\includegraphics[width=3.4cm]{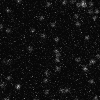}
	\includegraphics[width=3.4cm]{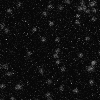}

  \centering
	\includegraphics[width=3.4cm]{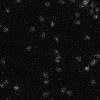}
	\includegraphics[width=3.4cm]{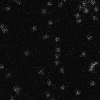}
	\hspace{0.1cm}
	\includegraphics[width=3.4cm]{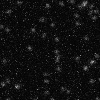}
	\includegraphics[width=3.4cm]{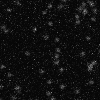}

  \centering
	\includegraphics[width=3.4cm]{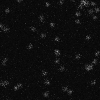}
	\includegraphics[width=3.4cm]{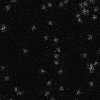}
	\hspace{0.1cm}
	\includegraphics[width=3.4cm]{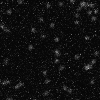}
	\includegraphics[width=3.4cm]{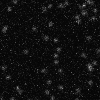}

  \centering
	\includegraphics[width=3.4cm]{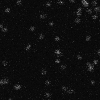}
	\includegraphics[width=3.4cm]{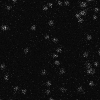}
	\hspace{0.1cm}
	\includegraphics[width=3.4cm]{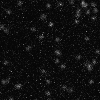}
	\includegraphics[width=3.4cm]{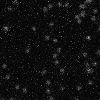}

  \caption{Example images of continuous $10$ frames ($100 \times 100$ pixels).}
  \label{fig;images_tirfm_simple_model_03}
\end{figure}

\newpage

\subsection{LSCM simulation module}
\begin{figure}[!h]
  \centering
      \includegraphics[width=9cm]{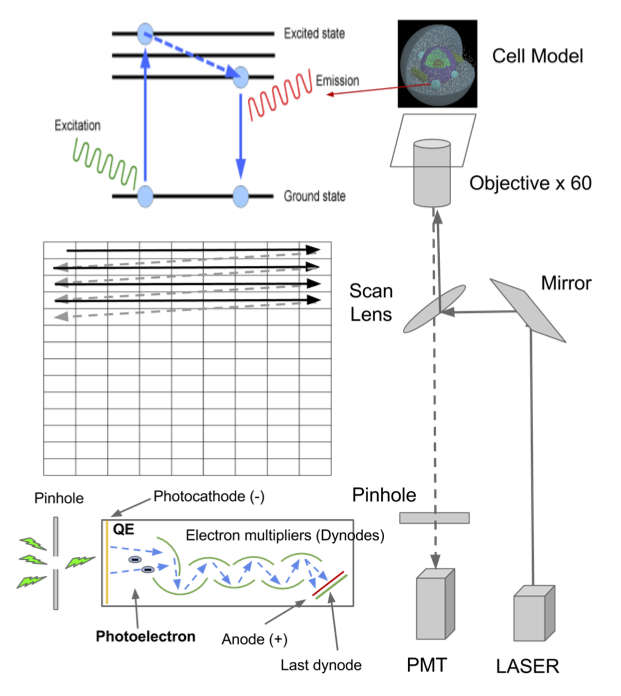}
  \caption[Optical configurations]{Optical configurations of the LSCM simulation module}
  \label{fig;config_lscm}
\end{figure}

\paragraph{}
The LSCM simulation module enables a selective visualization of  focal regions of cell model. Its optical configuration is shown in Figure \ref{fig;config_lscm}. Implementation assumptions are summarized in Table \ref{tab:char_lscm}. 

\begin{table}[!h]
    \centering
    \begin{tabular}{|c||c|}
    \hline
    \multirow{2}{*}{Principle} & Photon-counting \\
    & 1st-order paraxial approximation (Linear term) \\ \hline
    \multirow{2}{*}{Illumination} & Gaussian beam profile \\
    & Continuous / Gaussian / Unpolarized \\ \hline
    \multirow{2}{*}{Fluorescence} & Linear convertion ($\times 10^{-6}$)\\
    & Cross-section ($\sigma \cong 10^{-14}\ {\rm cm}^2$) \\ \hline
    \multirow{2}{*}{Image-forming} & 3-D Airy PSF (Unpolarized form) \\
    & PMT \\ \hline
    \end{tabular}
    \caption{Implementation assumption for the LSCM simulation module. Detection process for the PMT is performed with Monte Carlo simulation.}
    \label{tab:char_lscm}
\end{table}

\subsubsection{Illumination system}
\paragraph{}
Uncertainty sources of the illumination system are ruled by specification and conditions of Gaussian beam profile. We assume ideal Gaussian laser-beam intensity profile, which corresponds to the theoretical $TEM_{00}$ mode. The Gaussian beam wavefront of excitation wavelength continuously illuminate the specimen, and propagate perfectly flat with all elements moving precisely in the parallel direction. The wavefront quickly generate the $1/e^2$ irradiance contour at the plane after the wavefront has propagated a distance $z$. The contour spreads in the form of
\begin{eqnarray}
w(z) & = & w_0 \sqrt{1 + \left( \frac{\lambda z}{\pi w_0^2} \right)^2}
\end{eqnarray}
where $w_0$ is the beam waist radius at the focal plane where the wavefront is assumed to be flat. $z$ is the distance of propagation from the focal plane where the wavefront is assumed to be flat. $\lambda$ is the wavelength of excitation. The intensity of the Gaussian ${\rm TEM_{00}}$ beam is written in the form of
\begin{eqnarray}
I(r, z) & = & \frac{2 P'}{\pi w(z)^2} \exp{\left( -2r^2/w(z)^2 \right)} 
\end{eqnarray}
where $P'$ is the beam flux at the level of photon-counting unit ${\rm [\# photons/sec]}$. More details are described in ref. \cite{pawley2008, cmg2009}.

\subsubsection{Image-forming system}
\paragraph{}
The Monte Carlo simulation for the photomultipliers tube (PMT) includes a statistical model of noise source. Emitted photons of longer wavelengths are distributed as the sum of the PSFs shown in Eq. (\ref{eqn:bw_psf_model}). As the incident beam is scanned across the cell model in horizontal and vertical axes, a digital image that closely represents the actual confocal image can be obtained pixel-by-pixel. Details of the PMT simulation are described as follows.

\begin{figure}[!h]
  \centering
	\includegraphics[width=7.5cm]{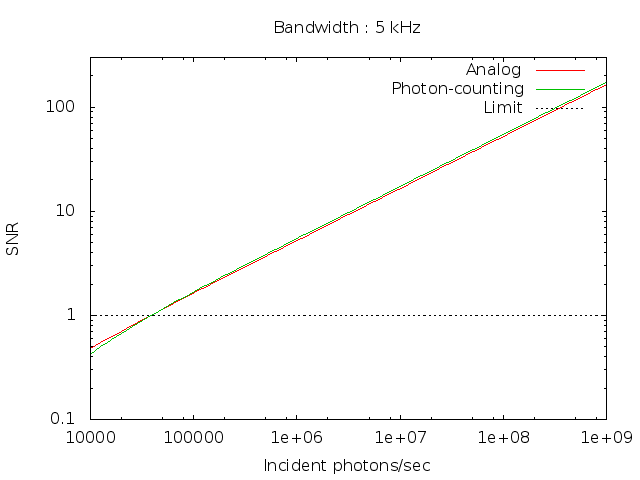}
	\includegraphics[width=7.5cm]{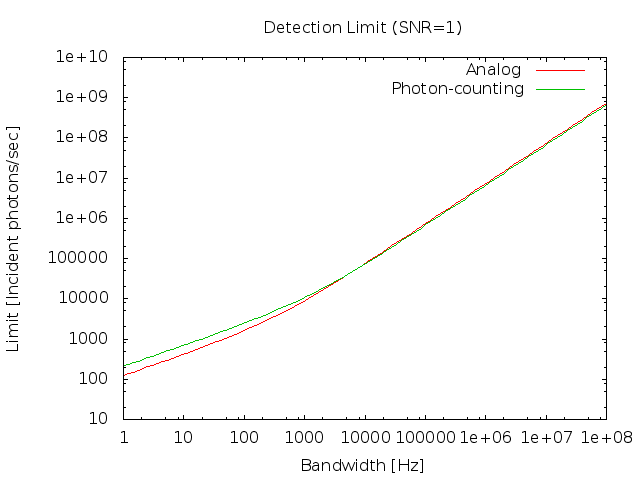}
  \caption{SNR (left) and detection limit when $SNR = 1$  (right)}
  \label{fig;snr_pmt}
\end{figure}

\begin{enumerate}
\item[(1)] Uncertainty sources \cite{pawley2008, pmt2007}: Uncertainty sources of the PMT system are ruled by PMT specifications and conditions shown in Table S4. First, shot noise arises in the number of photons incident to the PMT. When the incident photons interact with the photocathode placed on the head part of a PMT, photoelectrons are emitted. These photoelectrons are multiplied by the cascade process of secondary emission through a series of dynodes and finally reach the anode connected to an output processing circuit. The methods of readout processing the output signal of a PMT can be broadly divided into the analog and digital (photon counting) modes, depending on the number of incident photons and the bandwidth of the output processing circuit. If the output pulse-to-pulse interval is narrower than each pulse width or if the signal processing circuit is not sufficiently fast, then the actual output pulses overlap and become a direct current with shot noise fluctuations. This method is in the analog mode. In contrast, if the output pulse intervals are separated from noise pulses, discrete output pulses can then be detected by the photon counting method.

\begin{table}[!h]
\centering
\begin{tabular}{|l||p{5.0cm}|p{5.0cm}|}
\hline
PMT mode & Photon-counting & Analog \\ \hline
QE & \multicolumn{2}{c|}{$30\ \%$} \\ \hline
Dynode & \multicolumn{2}{c|}{$11$ stages} \\ \hline
Average gain & \multicolumn{2}{c|}{$\times 10^6$} \\ \hline
Readout noise & $0\ {\rm counts/sec}$ & $0\ {\rm mA}$ \\ \hline
Excess noise & N/A & $1.1$ \\ \hline
Pair-pulse time & $18\ {\rm nsec}$ & N/A\\ \hline
Optical background & \multicolumn{2}{c|}{$0.00\ {\rm photons/sec}$} \\ \hline
\end{tabular}
\vspace{0.3cm}
\caption{PMT specifications and condition.}
\label{tab;spec_pmt}
\end{table}

\newpage

\item[(2)] Probability density function (PDF) \cite{tan1982} :
An approximate PDF at the output of the PMT is written in the form of
\begin{eqnarray}
q{\left( S | E \right)} & = & e^{(E (e^{-A/B}) - 1)} \delta(S) + \frac{\sqrt{A/S} e^{-(E + S/B)}}{B} \sum^{\infty}_{n = 0} \frac{\sqrt{n} (E e^{-A/B})^n}{n!} I_{1} \left( \frac{2 \sqrt{n A S}}{B} \right)
\end{eqnarray}
where $I_1$ is the modified Bessel function of the first kind of order one. The PMT is characterised by its average gain $A$ and the number of dynode stages $\nu$. The variance of the PMT output is $2AB$ where $B = 1/2 (A - 1)/(A^{1/\nu}-1)$. Assuming $A = 10^6$ and $\nu = 11$. $E$ is the number of photoelectrons emitted at photocathode (expectation). Approximated formulas are described in ref. \cite{stokey1983}. The PDFs in the photon-counting mode and analog mode are shown in Figure \ref{fig;pdf_pmt}.

\begin{figure}[!h]
  \centering
	\includegraphics[width=7.5cm]{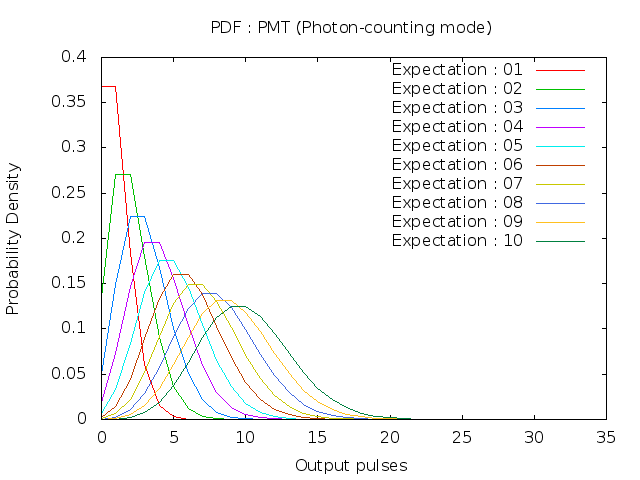}
	\includegraphics[width=7.5cm]{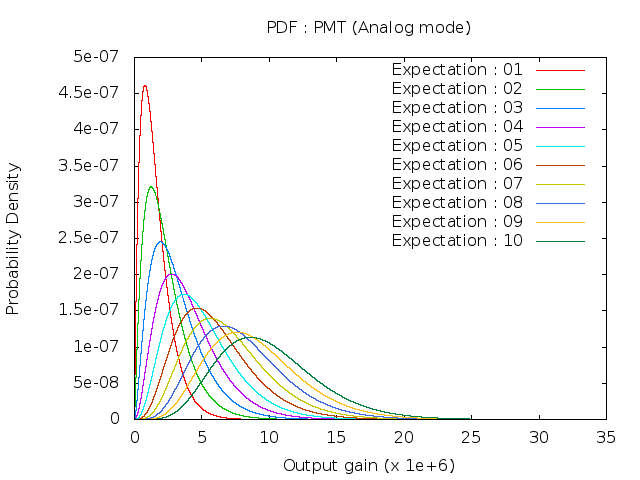}
  \caption{Probability density function for photon-counting mode (left) and analog mode  (right)}
  \label{fig;pdf_pmt}
\end{figure}

\item[(3)] Count rate and linearity \cite{pmt2007}:
The photon counting mode offers excellent linearity over a wide range. The lower limit is determined by the number of dark current pulses. Maximum count rate is limited by the pair-pulse time resolution where two pulses can be separated at a minimum time interval. The measured count rate is given as
\begin{eqnarray}
M = \frac{N_s}{1 + N_s \delta t}
\end{eqnarray}
where $N_s$ is the input photon flux, and $\delta t$ is the pair-pulses time resolution ($\sim 18\ {\rm nsec}$). Linearity is shown in Figure \ref{fig;linearity_pmt}.

\begin{figure}[!h]
  \centering
	\includegraphics[width=7.5cm]{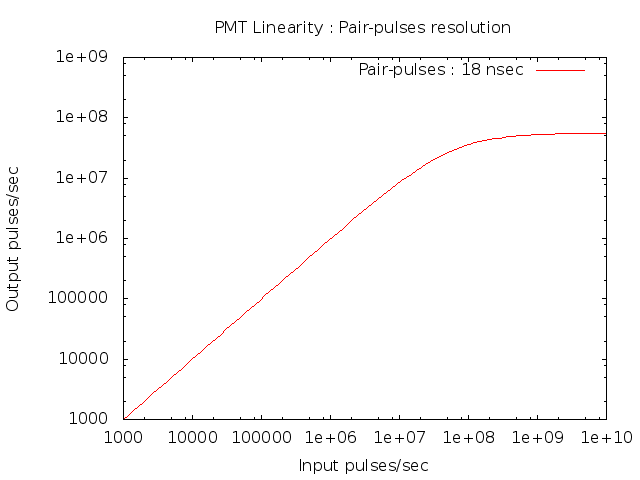}
  \caption{Linearity for photon-counting mode}
  \label{fig;linearity_pmt}
\end{figure}

\item[(3)] SNR per pixel \cite{pmt2007} :
The variance of the PMT output is given by the sum of the variances of all noise sources. The SNR and detection limits are plotted in Figure \ref{fig;snr_pmt}. The SNR in the analog mode is written in the form of
\begin{eqnarray}
SNR & = & \frac{I_k}{\sqrt{2 e B F \left( I_k + 2 \left(I_d + I_b \right)\right) + (N_A/G)^2}} 
\end{eqnarray}
where $I_k  = e QE N_{p}$ and $N_p$ is the number of incident photons/sec. $I_d$ is dark current. $B$ is bandwidth in Hz ($B = 1/(2 T)$) and $T$ is the observational period. $G$ is gain factor ($\sim 10^6$). $F$ is the excess noise ($F \approx \delta/(\delta - 1)$) and $\delta$ is the number of dynode stages. Detection limits ($SNR = 1$) as a function of bandwidth is given by the following equation.
\begin{eqnarray}
N_p^{limit} & = & \frac{e B F + \sqrt{\left( e B F \right)^2 + \left( 4 e B F  I_d \right)}}{e QE}
\end{eqnarray}

The SNR in the photon-counting mode is written in the form of
\begin{eqnarray}
SNR & = & \frac{N_s}{\sqrt{\left(N_s + 2 \left(N_d + N_b \right)\right)/T + N_A^2}}
\end{eqnarray}
where $N_s = QE \cdot N_p$ and $N_d$ is the dark count/sec. The detection limit is also given by the following equation.
\begin{eqnarray}
N_p^{limit} & = & \frac{B + \sqrt{B^2 + 4 B N_d}}{QE}
\end{eqnarray}

\end{enumerate}

\subsubsection{Examples of outputs}
\begin{figure}[!h]
  \centering
  	{\bf Photon-counting mode} \hspace{3.0cm} {\bf Analog mode}
	\vspace{0.1cm}

  \centering
	\includegraphics[width=2.8cm]{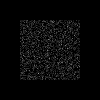}
	\includegraphics[width=3.8cm]{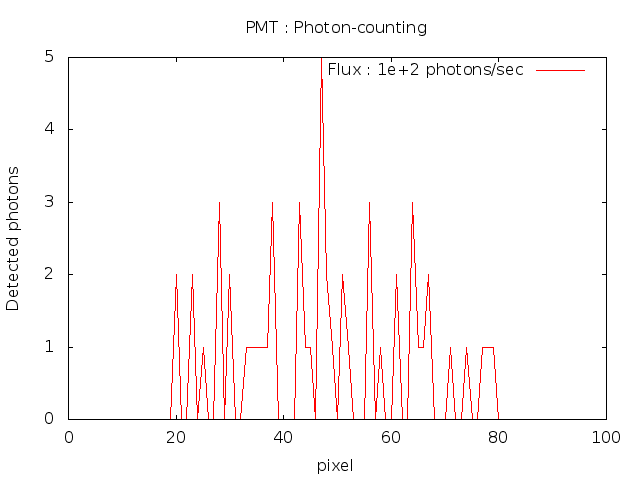}
	\includegraphics[width=2.8cm]{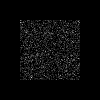}
	\includegraphics[width=3.8cm]{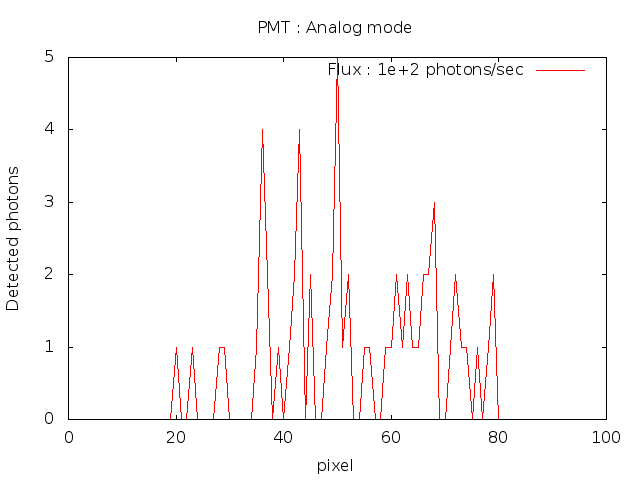}
	
  \centering
	\includegraphics[width=2.8cm]{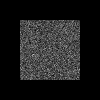}
	\includegraphics[width=3.8cm]{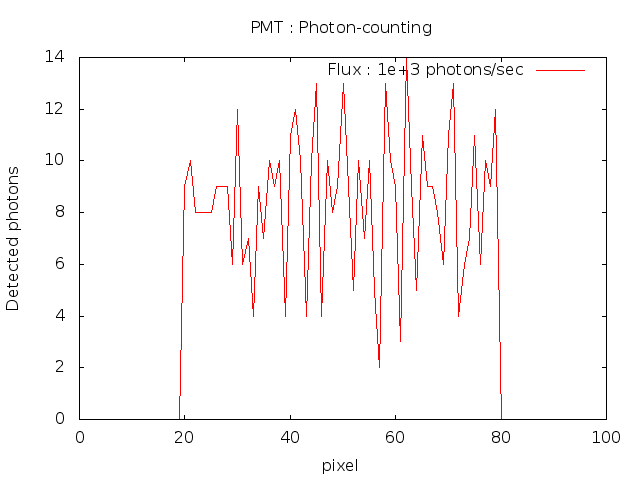}
	\includegraphics[width=2.8cm]{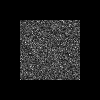}
	\includegraphics[width=3.8cm]{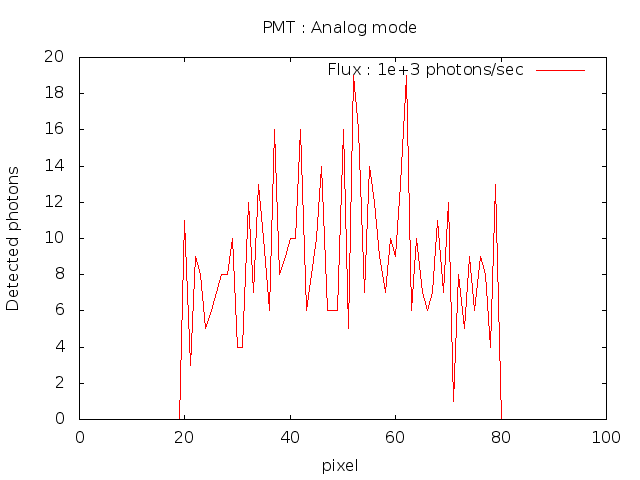}
	
  \centering
	\includegraphics[width=2.8cm]{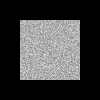}
	\includegraphics[width=3.8cm]{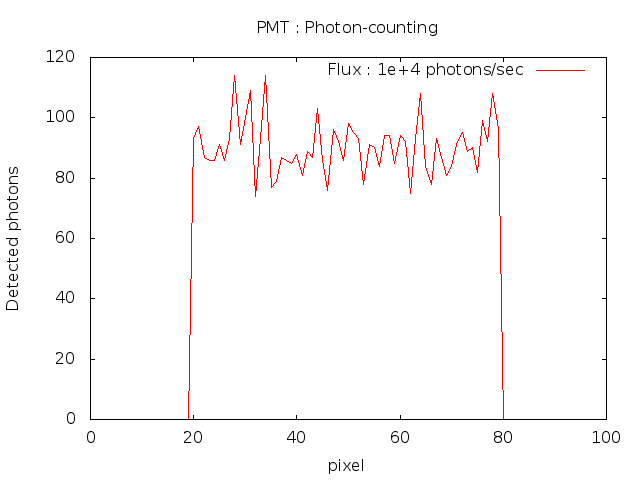}
	\includegraphics[width=2.8cm]{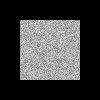}
	\includegraphics[width=3.8cm]{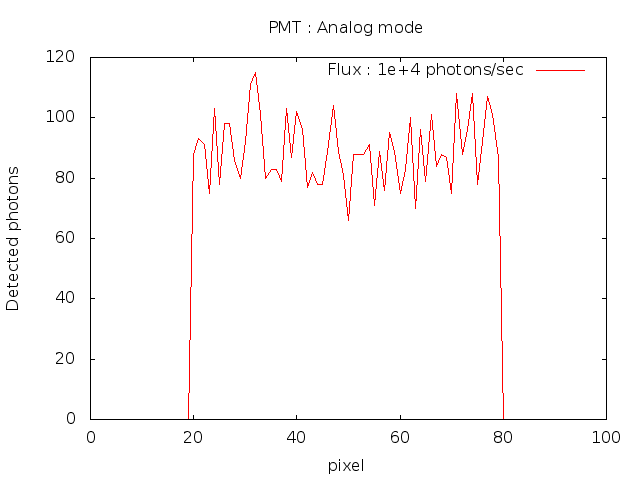}
	
  \centering
	\includegraphics[width=2.8cm]{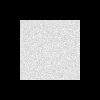}
	\includegraphics[width=3.8cm]{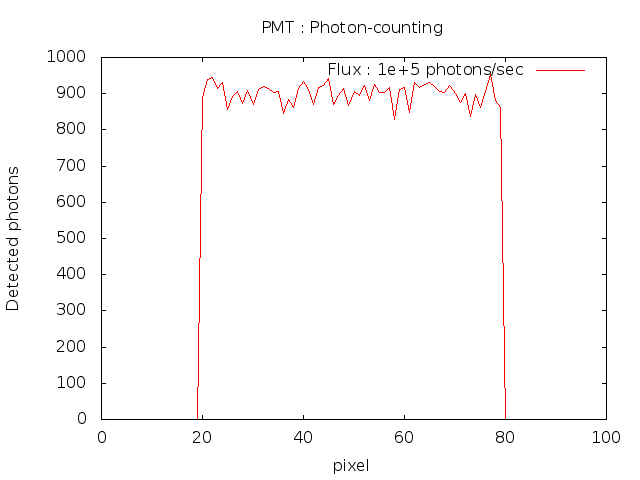}
	\includegraphics[width=2.8cm]{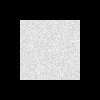}
	\includegraphics[width=3.8cm]{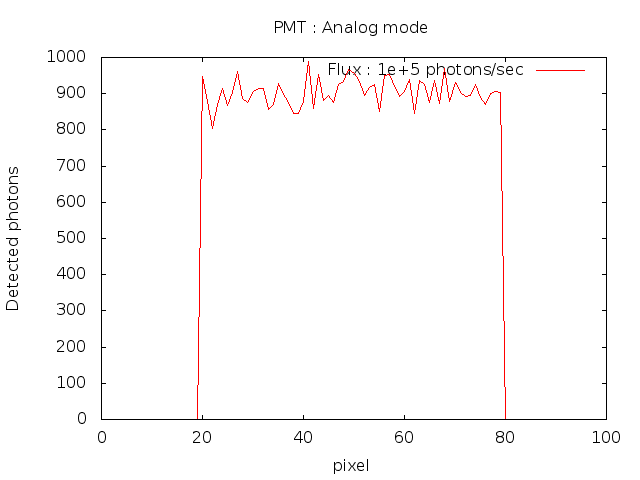}
	
  \centering
	\includegraphics[width=2.8cm]{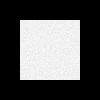}
	\includegraphics[width=3.8cm]{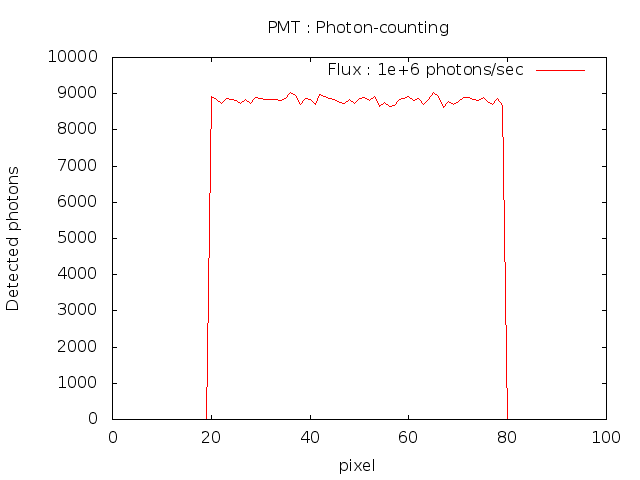}
	\includegraphics[width=2.8cm]{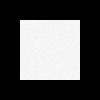}
	\includegraphics[width=3.8cm]{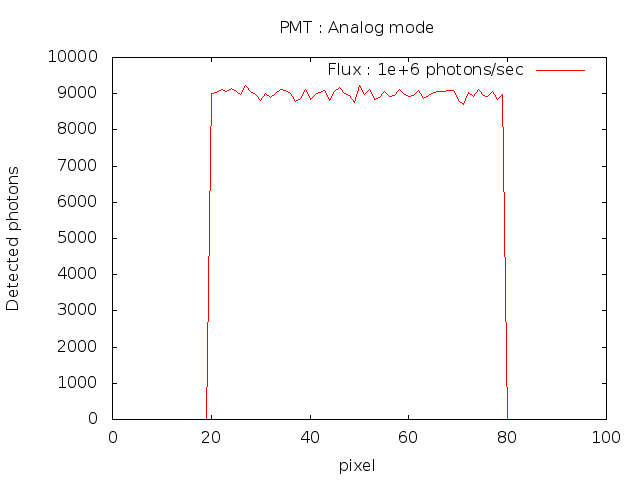}

  \centering
	\includegraphics[width=2.8cm]{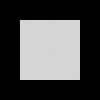}
	\includegraphics[width=3.8cm]{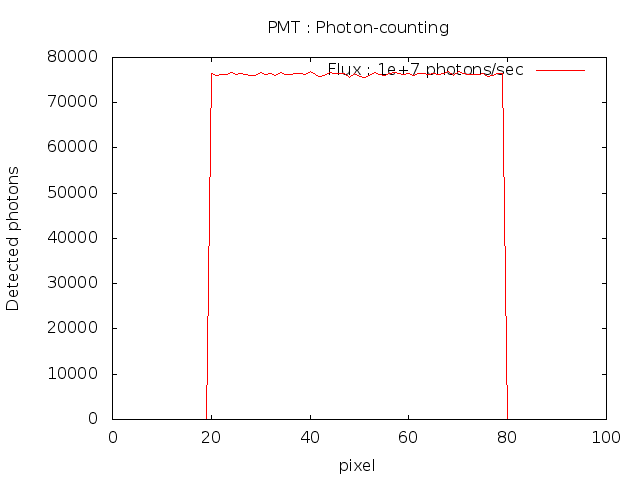}
	\includegraphics[width=2.8cm]{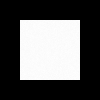}
	\includegraphics[width=3.8cm]{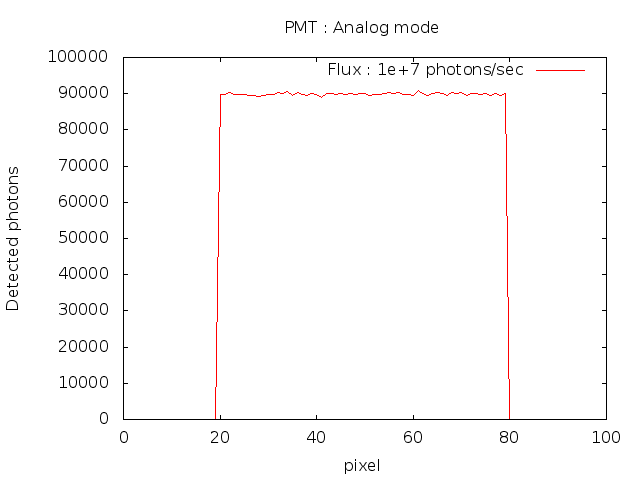}
	
  \caption{Image output of PMT. No dark count rate}
  \label{fig;pmt_images_nodc}
\end{figure}

\begin{figure}[!h]
  \centering
  	{\bf Photon-counting mode} \hspace{3.0cm} {\bf Analog mode}
	\vspace{0.1cm}

  \centering
	\includegraphics[width=2.8cm]{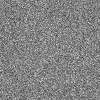}
	\includegraphics[width=3.8cm]{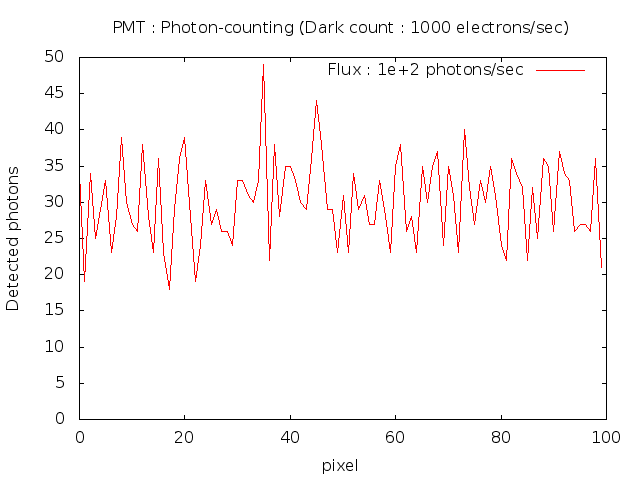}
	\includegraphics[width=2.8cm]{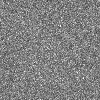}
	\includegraphics[width=3.8cm]{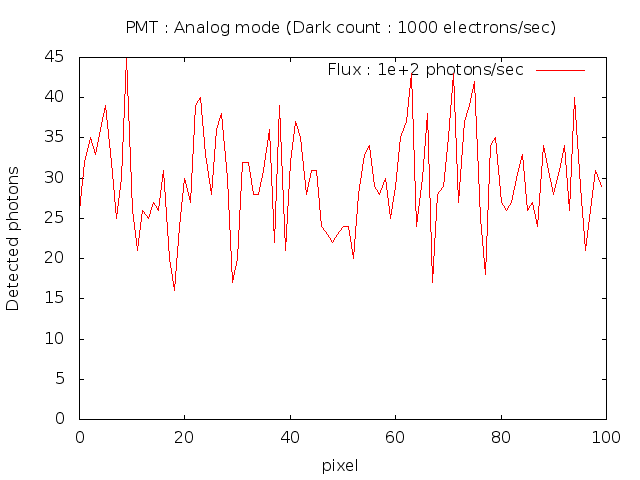}

  \centering
	\includegraphics[width=2.8cm]{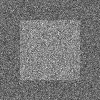}
	\includegraphics[width=3.8cm]{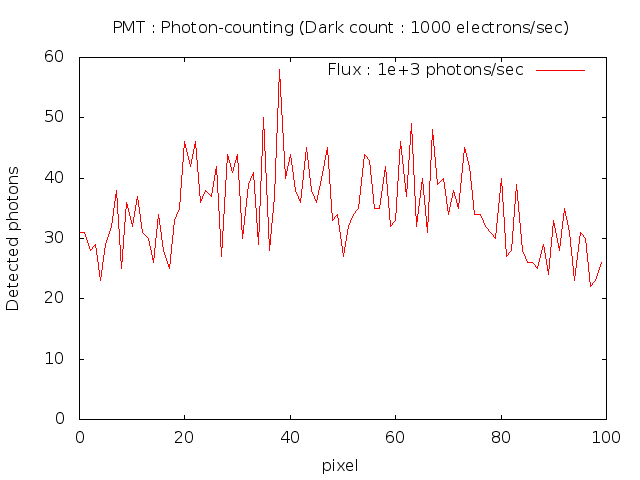}
	\includegraphics[width=2.8cm]{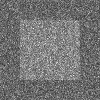}
	\includegraphics[width=3.8cm]{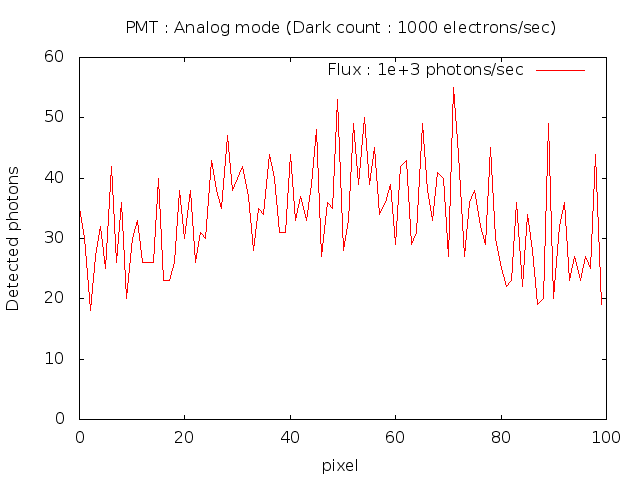}

  \centering
	\includegraphics[width=2.8cm]{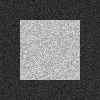}
	\includegraphics[width=3.8cm]{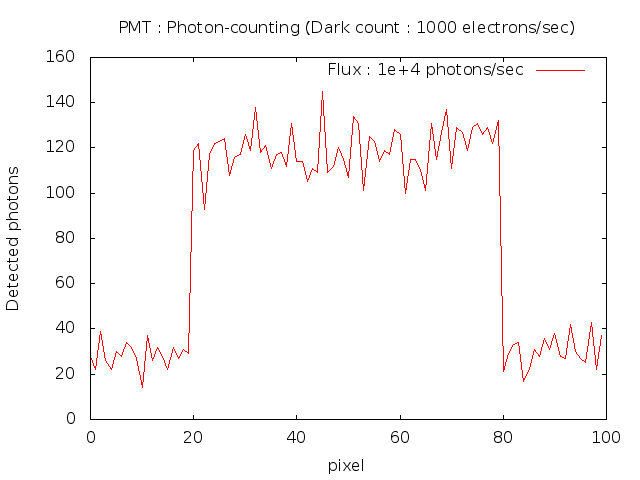}
	\includegraphics[width=2.8cm]{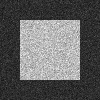}
	\includegraphics[width=3.8cm]{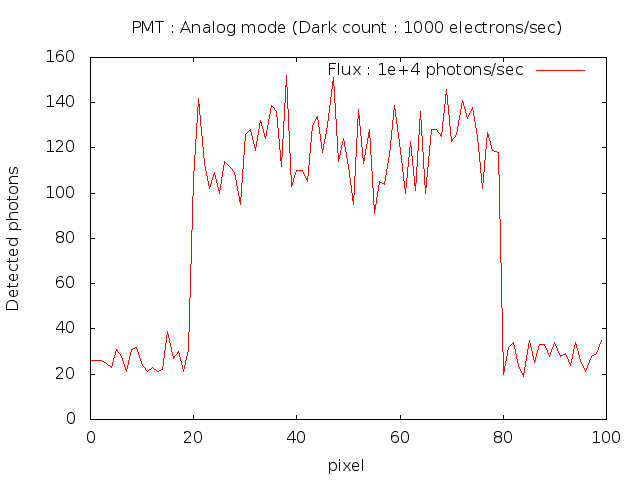}

  \centering
	\includegraphics[width=2.8cm]{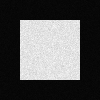}
	\includegraphics[width=3.8cm]{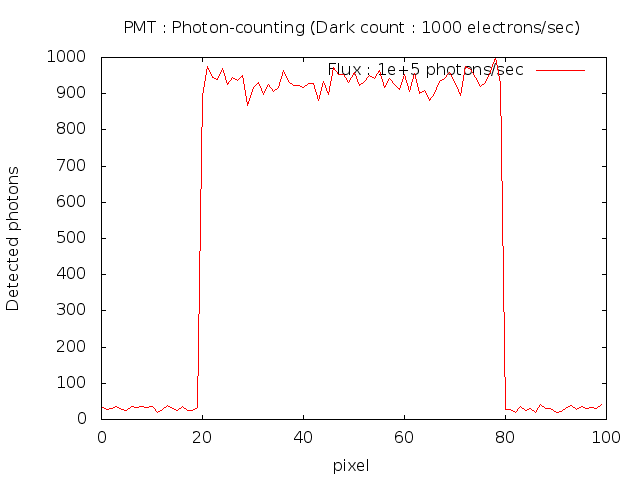}
	\includegraphics[width=2.8cm]{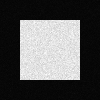}
	\includegraphics[width=3.8cm]{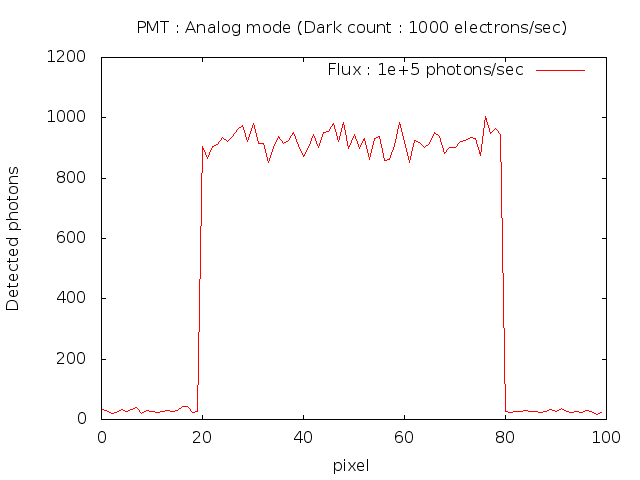}

  \centering
	\includegraphics[width=2.8cm]{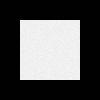}
	\includegraphics[width=3.8cm]{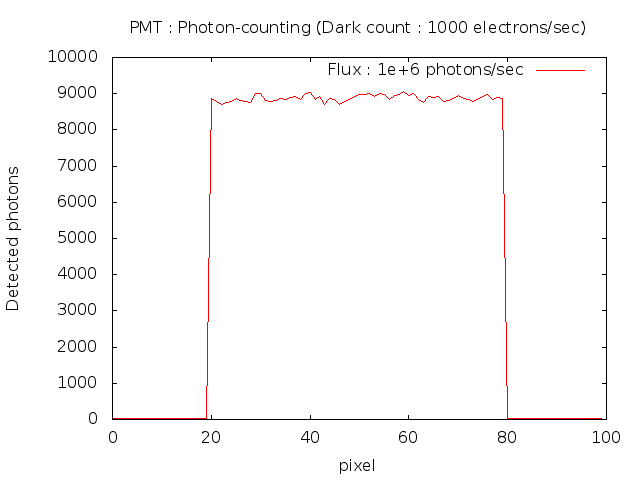}
	\includegraphics[width=2.8cm]{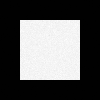}
	\includegraphics[width=3.8cm]{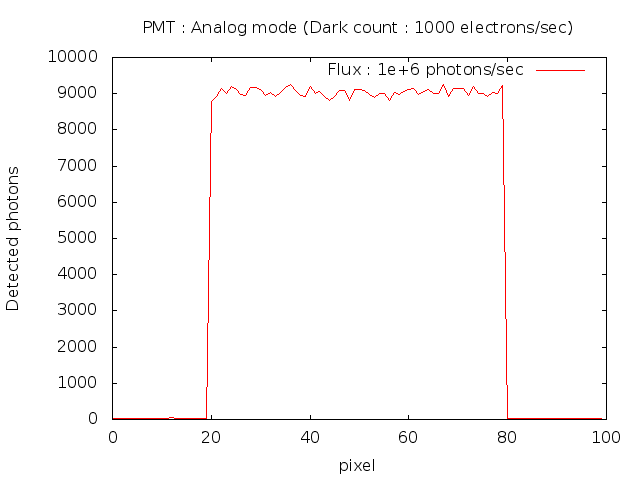}

  \centering
	\includegraphics[width=2.8cm]{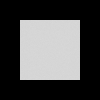}
	\includegraphics[width=3.8cm]{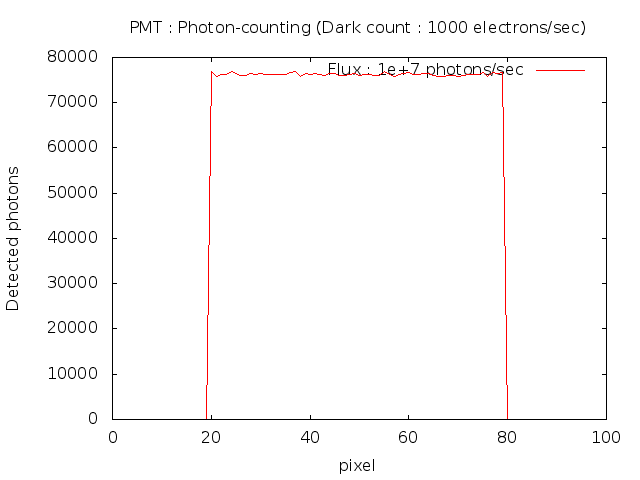}
	\includegraphics[width=2.8cm]{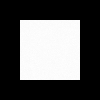}
	\includegraphics[width=3.8cm]{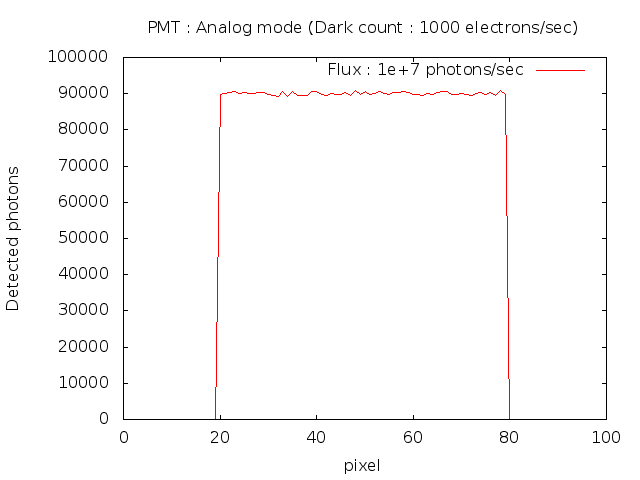}
	
  \caption{Image output of PMT. Background count rate is 1000 electrons/sec.}
  \label{fig;pmt_images_1000dc}
\end{figure}

\paragraph{}
Despite the absence of the scanning process, Figure \ref{fig;pmt_images_nodc} and \ref{fig;pmt_images_1000dc} show images of outputs and the graphs showing the signal intensity and noise of the horizontal line at vertical center. From the top to bottom rows in each figure, $10^3, 10^4, 10^5, 10^6, $ and $10^7$ incident photon fluxes are expected in $80 \times 80$ pixel squares at the image center.

\newpage

\newpage

\newpage

\begin{figure}[!h]
  \centering
	\includegraphics[width=10cm]{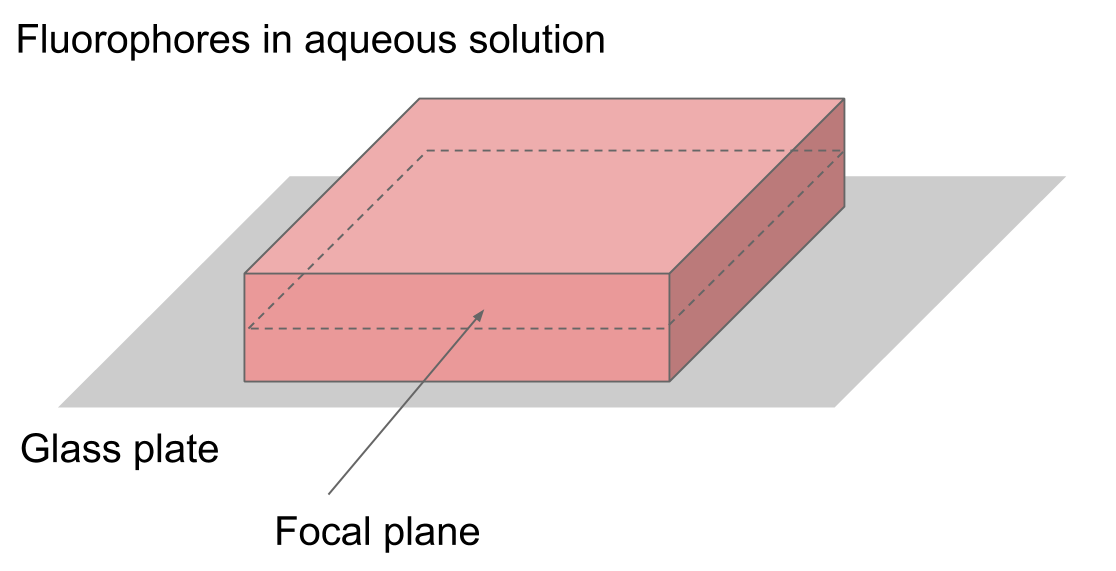}

  \caption{Fluorophores in aqueous solution.}
  \label{fig;simple_model_01}
\end{figure}

\begin{figure}[!h]
  \leftline{\hspace{6.0cm} \bf Analog \hspace{0.9cm} \bf Photon-counting}
	\vspace{0.1cm}

  \centering
	\includegraphics[width=3.4cm]{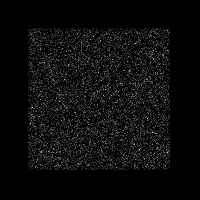}
	\includegraphics[width=3.4cm]{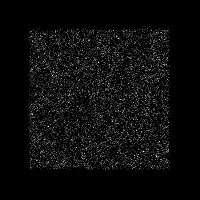}

  \centering
	\includegraphics[width=3.4cm]{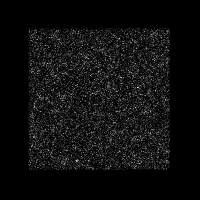}
	\includegraphics[width=3.4cm]{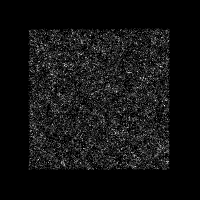}

  \centering
	\includegraphics[width=3.4cm]{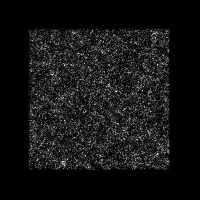}
	\includegraphics[width=3.4cm]{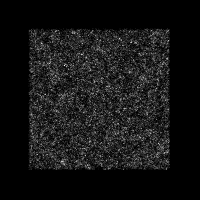}

  \centering
	\includegraphics[width=3.4cm]{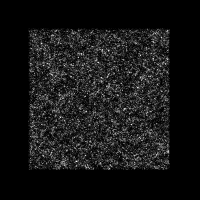}
	\includegraphics[width=3.4cm]{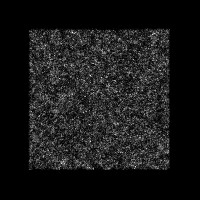}

  \centering
	\includegraphics[width=3.4cm]{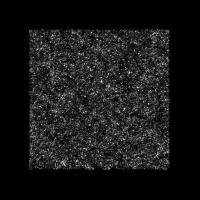}
	\includegraphics[width=3.4cm]{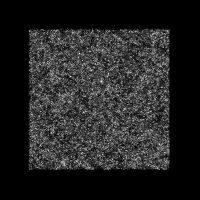}

  \caption{Image comparison ($200 \times 200$ pixels at image center)}
  \label{fig;images_lscm_simple_model_01}
\end{figure}

\paragraph{Simple model :}
We constructed relatively simple particle model of TMR in aqueous solution as shown in Figure \ref{fig;simple_model_01}. We assumed that $19,656$ TMR molecules are distributed in the solution box ($30\times30\times6\ {\rm \mu m^3}$), and diffuse with $100\ {\rm \mu m^2/sec}$. Images are simulated for the optical specification and condition of the LSCM simulation module shown in Table \ref{tab;spec_lscm_simple_model_01}. Results are shown in Figure \ref{fig;images_lscm_simple_model_01}. Figures from top row to bottom one correspond to the beam inputs of $10, 30, 70, 100$ and $300\ {\rm \mu W}$, respectively.

\begin{table}[!h]
\centering
\begin{tabular}{|l||p{5.0cm}|p{5.0cm}|}
\hline
Beam flux & \multicolumn{2}{c|}{$30,\ 70,\ 100,\ 300\ {\rm \mu W}$} \\ \hline
Beam wavelength & \multicolumn{2}{c|}{$488\ {\rm nm}$} \\ \hline
Beam waist & \multicolumn{2}{c|}{$200\ {\rm nm}$ (Assumed)} \\ \hline
Fluorophore & \multicolumn{2}{c|}{mEGFP\ (${\rm Abs.}\ 484\ {\rm nm} /\ {\rm Em.}\ 507\ {\rm nm}$)} \\ \hline
Objective & \multicolumn{2}{c|}{$\times\ 60\ /\ {\rm N.A.}\ 1.49$} \\ \hline
Scan lens & \multicolumn{2}{c|}{$\times\ 1$} \\ \hline
Pinhole & \multicolumn{2}{c|}{$57.6\ {\rm \mu m}$ diameter ($2\ {\rm A.U}$)} \\ \hline
Optical magnification & \multicolumn{2}{c|}{$\times\ 60$} \\ \hline
Linear conversion & \multicolumn{2}{c|}{$10^{-6}$} \\ \hline
Scan time & \multicolumn{2}{c|}{$1.1\ {\rm \mu sec/pixel}$} \\ \hline
Pixel length & \multicolumn{2}{c|}{$210\ {\rm nm/pixel}$} \\ \hline
Image size & \multicolumn{2}{c|}{$1024 \times 1024$} \\ \hline
PMT mode & Photon-counting & Analog \\ \hline
A/D Converter  & $12$-bit & $12$-bit \\ \hline
QE & $30\ \%$ & $30\ \%$ \\ \hline
Readout noise & $0\ {\rm counts/sec}$ & $0\ {\rm mA}$ \\ \hline
Excess noise & N/A & $1.1$ \\ \hline
Optical background & \multicolumn{2}{c|}{$0.10\ {\rm photons/sec}$} \\ \hline
\end{tabular}
\vspace{0.3cm}
\caption{LSCM specifications and condition to image the simple particle model of fluorescent molecules.}
\label{tab;spec_lscm_simple_model_01}
\end{table}

\newpage

\newpage

\section{Comparison of {\it in vitro} images}
\subsection{HaloTag-TMR molecules on glass surface}
\paragraph{{\it In vitro} \bf Experiment :}
HaloTag-TMR molecules were provided by Dr. Masahiro Ueda, laboratory for cell signaling dynamics, RIKEN QBiC. Data was taken by Dr. Satomi Matsuoka, laboratory for cell signaling dynamics, RIKEN QBiC. The molecules were distributed on glass surface, and observed using total internal reflection microscopy with 60X/1.40NA objective (Nikon). Fluorescent images of the HaloTag-TMR molecules are acquired with an EMCCD camera (iXon+, Andor). The images were obtained at a 30 msec exposure time. 

\paragraph{\bf Particle model :}
We constructed simple model of $100$ stationary HaloTag tetramethyl rhodamine (TMR) molecules distributed on glass surface. 

\paragraph{\bf Simulated imaging :}
We simulated imaging the basal region of the particle model for the pecification and condition of the TIRFM simulation module shown in Table \ref{tab;tirfm_tmr}. 

\begin{table}[!h]
\centering
\begin{tabular}{|l||C{10cm}|}
\hline
Beam flux density & $20,\ 30,\ 40,\ 50\ {\rm W/cm^2}$ \\ \hline
Beam wavelength & $488 {\rm nm}$ \\ \hline
Refraction index & $1.33\ ({\rm glass})$, $1.27\ ({\rm water})$ \\ \hline
Critical angle & $65.6^{\circ}$ \\ \hline
Fluorophore & HaloTag TMR ligand\ (${\rm Abs.}\ 555\ {\rm nm} /\ {\rm Em.}\ 585\ {\rm nm}$) \\ \hline
Objective & $\times\ 60\ /\ {\rm N.A.}\ 1.40$ \\ \hline
Dichroic mirror & Semrok FF-562-Di03 \\ \hline
Emission filter & Semrok FF-593-25/40 \\ \hline
Linear conversion & $10^{-6}$ \\ \hline
Tube lens & $\times\ 3.3$ \\ \hline
Optical magnification & $\times\ 198$ \\ \hline
Camera type & EMCCD (iXon+ Andor) \\ \hline
Image size & $512 \times 512$ \\ \hline
Pixel size & $16\ {\rm \mu m}$ \\ \hline
QE & $92\ \%$ \\ \hline
EM Gain & $\times\ 300$ \\ \hline
Exposure time & $30\ {\rm msec}$ \\ \hline
Readout noise & $100\ {\rm electrons}$ \\ \hline
Full well & $180,000\ {\rm electrons}$ \\ \hline
Dynamic range & $71.1\ {\rm dB}$ \\ \hline
Excess noise & $\sqrt{2}$ \\ \hline
A/D Converter & $16$-bit \\ \hline
Gain & $11.1\ {\rm electrons/count}$ \\ \hline
Offset  & $100\ {\rm counts}$ \\ \hline
Optical background & $1.0\ {\rm photons/pixel}$ \\ \hline
\end{tabular}
\vspace{0.3cm}
\caption{TIRFM specifications and condition to image the simple particle model of fluorescent molecules.}
\label{tab;tirfm_tmr}
\end{table}

\newpage

\begin{figure}[!h]
  \centering
  	{\bf Expected images} \hspace{5.0cm} {\bf Simulated images}
	
  \vspace{0.3cm}
  \leftline{\bf Beam flux density : $20\ {\rm W/cm^2}$}
  \vspace{0.1cm}
  \centering
	\includegraphics[width=3.8cm]{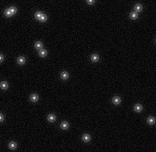}
	\includegraphics[width=4.5cm]{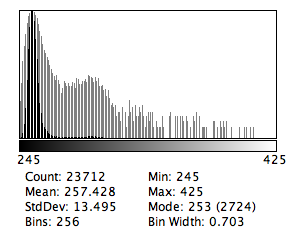}
	\includegraphics[width=3.8cm]{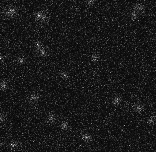}
	\includegraphics[width=4.5cm]{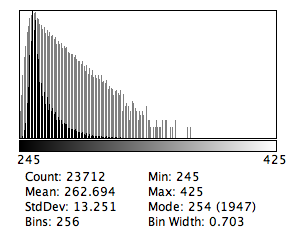}

  \vspace{0.2cm}
  \leftline{\bf Beam flux density : $30\ {\rm W/cm^2}$}
  \vspace{0.1cm}
  \centering
	\includegraphics[width=3.8cm]{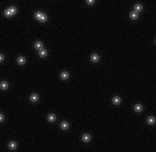}
	\includegraphics[width=4.5cm]{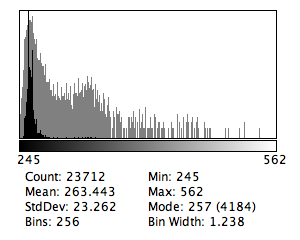}
	\includegraphics[width=3.8cm]{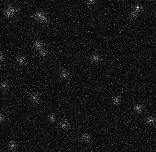}
	\includegraphics[width=4.5cm]{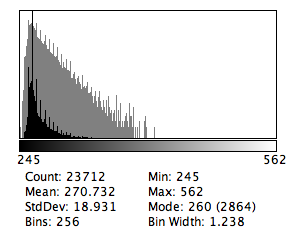}

  \vspace{0.2cm}
  \leftline{\bf Beam flux density : $40\ {\rm W/cm^2}$}
  \vspace{0.1cm}
  \centering
	\includegraphics[width=3.8cm]{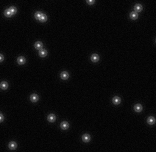}
	\includegraphics[width=4.5cm]{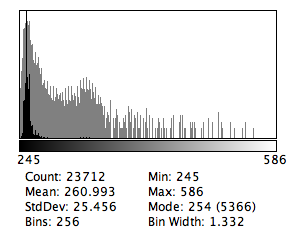}
	\includegraphics[width=3.8cm]{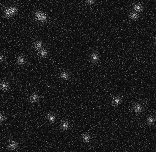}
	\includegraphics[width=4.5cm]{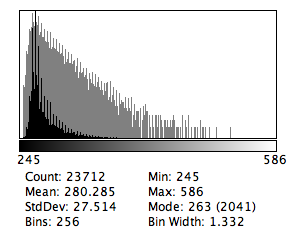}

  \vspace{0.2cm}
  \leftline{\bf Beam flux density : $50\ {\rm W/cm^2}$}
  \vspace{0.1cm}
  \centering
	\includegraphics[width=3.8cm]{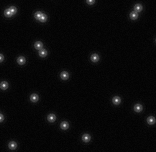}
	\includegraphics[width=4.5cm]{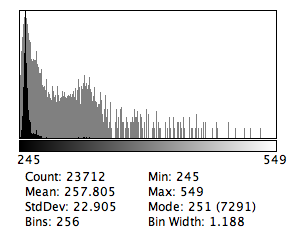}
	\includegraphics[width=3.8cm]{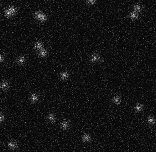}
	\includegraphics[width=4.5cm]{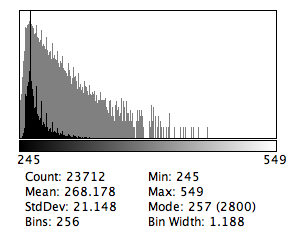}

  \caption{Comparison of {\it in vitro} images ($156 \times 152\ {\rm pixels}$) and intensity histograms. Log-scaled intensity histograms are shown in grey color.}
  \label{fig;comparison_hist_tirfm_expected}
\end{figure}

\newpage

\begin{figure}[!h]
  \centering
  	{\bf Simulated images} \hspace{5.0cm} {\bf Actual images}
	
  \vspace{0.3cm}
  \leftline{\bf Beam flux density : $20\ {\rm W/cm^2}$}
  \vspace{0.1cm}
  \centering
	\includegraphics[width=3.8cm]{figures_SI/images_comparison/images_tmr_tirfm_sim/image_0000000_20w.png}
	\includegraphics[width=4.5cm]{figures_SI/images_comparison/images_tmr_tirfm_sim/hist_log_0000000_20w.png}
	\includegraphics[width=3.8cm]{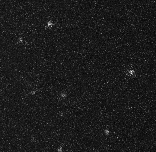}
	\includegraphics[width=4.5cm]{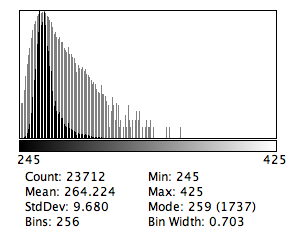}

  \vspace{0.2cm}
  \leftline{\bf Beam flux density : $30\ {\rm W/cm^2}$}
  \vspace{0.1cm}
  \centering
	\includegraphics[width=3.8cm]{figures_SI/images_comparison/images_tmr_tirfm_sim/image_0000000_30w.png}
	\includegraphics[width=4.5cm]{figures_SI/images_comparison/images_tmr_tirfm_sim/hist_log_0000000_30w.png}
	\includegraphics[width=3.8cm]{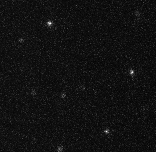}
	\includegraphics[width=4.5cm]{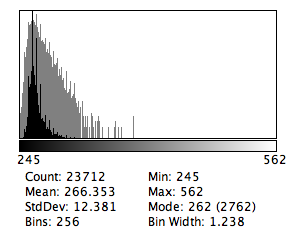}

  \vspace{0.2cm}
  \leftline{\bf Beam flux density : $40\ {\rm W/cm^2}$}
  \vspace{0.1cm}
  \centering
	\includegraphics[width=3.8cm]{figures_SI/images_comparison/images_tmr_tirfm_sim/image_0000000_40w.png}
	\includegraphics[width=4.5cm]{figures_SI/images_comparison/images_tmr_tirfm_sim/hist_log_0000000_40w.png}
	\includegraphics[width=3.8cm]{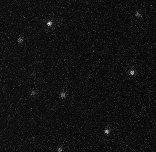}
	\includegraphics[width=4.5cm]{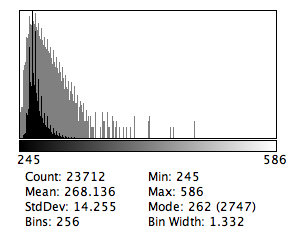}

  \vspace{0.2cm}
  \leftline{\bf Beam flux density : $50\ {\rm W/cm^2}$}
  \vspace{0.1cm}
  \centering
	\includegraphics[width=3.8cm]{figures_SI/images_comparison/images_tmr_tirfm_sim/image_0000000_50w.png}
	\includegraphics[width=4.5cm]{figures_SI/images_comparison/images_tmr_tirfm_sim/hist_log_0000000_50w.png}
	\includegraphics[width=3.8cm]{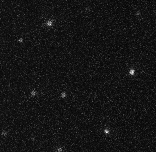}
	\includegraphics[width=4.5cm]{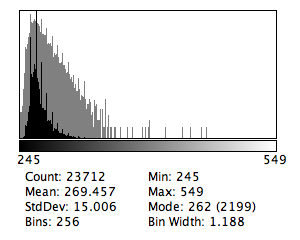}

  \caption{Comparison of {\it in vitro} images ($156 \times 152\ {\rm pixels}$) and intensity histograms. Log-scaled intensity histograms are shown in grey color.}
  \label{fig;comparison_hist_tirfm}
\end{figure}

\newpage

\begin{figure}[!h]
  \centering
	\includegraphics[width=12cm]{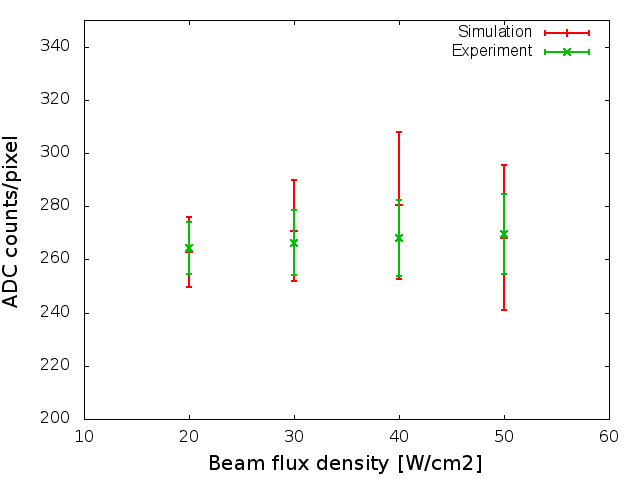}

  \caption{Linearity is shown for various beam flux density. Experiment (green) and simulation (red).}
  \label{fig;comparison_linearity_tirfm}
\end{figure}

\newpage

\subsection{HaloTag-TMR molecules in aqueous solution}
\paragraph{{\it In vitro} \bf Experiment :}
HaloTag-TMR molecules were provided by Dr. Masahiro Ueda, laboratory for cell signaling dynamics, RIKEN QBiC. Data was taken by Dr. Satomi Matsuoka, laboratory for cell signaling dynamics, RIKEN QBiC. $5\ {\rm nM}$ concentration of HaloTag-TMR molecules in aqueous solution were observed using a laser scanning confocal microscope (A1; Nikon, Japan) with 60X/1.40NA objective (Nikon). Images of the HaloTag-TMR molecules were obtained at a time resolution of 1 sec. 

\paragraph{\bf Particle model :}
We constructed the particle model of $19,656$ HaloTag-TMR molecules fast diffusing with $100\ {\rm \mu m^2/sec}$ and distributed in $30 \times 30 \times 6\ {\rm \mu m^3}$ box of aqueous solutions.

\paragraph{\bf Simulated imaging :}
We simulated imaging the middle region of the particle model for the specification and condition of the LSCM simulation module shown in Table \ref{tab;lscm_tmr}. 

\begin{table}[!h]
\centering
\begin{tabular}{|l||C{10cm}|}
\hline
Beam flux & $5,\ 10,\ 30,\ 50,\ 100\ {\rm \mu W}$ \\ \hline
Beam wavelength & $512\ {\rm nm}$ \\ \hline
Beam waist & $400\ {\rm nm}$ (Assumed) \\ \hline
Fluorophore & HaloTag TMR ligand\ (${\rm Abs.}\ 555\ {\rm nm} /\ {\rm Em.}\ 585\ {\rm nm}$) \\ \hline
Objective & $\times\ 60\ /\ {\rm N.A.}\ 1.49$ \\ \hline
Scan lens & $\times\ 1$ \\ \hline
Pinhole & $57.6\ {\rm \mu m}$ diameter ($2\ {\rm A.U}$) \\ \hline
Optical magnification & $\times\ 60$ \\ \hline
Linear conversion & $10^{-6}$ \\ \hline
Scan time & $0.95\ {\rm \mu sec/pixel}$ \\ \hline
Pixel length & $207.16\ {\rm nm/pixel}$ \\ \hline
Image size & $1024 \times 1024$ \\ \hline
PMT mode & Photon-counting \\ \hline
A/D Converter  & $12$-bit \\ \hline
Gain & $1.025\ {\rm electrons/count}$ \\ \hline
Offset & $100\ {\rm counts}$ \\ \hline
Readout noise & $0\ {\rm counts/sec}$ \\ \hline
Excess noise & N/A \\ \hline
Optical background & $0$-$5\ {\rm photons}$ \\ \hline
\end{tabular}
\vspace{0.3cm}
\caption{LSCM specifications and condition to image the simple particle model of fluorescent molecules.}
\label{tab;lscm_tmr}
\end{table}

\newpage

\begin{figure}[!h]
  \centering
  	{\bf Expected images} \hspace{5.0cm} {\bf Simulated images}
	
  \vspace{0.1cm}
  \leftline{\bf Beam flux : $5\ {\rm \mu W}$}
  \vspace{0.1cm}
  \centering
	\includegraphics[width=3.4cm]{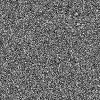}
	\includegraphics[width=4.0cm]{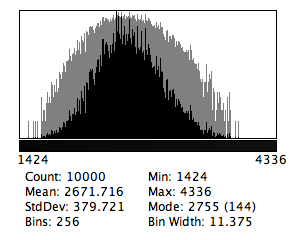}
	\includegraphics[width=3.4cm]{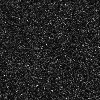}
	\includegraphics[width=4.0cm]{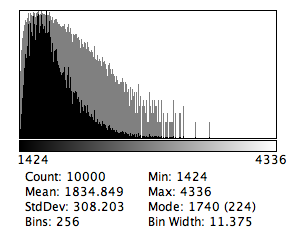}
	
  \vspace{0.1cm}
  \leftline{\bf Beam flux : $10\ {\rm \mu W}$}
  \vspace{0.1cm}
  \centering
	\includegraphics[width=3.4cm]{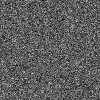}
	\includegraphics[width=4.0cm]{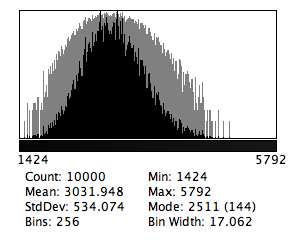}
	\includegraphics[width=3.4cm]{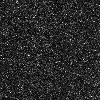}
	\includegraphics[width=4.0cm]{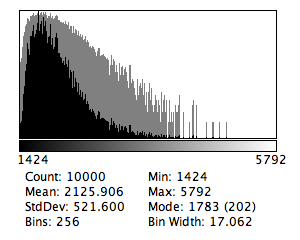}
	
  \vspace{0.1cm}
  \leftline{\bf Beam flux : $30\ {\rm \mu W}$}
  \vspace{0.1cm}
  \centering
	\includegraphics[width=3.4cm]{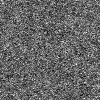}
	\includegraphics[width=4.0cm]{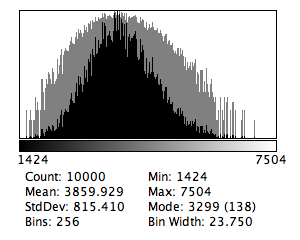}
	\includegraphics[width=3.4cm]{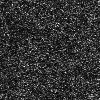}
	\includegraphics[width=4.0cm]{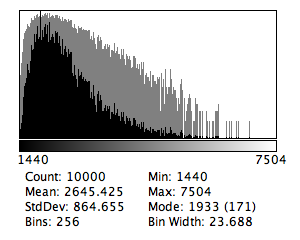}
	
  \vspace{0.1cm}
  \leftline{\bf Beam flux : $50\ {\rm \mu W}$}
  \vspace{0.1cm}
  \centering
	\includegraphics[width=3.4cm]{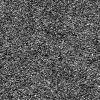}
	\includegraphics[width=4.0cm]{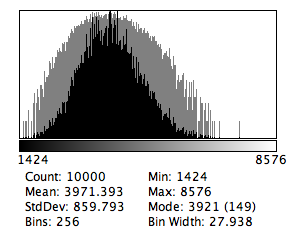}
	\includegraphics[width=3.4cm]{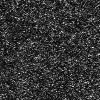}
	\includegraphics[width=4.0cm]{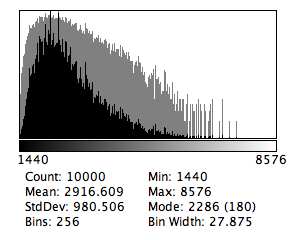}
	
  \vspace{0.1cm}
  \leftline{\bf Beam flux : $100\ {\rm \mu W}$}
  \vspace{0.1cm}
  \centering
	\includegraphics[width=3.4cm]{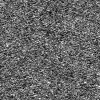}
	\includegraphics[width=4.0cm]{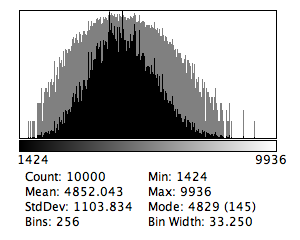}
	\includegraphics[width=3.4cm]{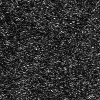}
	\includegraphics[width=4.0cm]{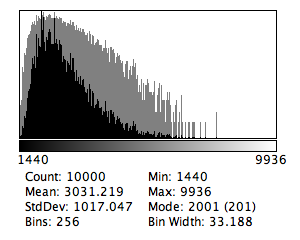}
	
  \caption{Comparison of {\it in vitro} images ($100 \times 100\ {\rm pixels}$) and intensity histograms. Log-scaled intensity histograms are shown in grey color. The PMT dark current has not been simulated yet.}
  \label{fig;comparison_hist_lscm_expected}
\end{figure}

\newpage

\begin{figure}[!h]
  \centering
  	{\bf Simulated images} \hspace{5.0cm} {\bf Actual images}
	
  \vspace{0.1cm}
  \leftline{\bf Beam flux : $5\ {\rm \mu W}$}
  \vspace{0.1cm}
  \centering
	\includegraphics[width=3.4cm]{figures_SI/images_comparison/images_tmr_lscm_sim/image_0000000_005uW.png}
	\includegraphics[width=4.0cm]{figures_SI/images_comparison/images_tmr_lscm_sim/hist_log_0000000_005uW.png}
	\includegraphics[width=3.4cm]{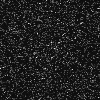}
	\includegraphics[width=4.0cm]{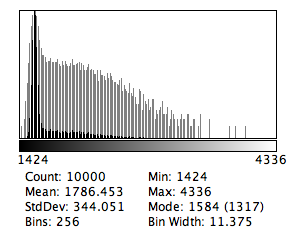}
	
  \vspace{0.1cm}
  \leftline{\bf Beam flux : $10\ {\rm \mu W}$}
  \vspace{0.1cm}
  \centering
	\includegraphics[width=3.4cm]{figures_SI/images_comparison/images_tmr_lscm_sim/image_0000000_010uW.png}
	\includegraphics[width=4.0cm]{figures_SI/images_comparison/images_tmr_lscm_sim/hist_log_0000000_010uW.png}
	\includegraphics[width=3.4cm]{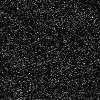}
	\includegraphics[width=4.0cm]{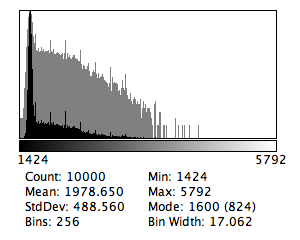}
	
  \vspace{0.1cm}
  \leftline{\bf Beam flux : $30\ {\rm \mu W}$}
  \vspace{0.1cm}
  \centering
	\includegraphics[width=3.4cm]{figures_SI/images_comparison/images_tmr_lscm_sim/image_0000000_030uW.png}
	\includegraphics[width=4.0cm]{figures_SI/images_comparison/images_tmr_lscm_sim/hist_log_0000000_030uW.png}
	\includegraphics[width=3.4cm]{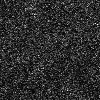}
	\includegraphics[width=4.0cm]{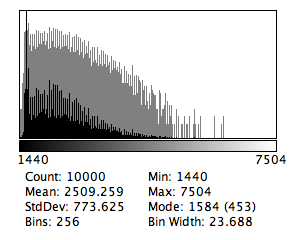}
	
  \vspace{0.1cm}
  \leftline{\bf Beam flux : $50\ {\rm \mu W}$}
  \vspace{0.1cm}
  \centering
	\includegraphics[width=3.4cm]{figures_SI/images_comparison/images_tmr_lscm_sim/image_0000000_050uW.png}
	\includegraphics[width=4.0cm]{figures_SI/images_comparison/images_tmr_lscm_sim/hist_log_0000000_050uW.png}
	\includegraphics[width=3.4cm]{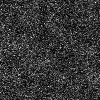}
	\includegraphics[width=4.0cm]{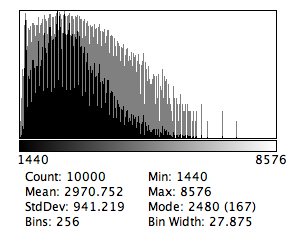}
	
  \vspace{0.1cm}
  \leftline{\bf Beam flux : $100\ {\rm \mu W}$}
  \vspace{0.1cm}
  \centering
	\includegraphics[width=3.4cm]{figures_SI/images_comparison/images_tmr_lscm_sim/image_0000000_100uW.png}
	\includegraphics[width=4.0cm]{figures_SI/images_comparison/images_tmr_lscm_sim/hist_log_0000000_100uW.png}
	\includegraphics[width=3.4cm]{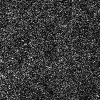}
	\includegraphics[width=4.0cm]{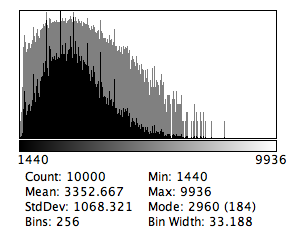}
	
  \caption{Comparison of {\it in vitro} images ($100 \times 100\ {\rm pixels}$) and intensity histograms. Log-scaled intensity histograms are shown in grey color. The PMT dark current has not been simulated yet.}
  \label{fig;comparison_hist_lscm}
\end{figure}

\newpage

\begin{figure}[!h]
  \centering
	\includegraphics[width=12cm]{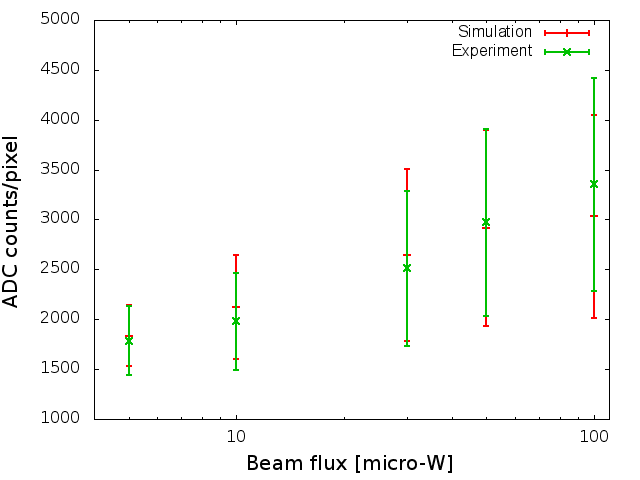}

  \caption{Linearity is shown for various beam flux. Experiment (green) and simulation (red).}
  \label{fig;comparison_linearity_lscm}
\end{figure}

\newpage

\section{Comparison of {\it in vivo} images}
\subsection{ERK nuclear translocation model of EGF signaling pathway}
\paragraph{\bf Cell preparation :}
Rat PC12 pheochromocytoma cells stably expressing mEGFP-tagged ERK2 were provided by Dr. Yasushi Sako, Cellular Informatics Laboratory, RIKEN. Data was taken by Yuki Shindo, laboratory for biochemical simulation, RIKEN QBiC. Cells were plated on poly-L-lysine coated coverslips and cultured for 12 h in Dulbecco's modified Eagle's medium (DMEM) supplemented with $10 \%$ hrs serum and $5 \%$ fetal bovine serum. Then, cells were serum-starved for 16 h in DMEM without fenol-red supplemented with $1 \%$ BSA (DMEM-BSA). Before microcopy experiments, the medium was changed to DMEM-BSA containing 5 mM PIPES (pH $7.2$).

\paragraph{\bf Timelapse imaging :}
mEGFP-ERK2 proteins in living PC12 cells were observed using a laser scanning confocal microscope (A1; Nikon, Japan) with 60X/1.49NA objective (Nikon). Cells were stimulated with epidermal growth facotr (EGF) (5 ng ml$^{-1}$ final concentration) on the microscope at room temperature. Timelapse movies were obtained at a time resolution of 1 min.

\paragraph{\bf Cell model :}
A particle detailed ERK nuclear translocation model of the EGF signalling pathway is constructed by Dr. Kazunari Iwamoto, laboratory for biochemical simulation, RIKEN QBiC. The model consists of 73 chemical species, 144 reactions and 85 kinetic parameters. The EGF signalling pathway regulates cellular proliferation, differentiation and apoptosis \cite{yarden2001}. EGF ligands bind to EGF receptors, which are dimerized and subsequently autophosphorylated. Adaptor proteins, Shc and Grb2, bind to the phosphorylated receptors to form a signalling complex. Sos binds to the signaling complex and then promotes the Ras-GDP/Ras-GTP exchange \cite{corbalan1998}. Although both Ras-GDP and Ras-GTP bind to Raf protein at cellular membrane, only Ras-GTP can activate Raf \cite{hibino2011}. Activated Raf doubly phosphorylates and activates MEK at cytoplasm. Active MEK also doubly phosphorylates ERK, followed by the translocation of phosphorylated ERK from cytoplasm into nucleus \cite{fujioka2006, cohen2009}. Phosphorylated ERK negatively regulates the signaling complex through the phosphorylation of Sos \cite{sturm2010}. We simulated the cell model using the Spatiocyte method. 


\paragraph{\bf Simulated imaging :}
We simulated imaging the middle region of the cell model for the specification and condition of the LSCM simulation module shown in Table \ref{tab;lscm_egf}.

\newpage

\begin{table}[!h]
\centering
\begin{tabular}{|l||C{10cm}|}
\hline
Beam flux & $10\ {\rm \mu W}$ (Assumed) \\ \hline
Beam wavelength & $488\ {\rm nm}$ \\ \hline
Beam waist & $200\ {\rm nm}$ (Assumed) \\ \hline
Fluorophore & mEGFP (${\rm Abs.}\ 484\ {\rm nm} /\ {\rm Em.}\ 507\ {\rm nm}$) \\ \hline
Objective & $\times\ 60\ /\ {\rm N.A.}\ 1.49$ \\ \hline
Scan lens & $\times\ 1$ \\ \hline
Pinhole & $57.6\ {\rm \mu m}$ diameter ($2\ {\rm A.U}$) \\ \hline
Optical magnification & $\times\ 60$ \\ \hline
Linear conversion & $10^{-6}$ \\ \hline
Scan time & $1.15\ {\rm \mu sec/pixel}$ \\ \hline
Pixel length & $207.16\ {\rm nm/pixel}$ \\ \hline
Image size & $1024 \times 1024$ \\ \hline
PMT mode & Analog \\ \hline
A/D Converter  & $12$-bit \\ \hline
QE & $30\ \%$ \\ \hline
Gain & $\times 10^6$ \\ \hline
Dynode & $11$ stages \\ \hline
Readout noise & $0\ {\rm mA}$ \\ \hline
Excess noise & $1.1$ \\ \hline
Optical background & $0\ {\rm photons/sec}$ \\ \hline
\end{tabular}

\vspace{0.3cm}
\caption{LSCM specifications and condition to image the ERK nuclear translocation model of EGF signaling pathway.}
\label{tab;lscm_egf}
\end{table}

\newpage

\subsection{Self-organizing wave model for the chemotactic pathway of {\it D. discoideum}}
\paragraph{\bf Cell preparation :}
{\it Dictyostelium discoideum} cells were provided by Dr. Masahiro Ueda, laboratory for cell signaling dynamics, RIKEN QBiC. Data was taken by Seiya Fukushima, Graduate School of Frontier Bioscience, Osaka University. Cell preparation and growth conditions were described in ref. \cite{arai2010, shibata2012}. 

\paragraph{\bf Timelapse imaging :}
PTEN-TMR and PH-EGFP in living {\it Dictyostelium discoiduem} cells were observed using a laser scanning confocal microscope (A1; Nikon, Japan) with 60X/1.49NA objective (Nikon). Images of PH-EGFP and PTEN-TMR-expressing cells were obtained at a time resolution of 5 sec. 

\paragraph{\bf Cell model :}
{\it Dictyostelium discoideum} migrates toward the elevated side of 3'-5'-cyclic adenosine monophosphate (cAMP) external gradient by extending pseudopodia. The accumulation of phosphatidylinositol 3,4,5-trisphosphate (PIP3) lipid and F-actin at the leading edge of the cell is necessary for the pseudopodia formation. When F-actin polymerization is inhibited in the absence of chemoattractant, the cells maintain their disc-like shape without triggering protrusions. Despite the absence of F-actin membrane accumulation, self-organized waves of PIP3 are spontaneously generated on the membrane of these cells. The waves are regulated by phosphatase and tensin homolog (PTEN) and phosphoinositide-3-kinase (PI3K). A detailed particle model of the waves was constructed by Dr. Satya N. V. Arjunan, laboratory for biochemical simulation, RIEKN QBiC. The model consists of 8 chemical species, 12 reactions and 17 kinetic parameters. On the membrane, PI3K phosphorylates phosphatidylinositol 3,4,5-biphosphate (PIP2) into PIP3, whereas PTEN dephosphorylates PIP3 into PIP2. Cytosolic PTEN is recruited to the membrane regions containing PIP2. Nonetheless, PIP3 can dislodge PTEN from PIP2 into the cytosol when it comes in contact. This last reaction acts a positive feedback for PIP3 accumulation. 


\paragraph{\bf Simulated imaging :}
We simulated imaging the middle region of the cell model for the specification and condition of the LSCM simulation module are shown in Table \ref{tab;lscm_dicty}.

\newpage

\begin{table}[!h]
\centering
\begin{tabular}{|l||C{10cm}|}
\hline
Beam flux 1 & $10\ {\rm \mu W}$ (Assumed) \\ \hline
Beam wavelength 1 & $488\ {\rm nm}$ \\ \hline
Beam waist 1 & $200\ {\rm nm}$ (Assumed) \\ \hline
Fluorophore 1 & EGFP\ (${\rm Abs.}\ 384\ {\rm nm} /\ {\rm Em.}\ 509\ {\rm nm}$) \\ \hline
Beam flux 2 & $10\ {\rm \mu W}$ \\ \hline
Beam wavelength 2 & $561\ {\rm nm}$ \\ \hline
Beam waist 2 & $200\ {\rm nm}$ (Assumed) \\ \hline
Fluorophore 2 & TRITC\ (${\rm Abs.}\ 584\ {\rm nm} /\ {\rm Em.}\ 608\ {\rm nm}$) \\ \hline
Objective & $\times\ 60\ /\ {\rm N.A.}\ 1.49$ \\ \hline
Scan lens & $\times\ 1$ \\ \hline
Pinhole & $37\ {\rm \mu m}$ diameter ($2\ {\rm A.U}$) \\ \hline
Optical magnification & $\times\ 60$ \\ \hline
Linear conversion & $10^{-6}$ \\ \hline
Scan time & $4.27\ {\rm \mu sec/pixel}$ \\ \hline
Pixel length & $414.3\ {\rm nm/pixel}$ \\ \hline
Image size & $512 \times 512$ \\ \hline
Detector & PMT : Analog mode \\ \hline
A/D Converter  & $12$-bit \\ \hline
QE & $30\ \%$ \\ \hline
Readout noise & $0\ {\rm mA}$ \\ \hline
Gain & $\times 10^6$ \\ \hline
Dynode & $11$ stages\\ \hline
Excess noise & $1.1$ \\ \hline
Optical background & $0.00\ {\rm photons/pixel}$ \\ \hline 
\end{tabular}
\vspace{0.3cm}
\caption{2-color LSCM specifications and condition to image the self-organizing wave model of {\it Dictyostelium discoiduem} cell.}
\label{tab;lscm_dicty}
\end{table}

%
%
%
%

\end{document}